\documentclass[a4paper,noindent]{JHEP3}
\usepackage[dvips]{graphicx}
\usepackage{amsmath}
\usepackage{multirow}
\usepackage{mathenv}
\usepackage{feynarts}
\usepackage{array}
\usepackage{cite}
\usepackage{color}
\let\normalcolor\relax

\newcommand{\be}{\begin{eqnarray*}}
\newcommand{\ee}{\end{eqnarray*}}
\newcommand{\bee}{\begin{eqnarray}}
\newcommand{\eee}{\end{eqnarray}}
\renewcommand{\eqref}[1]{Eq.~(\ref{#1})}
\newcommand{\eqsref}[2]{Eqs.~(\ref{#1}) and (\ref{#2})}
\newcommand{\figref}[1]{Figure~\ref{#1}}

\newcommand{\cf}{cf.\/}
\newcommand{\eg}{e.\,g.\/}
\newcommand{\ie}{i.\,e.\/}

\newcommand{\alphas}{\alpha_s}
\newcommand{\Li}[1]{\text{Li}_2\left(#1\right)}
\newcommand{\pT}{{\rm p}_{\rm T}}

\def \cm{c.\,m.\,}

\def \mm{\mathcal{M}}
\def \Order{\mathcal{O}}

\def \msbar{\overline{\text{MS}}}

\renewcommand{\Re}{\mbox{Re}}

\newcommand{\squark}{\tilde{q}}
\newcommand{\Squark}{\tilde{q}}
\newcommand{\Squarka}{\tilde{q}_{\alpha}}
\newcommand{\Squarkb}{\tilde{q}_{\beta}}
\newcommand{\squarka}{\tilde{q}_{\alpha}}
\newcommand{\squarkb}{\tilde{q}_{\beta}}
\newcommand{\Sup}{\tilde{u}}
\newcommand{\Sdown}{\tilde{d}}
\newcommand{\Scharm}{\tilde{c}}
\newcommand{\Sstrange}{\tilde{s}}
\newcommand{\Stop}{\tilde{t}}

\newcommand{\cha}{\tilde{\chi}^{\pm}}
\newcommand{\neu}{\tilde{\chi}^{0}}
\newcommand{\gluino}{\tilde{g}}
\newcommand{\GeV}{{\rm GeV}}

\newcommand{\red}[1]{\textcolor{red}{#1}}
\newcommand{\blue}[1]{\textcolor{blue}{#1}}

\def\Dzero{D\O\ }

\newcommand{\trule}{\rule[-1.5mm]{0mm}{6mm}}

\title{Hadronic production of squark--squark pairs: \\ The electroweak contributions}
\author{Jan Germer, Wolfgang Hollik\\
Max-Planck-Institut f\"ur Physik, F\"ohringer Ring 6, D-80805 M\"unchen, Germany\\
Email: \email{germer@mppmu.mpg.de, hollik@mppmu.mpg.de}}
\author{Edoardo Mirabella\\
Institut de Physique Th\'eorique, CEA-Saclay, F-91191, Gif-sur-Yvette
cedex, France \\
Email: \email{edoardo.mirabella@cea.fr}}
\author{Maike K. Trenkel \\
University of Wisconsin, Madison, WI, 53706, USA\\
Email: \email{trenkel@hep.wisc.edu}}

\abstract{We compute the electroweak (EW) contributions to
squark--squark pair production processes at the LHC within the
framework of the Minimal Supersymmetric Standard Model (MSSM). Both
tree-level EW contributions, of $\Order(\alpha_s \alpha + \alpha^2)$,
and next-to-leading order (NLO) EW corrections, of $\Order(\alpha_s^2
\alpha)$, are calculated.
Depending on the flavor and chirality of the produced quarks, many
interferences between EW-mediated and QCD-mediated diagrams give
non-zero contributions at tree-level and NLO.  We discuss the
computational techniques and present an extensive numerical analysis
for inclusive squark--squark production as well as for subsets and
single processes.  While the tree-level EW contributions to the
integrated cross sections can reach the $20\%$ level, the NLO EW
corrections typically lower the LO prediction by a few percent.}

\preprint{MPP-2010-1\\IPhT-t10/007}

\begin{document}

\section{Introduction}
\label{sect_Intro}

Supersymmetry (SUSY)~\cite{Wess:1974tw} is one of the most appealing
scenarios for physics beyond the Standard Model (SM). Being proposed
as the only nontrivial extension to the space-time symmetry, it was
found to have many nice phenomenological properties which overcome
weaknesses of the SM. With SUSY particles at the TeV scale or below,
the technical part of the hierarchy problem is solved by stabilizing
the electroweak scale, in particular the mass of the Higgs boson. In
addition, many supersymmetric extensions of the SM offer a dark matter
candidate that gives the observed dark matter relict
density~\cite{Dunkley:2008ie}. Of particular interest is the Minimal
Supersymmetric Standard Model
(MSSM)~\cite{Nilles:1983ge,Haber:1984rc,Barbieri:1987xf} which
provides a fit to electroweak precision observables and B--physics
data with a $\chi^2$ comparable to the SM and a naturally light Higgs
boson~\cite{Ellis:2007fu,Buchmueller:2007zk,Buchmueller:2009fn} and
can explain the measured value of the anomalous magnetic
moment~\cite{Bennett:2002jb,Bennett:2004pv}.\par

If supersymmetry is realized at the TeV scale, it will be probed at
the Large Hadron Collider (LHC) at CERN. Indeed, the $95\%$ C.L.
area of the $(m_0,m_{1/2})$ plane of the Constrained MSSM (CMSSM),
lies largely within the region that can be explored within $1\text{fb}^{-1}$ of integrated luminosity~\cite{Buchmueller:2008qe}. Within the framework of the MSSM, imposing R-parity conservation, SUSY particles can only be produced in pairs.
Among the potential SUSY discovery channels, the
direct production of pairs of color-charged SUSY particles is of particular importance 
at hadron colliders since it proceeds via the strong interaction. 
Many searches for squarks and gluinos have thus already been performed at high-energy colliders.
Results of the \Dzero and CDF collaboration can be
found in e.g.~\cite{Portell:2007zz}. Studies for the LHC are based on 
 Monte Carlo simulations. It has been shown that there is the possibility of early SUSY
discovery within
$1\text{fb}^{-1}$ of data in the inclusive jets plus missing energy channel, provided that the SUSY particles are not too heavy
\cite{deJong:2008uj}.\par

First theoretical cross section predictions for squark and gluino pair production processes
  based on leading order (LO) calculations  were made already many years ago~\cite{Harrison:1982yi,Reya:1984yz,Dawson:1983fw,Baer:1985xz}.
Later calculations of next-to-leading order (NLO) in perturbative QCD~\cite{Beenakker:1996ch, Beenakker:1997ut} 
could reduce theoretical uncertainties considerably and revealed corrections of typically $20-30\%$. 
Recently also results beyond the one-loop level in QCD have become
available~\cite{Langenfeld:2009eg,Kulesza:2008jb,Kulesza:2009kq,Beenakker:2009ha},
increasing the cross section by another $2-10\%$, and stabilizing the
prediction considerably.

\medskip
For a reliable cross section prediction, also electroweak
(EW) contributions have to be taken into account which are of the same
size as the NNLO QCD contributions. The contributing processes are
manifold and their interplay is nontrivial, in particular if not only
QCD-mediated but also EW-mediated production channels exist at
tree-level.

The latter arise from $q \bar q$ annihilation or $qq$ scattering, as well as photon-induced 
processes, and contribute at $\mathcal{O}(\alphas\alpha+\alpha^2)$. 
For squark--(anti)-squark production processes, the $q \bar q$ ($qq$)-channels can rise the 
LO cross section by up to
$20\%$ \cite{Bornhauser:2007bf,Arhrib:2009sb}.
The photon-induced channels are typically more important for the pair production 
of lighter stops~\cite{Hollik:2007wf}, and have a reduced impact on 
squark--anti-squark~\cite{Hollik:2008yi} and gluino--squark production~\cite{Hollik:2008vm}.

NLO EW corrections contribute at $\mathcal{O}(\alphas^2\alpha)$ and
have been investigated for
stop--anti-stop~\cite{Hollik:2007wf,Beccaria:2008mi},
squark--anti-squark~\cite{Hollik:2008yi},
gluino--squark~\cite{Hollik:2008vm} and
gluino--gluino~\cite{Mirabella:2009ap} production processes. In this
paper, we provide the yet-missing NLO EW corrections to squark--squark
production.
We give details on the NLO computation and present an elaborate
numerical analysis of all squark--squark production processes,
including the anti-particles.

In the context of all squark and gluino production processes,
squark--squark~production is of particular interest at the
proton-proton collider LHC.  The partonic process proceeds at LO from
$qq$-induced diagrams only.  Squark--anti-squark and
gluino--gluino~production require $q\bar{q}$ or $gg$~initial states
instead. Since the final-state SUSY particles are very massive, an
important contribution to the hadronic cross sections arises from the
high-$x$ region where valence-quark densities dominate.
As a result, squark--squark~production
has generally a higher tree-level yield than
squark--anti-squark~production and can be comparable to
gluino--gluino~production depending on the precise squark--gluino mass
configuration.  


The outline of this paper is as follows: In section~\ref{sect_tree} we
review the various tree-level contributions to squark--squark
production and introduce some notation used throughout this
paper. Section~\ref{sect_NLO} shows the details to the NLO calculation
of $\mathcal{O}(\alphas^2 \alpha)$ and the strategy of the
calculation. In Section~\ref{sect_results} we list the input
parameters used in our numerical analysis and show hadronic cross
sections and distributions for squark--squark production in proton--proton collisions at the 
LHC. Different SUSY scenarios are considered and a scan over squark
and gluino masses is performed. Analytic formulas for the tree-level
cross section and expressions for the cross section in the soft and
collinear singular region, as well as the Feynman diagrams 
for the NLO calculation are collected in the Appendix.

\section{Classification of processes and tree-level cross sections}
\label{sect_tree}

We consider the pair production of two squarks or two anti-squarks,
\bee
   PP \rightarrow \Squarka \Squarkb^{\prime}, \quad
   PP \rightarrow \Squarka^* \Squarkb^{\prime*}, \qquad\qquad
  q,q' = \{u,d,c,s\};
    \hspace*{-3cm}
\label{eq_process}
\eee
where $\alpha, \beta =\{ L,R \}$ label the chirality of the
squarks, neglecting left-right mixing. At lowest order in QCD there is only 
one partonic channel for each process,
\begin{align}
  q(p_1) ~q'(p_2) ~&\rightarrow ~ \Squarka(p_3) ~\Squarkb^{\prime}(p_4), \qquad
  \bar{q}(p_1) ~\bar{q}'(p_2) ~\rightarrow ~\Squarka^*(p_3) ~\Squarkb^{\prime*}(p_4), \quad
\label{eq_LOmom}
\end{align}
where the initial-state quarks and the final-state squarks have to have the same 
flavor.  We thus do not consider the production of top (bottom)
squarks due to the vanishing (small) density of the corresponding
quarks inside the proton. 
Moreover, b--squark production has specific different features and will
be discussed separately.
The unpolarized cross sections for squark--squark and anti-squark--anti-squark 
production are related by charge-conjugation. In the following we will refer 
to squark--squark production only, while the charge conjugated processes 
are properly taken into account in the numerical results. 
Since the electroweak interaction is sensitive to flavor and chirality,
one has to treat processes with final-state squarks of different chiralities
or of different isospin separately, even in the limit of degenerate
squark masses.
CKM mixing effects are neglected in our discussion.

\medskip

\FIGURE[t]{
  \begin{tabular}{m{2.5cm}|cm{.5cm}l}
    Class & QCD diagram(s) && ~~EW diagram(s) \\
    \hline
    \parbox{2.5cm}{\vspace{-1.5cm}${{PP} \rightarrow \squark_\alpha \squark_\beta}$
    \\{\footnotesize same flavor}}
    &
    \begin{feynartspicture}(100,50)(2,1)
      \FADiagram{}
      \FAProp(0.,15.)(10.,15.)(0.,){/Straight}{1}
      \FAProp(0.,5.)(10.,5.)(0.,){/Straight}{1}
      \FAProp(20.,15.)(10.,15.)(0.,){/ScalarDash}{-1}
      \FAProp(20.,5.)(10.,5.)(0.,){/ScalarDash}{-1}
      \red{\FAProp(10.,15.)(10.,5.)(0.,){/Straight}{0}
        \FAProp(9.7,15.)(9.7,5.)(0.,){/Cycles}{0}
        \FALabel(9.18,10.)[r]{${\scriptstyle \tilde g}$}}
      \FAVert(10.,15.){0}
      \FAVert(10.,5.){0}
      
      \FADiagram{}
      \FAProp(0.,15.)(10.,15.)(0.,){/Straight}{1}
      \FAProp(0.,5.)(10.,5.)(0.,){/Straight}{1}
      \FAProp(14.,9.)(10.,5.)(0.,){/ScalarDash}{0}
      \FAProp(14.,11.)(10.,15.)(0.,){/ScalarDash}{0}
      \FAProp(14.,9.)(20.,15.)(0.,){/ScalarDash}{1}
      \FAProp(14.,11.)(20.,5.)(0.,){/ScalarDash}{1}
      \red{\FAProp(10.,15.)(10.,5.)(0.,){/Straight}{0}
        \FAProp(9.7,15.)(9.7,5.)(0.,){/Cycles}{0}
        \FALabel(7.,10.)[r]{${\scriptstyle \tilde g}$}}
      \FAVert(10.,15.){0}
      \FAVert(10.,5.){0}
    \end{feynartspicture}
    &\parbox{.5cm}{\vspace{-1.5cm}~+}&
    \begin{feynartspicture}(100,50)(2,1)
      \FADiagram{}
      \FAProp(0.,15.)(10.,15.)(0.,){/Straight}{1}
      \FAProp(0.,5.)(10.,5.)(0.,){/Straight}{1}
      \FAProp(20.,15.)(10.,15.)(0.,){/ScalarDash}{-1}
      \FAProp(20.,5.)(10.,5.)(0.,){/ScalarDash}{-1}
      \blue{\FAProp(10.,15.)(10.,5.)(0.,){/Straight}{0}
        \FAProp(10.,15.)(10.,5.)(0.,){/Sine}{0}
        \FALabel(9.18,10.)[r]{${\scriptstyle \tilde \chi^0}$}}
      \FAVert(10.,15.){0}
      \FAVert(10.,5.){0}
      
      \FADiagram{}
      \FAProp(0.,15.)(10.,15.)(0.,){/Straight}{1}
      \FAProp(0.,5.)(10.,5.)(0.,){/Straight}{1}
      \FAProp(14.,9.)(10.,5.)(0.,){/ScalarDash}{0}
      \FAProp(14.,11.)(10.,15.)(0.,){/ScalarDash}{0}
      \FAProp(14.,9.)(20.,15.)(0.,){/ScalarDash}{1}
      \FAProp(14.,11.)(20.,5.)(0.,){/ScalarDash}{1}
      \blue{\FAProp(10.,15.)(10.,5.)(0.,){/Straight}{0}
        \FAProp(10.,15.)(10.,5.)(0.,){/Sine}{0}
        \FALabel(8.,10.)[r]{${\scriptstyle \tilde \chi^0}$}}
      \FAVert(10.,15.){0}
      \FAVert(10.,5.){0}
    \end{feynartspicture}\\
    \parbox{2.5cm}{\vspace{-1.5cm}${{ PP} \rightarrow \squark_{\alpha}
        \squark_{\beta}^\prime}$\\
      {\footnotesize different flavor,\\[-1ex]same doublet}}
    &
    \begin{feynartspicture}(50,50)(1,1)
      \FADiagram{}
      \FAProp(0.,15.)(10.,15.)(0.,){/Straight}{1}
      \FAProp(0.,5.)(10.,5.)(0.,){/Straight}{1}
      \FAProp(20.,15.)(10.,15.)(0.,){/ScalarDash}{-1}
      \FAProp(20.,5.)(10.,5.)(0.,){/ScalarDash}{-1}
      \red{\FAProp(10.,15.)(10.,5.)(0.,){/Straight}{0}
        \FAProp(9.7,15.)(9.7,5.)(0.,){/Cycles}{0}
        \FALabel(9.18,10.)[r]{${\scriptstyle \tilde g}$}}
      \FAVert(10.,15.){0}
      \FAVert(10.,5.){0}
    \end{feynartspicture}
    &\parbox{.5cm}{\vspace{-1.5cm}~+}&
    \begin{feynartspicture}(100,50)(2,1)
      \FADiagram{}
      \FAProp(0.,15.)(10.,15.)(0.,){/Straight}{1}
      \FAProp(0.,5.)(10.,5.)(0.,){/Straight}{1}
      \FAProp(20.,15.)(10.,15.)(0.,){/ScalarDash}{-1}
      \FAProp(20.,5.)(10.,5.)(0.,){/ScalarDash}{-1}
      \blue{\FALabel(9.18,10.)[r]{${\scriptstyle \tilde \chi^0}$}
        \FAProp(10.,15.)(10.,5.)(0.,){/Straight}{0}
        \FAProp(10.,15.)(10.,5.)(0.,){/Sine}{0}}
      \FAVert(10.,15.){0}
      \FAVert(10.,5.){0}
      
      \FADiagram{}
      \FAProp(0.,15.)(10.,15.)(0.,){/Straight}{1}
      \FAProp(0.,5.)(10.,5.)(0.,){/Straight}{1}
      \FAProp(14.,9.)(10.,5.)(0.,){/ScalarDash}{0}
      \FAProp(14.,11.)(10.,15.)(0.,){/ScalarDash}{0}
      \FAProp(14.,9.)(20.,15.)(0.,){/ScalarDash}{1}
      \FAProp(14.,11.)(20.,5.)(0.,){/ScalarDash}{1}
      \blue{\FAProp(10.,15.)(10.,5.)(0.,){/Straight}{0}
        \FAProp(10.,15.)(10.,5.)(0.,){/Sine}{0}
        \FALabel(8.,10.)[r]{${\scriptstyle \tilde \chi^\pm}$}}

      \FAVert(10.,15.){0}
      \FAVert(10.,5.){0}
    \end{feynartspicture}\\
    \parbox{2.5cm}{\vspace{-1.5cm}${{ PP} \rightarrow \squark_\alpha
          \squark_\beta^{\prime}}$\\
        {\footnotesize different flavor,\\[-1ex]different doublet}}
    &
    \begin{feynartspicture}(50,50)(1,1)
      \FADiagram{}
      \FAProp(0.,15.)(10.,15.)(0.,){/Straight}{1}
      \FAProp(0.,5.)(10.,5.)(0.,){/Straight}{1}
      \FAProp(20.,15.)(10.,15.)(0.,){/ScalarDash}{-1}
      \FAProp(20.,5.)(10.,5.)(0.,){/ScalarDash}{-1}
      \red{\FAProp(10.,15.)(10.,5.)(0.,){/Straight}{0}
        \FAProp(9.7,15.)(9.7,5.)(0.,){/Cycles}{0}
        \FALabel(9.18,10.)[r]{${\scriptstyle \tilde g}$}}
      \FAVert(10.,15.){0}
      \FAVert(10.,5.){0}
    \end{feynartspicture}
    &\parbox{.5cm}{\vspace{-1.5cm}~+}&
    \begin{feynartspicture}(50,50)(1,1)
      \FADiagram{}
      \FAProp(0.,15.)(10.,15.)(0.,){/Straight}{1}
      \FAProp(0.,5.)(10.,5.)(0.,){/Straight}{1}
      \FAProp(20.,15.)(10.,15.)(0.,){/ScalarDash}{-1}
      \FAProp(20.,5.)(10.,5.)(0.,){/ScalarDash}{-1}
      \blue{\FAProp(10.,15.)(10.,5.)(0.,){/Straight}{0}
        \FAProp(10.,15.)(10.,5.)(0.,){/Sine}{0}
        \FALabel(9.18,10.)[r]{${\scriptstyle \tilde \chi^0}$}}
      \FAVert(10.,15.){0}
      \FAVert(10.,5.){0}
    \end{feynartspicture}
    \parbox{10pt}{\vspace{-1.5cm}\bigg(}
    \begin{feynartspicture}(50,50)(1,1)
      \FADiagram{}
      \FAProp(0.,15.)(9.,15.)(0.,){/Straight}{1}
      \FAProp(0.,5.)(9.,5.)(0.,){/Straight}{1}
      \FAProp(17.,15.)(9.,15.)(0.,){/ScalarDash}{-1}
      \FAProp(17.,5.)(9.,5.)(0.,){/ScalarDash}{-1}
      \blue{\FAProp(9.,15.)(9.,5.)(0.,){/Straight}{0}
        \FAProp(9.,15.)(9.,5.)(0.,){/Sine}{0}
        \FALabel(8.18,10.)[r]{${\scriptstyle \tilde \chi^\pm}$}}
      \FAVert(9.,15.){0}
      \FAVert(9.,5.){0}
    \end{feynartspicture}
    \hspace*{-1.5ex}\parbox{10pt}{\vspace{-1.5cm}\bigg)}
  \end{tabular}
  \caption{Parton-level Feynman diagrams for the three classes of
    squark-squark production at tree-level, where $\alpha, \beta =\{
    L,R \}$.  The first class describes the production of two squarks
    of the same flavor, the second class that of two squarks of the
    same isospin doublet (but different flavor) and the third class
    refers to the production of two squarks belonging to different
    isospin doublets.  
	In the third class, the
    subprocess in brackets cannot interfere with other diagrams due to
    different initial state particles.
  In all three classes, the final-state squarks are of the same generation as
  the initial-state quarks. 
\label{tab_classes}}}


In total we distinguish 36 processes, resulting from the various
 combinations of squarks of different flavor or chirality in
the final state. They can be classified as follows:
\begin{subequations}
\begin{align}
\begin{split}
  - \text{production of two squarks } & \text{of the same flavor}, \\
     PP \rightarrow~
    \Sup_\alpha \Sup_\beta,~  \Sdown_\alpha \Sdown_\beta, &~ 
    \Scharm_\alpha \Scharm_\beta,~ \Sstrange_\alpha \Sstrange_\beta,
  \qquad\qquad~ \{\alpha \beta\} = \{LL,~RR,~LR\}.
\end{split}
\\
\begin{split}
  - \text{production of two squarks } & \text{%
    belonging to the same SU(2) doublet},\\
    PP\rightarrow~ 
    \Sup_{\alpha} \Sdown_{\beta},~ \Scharm_{\alpha} \Sstrange_{\beta},&
  \qquad\qquad\qquad\qquad\quad~~  \{\alpha\beta\} = \{LL,~RR,~LR,~RL\}.
\end{split}
\\
\begin{split}
  -\text{production of two squarks } & \text{in different SU(2)
    doublets},\\
    PP \rightarrow~ 
    \Sup_\alpha \Scharm_\beta,~ \Sup_\alpha \Sstrange_\beta, &~
    \Sdown_\alpha \Scharm_\beta,~ \Sdown_\alpha \Sstrange_\beta,
  \qquad\qquad  \{\alpha\beta\} = \{LL,~RR,~LR,~RL\}.
\end{split}
\end{align}
\label{eq_threeclasses}
\end{subequations}
The corresponding tree-level diagrams of both QCD and EW origin are listed in \figref{tab_classes}.
QCD diagrams are of $\Order(\alpha_s)$, mediated by gluino
exchange. EW diagrams are of $\Order(\alpha)$ and mediated by
neutralino or chargino exchange. Quarks and squarks are of the same
flavor, also in the EW diagrams. The only exception is given by the
two pure-EW chargino-mediated subprocesses $ud \to \Sdown_L \Scharm_L$
and $cd \to \Sup_L \Sstrange_L$ belonging to the third class, which
contribute to $\Sdown_L \Scharm_L$ and $\Sup_L \Sstrange_L$ final
states, respectively. Note that only $t$- and $u$-channel diagrams are present,
but no $s$-channel diagrams.

The appearance of both $t$- and $u$-channel diagrams for
chirality-diagonal $\Squarka\Squarka'$ production gives rise to nonzero
interferences between QCD and EW diagrams already at tree-level.%
\footnote{In the non-diagonal case, $\Squark_L \Squark_R'$ production,
  the interference contributions vanish as a consequence of the
  trivial squark mixing matrices in the limit of no L-R mixing, see
  also the discussion in Appendix~\ref{sect_App_treelevel}.}  The full
tree-level contributions to the cross section are thus given by the
$\mathcal{O}(\alphas^2)$ Born contribution and the
$\mathcal{O}(\alphas\alpha+\alpha^2)$ EW contributions.
Photon-induced squark--squark production is not possible at lowest
order from charge and color conservation.

\medskip
To keep track of the corresponding order in
perturbation theory of the various contributions,
we introduce the notation $d\hat{\sigma}^{a,\,b}$ $[\mm^{a,\,b}]$ 
in order to refer to the cross section [matrix element]  
at a given order $\Order(\alpha_s^a \alpha^b)$ in the strong and electroweak 
couplings, respectively.
Results are given in terms of the Mandelstam variables, defined as usual,
\begin{align}
    \hat s = (p_1 + p_2)^2, &\quad  \hat t = (p_1 - p_3)^2, \quad \hat u = (p_1 - p_4)^2.
\end{align}

The differential partonic cross section for a given subprocess $qq' \to \squarka\squarkb'$ at LO
can thus be written as
\begin{align}
    d\hat{\sigma}^{2,\,0}(\hat s)
  = 
    \overline{\sum}  \Bigl\lvert{\mm}^{1,\,0} \Bigr\rvert^2\, 
    \frac{d\hat{t}}{16 \pi \hat s^2}\,,
\label{eq_treeQCD}
\end{align}
in terms of the squared lowest-order matrix element, $\mm^{1,\,0}$,
averaged (summed) over initial (final) state spin and color.
Similarly, the pure EW differential cross section of $\Order(\alpha^2)$ 
and the EW--QCD $\Order(\alpha_s \alpha)$ interference contribution are given by
\begin{subequations}
\begin{align}
 d\hat{\sigma}^{0,\,2}(\hat s)
  &= \overline{\sum} 
    \Bigl\lvert{\mm}^{0,\,1} \Bigr\rvert^2\, 
    \frac{d\hat{t}}{16 \pi \hat s^2}\,,
\label{eq_treeEWEW}
\\
 d\hat{\sigma}^{1,\,1}(\hat s)
 & = 
	 \overline{\sum} 2\; \Re \Big\lbrace
        \Bigl({\cal M}^{0,\,1}\Bigr)^*
	\, {\cal M}^{1,\,0} \Big\rbrace \,
    \frac{d\hat{t}}{16 \pi \hat s^2}\,,
\label{eq_treeEWQCD}
\end{align}
\label{eq_treeEW}
\end{subequations}
where $\mm^{0,\,1}$ denotes the EW tree-level amplitude.
Explicit expressions for the squared matrix elements are given
in \cite{Beenakker:1996ch, Bornhauser:2007bf}. For completeness, 
we include a list of all tree-level differential cross sections in
Appendix~\ref{sect_App_treelevel}, 
correcting in particular a wrong color factor of \cite{Bornhauser:2007bf}.

The hadronic cross sections are obtained
from the partonic cross sections by convolution with the respective parton
luminosity function. At $\Order(\alpha_s^a \alpha^b)$, it is given by
\bee
  d\sigma^{a\,,b}(S) &=& 
  \int_{\tau_0}^1 d\tau \; \frac{dL_{qq^\prime}}{d\tau}
  d\hat{\sigma}^{a,\,b}(\hat s),
\label{eq_Lumi}
\\[1ex]
\text{with} \qquad
  \frac{dL_{qq^\prime}}{d\tau}&=&  \frac{1}{1+\delta_{qq^\prime}}
  \int_\tau^1   \frac{dx}{x} 
  \left[ f^A_q\left(\frac{\tau}{x},\mu_F\right)\, f^B_{q^\prime}(x, \mu_F) + 
        f^A_q(x, \mu_F)\,  f^B_{q^\prime}\left(\frac{\tau}{x}, \mu_F\right)\right].
\nonumber
\eee
Here $\tau_0= (m_{\squarka} + m_{\squarkb'})^2/S$ is the production threshold, 
determined by the masses of the two squarks  $m_{\squarka}$ and  $m_{\squarkb'}$.
The parton distribution functions (PDFs)  $f^{A}_q(x, \mu_F)$ 
give the probability to find a parton $q$ with momentum
fraction $x$ inside hadron $A$ at a factorization scale $\mu_F$. 
At the LHC, both hadrons $A,B$ are protons $P$.
 $S$ and $\hat{s}=\tau S$ are the squared
center-of-mass (\cm) energies of the hadronic and partonic processes,
respectively.

\section{Virtual and real corrections of
  $\boldsymbol{\mathcal{O}(\alphas^2\alpha)}$}
\label{sect_NLO}

At $\mathcal{O}(\alphas^2\alpha)$, squark--squark production gets
contributions from virtual corrections, real photon- and gluon
emission, as well as real quark radiation. Ultraviolet (UV) as well as
infrared (IR) and collinear singularities arise in the one-loop
diagrams, see Section~\ref{subsec_virt}.  The IR singularities cancel
in sufficiently inclusive observables once virtual and real photon and
gluon bremsstrahlung corrections are added (see
Section~\ref{subsec_real}).  Remaining collinear singularities are
universal and can be absorbed by redefining the PDFs, as described in
Section~\ref{subsec_fact}.
\par
Diagrams and corresponding amplitudes are generated using FeynArts
\cite{Hahn:2000kx,Hahn:2001rv}. The algebraic simplifications and
numerical evaluation is done with help of FormCalc and LoopTools
\cite{Hahn:2006qw,Hahn:2001rv}. IR and collinear singularities are
regularized by means of mass regularization, i.e. we introduce a fictitious 
mass for the photon and the gluon. Quarks are treated as massless, 
except where their masses are needed as regulators.

\subsection{Virtual corrections}
\label{subsec_virt}

\FIGURE[t]{
\begin{tabular}{ccm{.3cm}cm{.3cm}cm{.3cm}c}
  \parbox{.1\textwidth}{\vspace{-1.5cm}{$\left( a \right)$}} 
  &
  \begin{feynartspicture}(55,55)(1,1)
    \FADiagram{$\alpha_s$}
    \FAProp(0.,15.)(10.,15.)(0.,){/Straight}{1}
    \FAProp(0.,5.)(10.,5.)(0.,){/Straight}{1}
    \FAProp(20.,15.)(10.,15.)(0.,){/ScalarDash}{-1}
    \FAProp(20.,5.)(10.,5.)(0.,){/ScalarDash}{-1}
    \red{\FAProp(10.,15.)(10.,5.)(0.,){/Straight}{0}
      \FAProp(9.7,15.)(9.7,5.)(0.,){/Cycles}{0}
      \FALabel(9.18,10.)[r]{${\tiny \tilde g}$}}
    \FAVert(10.,15.){0}
    \FAVert(10.,5.){0}
  \end{feynartspicture}
  &\parbox{.3cm}{\vspace{-1.5cm}$\times$}&
  \parbox{.3cm}{\vspace{-1.5cm}\Bigg( }
  \begin{feynartspicture}(55,55)(1,1)
    \FADiagram{$\alpha_s\alpha$}
    \FAProp(0.,15.)(7.,15.)(0.,){/Straight}{1}
    \FAProp(0.,5.)(10.,5.)(0.,){/Straight}{1}
    \FAProp(20.,5.)(10.,5.)(0.,){/ScalarDash}{-1}
    \blue{
      \FAProp(13.,15.)(7.,15.)(0.,){/Straight}{0}
      \FAProp(7.,15.)(10.,10.5)(0.,){/Straight}{0}
      \FAProp(10.,10.5)(13.,15.)(0.,){/Straight}{0}
    }
    \FAProp(13.,15.)(20.,15.)(0.,){/ScalarDash}{1}
    \red{\FAProp(10.,5.)(10.,10.5)(0.,){/Straight}{0}
      \FAProp(9.7,10.5)(9.7,5.)(0.,){/Cycles}{0}
      \FALabel(7.3,8.)[r]{${\tiny \tilde g}$}}
    \blue{\FALabel(11.2,17.)[r]{${\scriptstyle EW}$}}
    \FAVert(10.,5.){0}
    \FAVert(10.,10.5){0}
    \FAVert(7.,15.){0}
      \FAVert(13.,15.){0}
  \end{feynartspicture}
  &\parbox{.3cm}{\vspace{-1.5cm}$+$}&
  \begin{feynartspicture}(55,55)(1,1)
    \FADiagram{$\alpha_s\alpha$}
    \FAProp(0.,15.)(5.,15.)(0.,){/Straight}{1}
    \FAProp(0.,5.)(5.,5.)(0.,){/Straight}{1}
    \FAProp(15.,5.)(20.,5.)(0.,){/ScalarDash}{1}
    \FAProp(15.,15.)(20.,15.)(0.,){/ScalarDash}{1}
    \blue{\FAProp(5.,15.)(15.,15.)(0.,){/Straight}{0}
      \FAProp(5.,5.)(15.,5.)(0.,){/Straight}{0}
      \FAProp(15.,15.)(15.,5.)(0.,){/Straight}{0}
      \FALabel(14.18,10.)[r]{${\scriptstyle EW}$}}
    \red{\FAProp(5.,15.)(5.,5.)(0.,){/Cycles}{0}
      \FAProp(5.,15.)(5.,5.)(0.,){/Straight}{0}
      \FALabel(2.8,10.)[r]{${\tiny \tilde g}$}}
    \FAVert(5.,15.){0}
    \FAVert(15.,15.){0}
    \FAVert(5.,5.){0}
    \FAVert(15.,5.){0}
  \end{feynartspicture}
  &\parbox{.3cm}{\vspace{-1.5cm}$+$}&
  \begin{feynartspicture}(55,55)(1,1)
    \FADiagram{$\alpha_s\alpha$}
    \FAProp(0.,15.)(6.,15.)(0.,){/Straight}{1}
    \FAProp(0.,5.)(6.,5.)(0.,){/Straight}{1}
    \FAProp(6.,5.)(13.,10.)(0.,){/ScalarDash}{0}
    \FAProp(6.,15.)(13.,10.)(0.,){/ScalarDash}{0}
    \FAProp(13.,10.)(20.,15.)(0.,){/ScalarDash}{1}
    \FAProp(13.,10.)(20.,5.)(0.,){/ScalarDash}{1}
    \red{\FAProp(6.,15.)(6.,5.)(0.,){/Straight}{0}
      \FAProp(6.,15.)(6.,5.)(0.,){/Cycles}{0}
      \FALabel(5.18,10.)[r]{${\tilde g}$}}
    \FAVert(6.,15.){0}
    \FAVert(6.,5.){0}
    \FAVert(13.,10.){0}
  \end{feynartspicture}
  \parbox{.3cm}{\vspace{-1.5cm}\Bigg) }
  \\
  \parbox{.1\textwidth}{\vspace{-1.5cm}{$\left( b \right)$}} 
  &
  \begin{feynartspicture}(55,55)(1,1)
    \FADiagram{$\alpha_s$}
    \FAProp(0.,15.)(10.,15.)(0.,){/Straight}{1}
    \FAProp(0.,5.)(10.,5.)(0.,){/Straight}{1}
    \FAProp(20.,15.)(10.,15.)(0.,){/ScalarDash}{-1}
    \FAProp(20.,5.)(10.,5.)(0.,){/ScalarDash}{-1}
    \red{\FAProp(10.,15.)(10.,5.)(0.,){/Straight}{0}
      \FAProp(9.7,15.)(9.7,5.)(0.,){/Cycles}{0}
      \FALabel(9.18,10.)[r]{${\tiny \tilde g}$}}
    \FAVert(10.,15.){0}
    \FAVert(10.,5.){0}
  \end{feynartspicture}
  &\parbox{.3cm}{\vspace{-1.5cm}$\times$}&
  \parbox{.3cm}{\vspace{-1.5cm}\Bigg( }
  \begin{feynartspicture}(55,55)(1,1)
    \FADiagram{$\alpha_s\alpha$}
    \FAProp(0.,15.)(7.,15.)(0.,){/Straight}{1}
    \FAProp(0.,5.)(10.,5.)(0.,){/Straight}{1}
    \FAProp(20.,5.)(10.,5.)(0.,){/ScalarDash}{-1}
    \red{
      \FAProp(13.,15.)(7.,15.)(0.,){/Straight}{0}
      \FAProp(7.,15.)(10.,10.5)(0.,){/Straight}{0}
      \FAProp(10.,10.5)(13.,15.)(0.,){/Straight}{0}
    }
    \FAProp(13.,15.)(20.,15.)(0.,){/ScalarDash}{1}
    \blue{\FAProp(10.,5.)(10.,10.5)(0.,){/Straight}{0}
      \FAProp(10.,5.)(10.,10.5)(0.,){/Sine}{0}
      \FALabel(7.5,8.)[r]{${\tiny \tilde \chi}$}}
    \red{\FALabel(11.5,17.)[r]{${\scriptstyle QCD}$}}
    \FAVert(10.,5.){0}
    \FAVert(10.,10.5){0}
    \FAVert(7.,15.){0}
    \FAVert(13.,15.){0}
  \end{feynartspicture}
  &\parbox{.3cm}{\vspace{-1.5cm}$+$}&
  \begin{feynartspicture}(55,55)(1,1)
    \FADiagram{$\alpha_s\alpha$}
    \FAProp(0.,15.)(5.,15.)(0.,){/Straight}{1}
    \FAProp(0.,5.)(5.,5.)(0.,){/Straight}{1}
    \FAProp(15.,5.)(20.,5.)(0.,){/ScalarDash}{1}
    \FAProp(15.,15.)(20.,15.)(0.,){/ScalarDash}{1}
    \red{\FAProp(5.,15.)(15.,15.)(0.,){/Straight}{0}
      \FAProp(5.,5.)(15.,5.)(0.,){/Straight}{0}
      \FAProp(15.,15.)(15.,5.)(0.,){/Straight}{0}
      \FALabel(15.,10.)[r]{${\scriptstyle QCD}$}}
    \blue{\FAProp(5.,15.)(5.,5.)(0.,){/Sine}{0}
      \FAProp(5.,15.)(5.,5.)(0.,){/Straight}{0}
      \FALabel(2.8,10.)[r]{${\tiny \tilde \chi}$}}
    \FAVert(5.,15.){0}
    \FAVert(15.,15.){0}
    \FAVert(5.,5.){0}
    \FAVert(15.,5.){0}
  \end{feynartspicture}
  &\parbox{.3cm}{\vspace{-1.5cm}$+$}&
  \begin{feynartspicture}(55,55)(1,1)
    \FADiagram{$\alpha_s\alpha$}
    \FAProp(0.,15.)(6.,15.)(0.,){/Straight}{1}
    \FAProp(0.,5.)(6.,5.)(0.,){/Straight}{1}
    \FAProp(6.,5.)(13.,10.)(0.,){/ScalarDash}{0}
    \FAProp(6.,15.)(13.,10.)(0.,){/ScalarDash}{0}
    \FAProp(13.,10.)(20.,15.)(0.,){/ScalarDash}{1}
    \FAProp(13.,10.)(20.,5.)(0.,){/ScalarDash}{1}
    \blue{\FAProp(6.,15.)(6.,5.)(0.,){/Straight}{0}
      \FAProp(6.,15.)(6.,5.)(0.,){/Sine}{0}
      \FALabel(5.18,10.)[r]{${\tilde \chi}$}}
    \FAVert(6.,15.){0}
    \FAVert(6.,5.){0}
    \FAVert(13.,10.){0}
  \end{feynartspicture}
  \parbox{.3cm}{\vspace{-1.5cm}\Bigg) }
  \\
  \parbox{.1\textwidth}{\vspace{-1.5cm}$\left( c \right)$} 
  &
  \begin{feynartspicture}(55,55)(1,1)
    \FADiagram{$\alpha$}
    \FAProp(0.,15.)(10.,15.)(0.,){/Straight}{1}
    \FAProp(0.,5.)(10.,5.)(0.,){/Straight}{1}
    \FAProp(20.,15.)(10.,15.)(0.,){/ScalarDash}{-1}
    \FAProp(20.,5.)(10.,5.)(0.,){/ScalarDash}{-1}
    \blue{\FAProp(10.,15.)(10.,5.)(0.,){/Straight}{0}
      \FAProp(10.,15.)(10.,5.)(0.,){/Sine}{0}
      \FALabel(9.18,10.)[r]{${\tiny \tilde \chi}$}}
    \FAVert(10.,15.){0}
    \FAVert(10.,5.){0}
  \end{feynartspicture}
  &\parbox{.3cm}{\vspace{-1.5cm}$\times$}&
  \parbox{.3cm}{\vspace{-1.5cm}\Bigg( }
  \begin{feynartspicture}(55,55)(1,1)
    \FADiagram{$\alpha_s^2$}
    \FAProp(0.,15.)(7.,15.)(0.,){/Straight}{1}
    \FAProp(0.,5.)(10.,5.)(0.,){/Straight}{1}
    \FAProp(20.,5.)(10.,5.)(0.,){/ScalarDash}{-1}
    \red{
      \FAProp(13.,15.)(7.,15.)(0.,){/Straight}{0}
      \FAProp(7.,15.)(10.,10.5)(0.,){/Straight}{0}
      \FAProp(10.,10.5)(13.,15.)(0.,){/Straight}{0}
    }
    \FAProp(13.,15.)(20.,15.)(0.,){/ScalarDash}{1}
    \red{\FAProp(10.,5.)(10.,10.5)(0.,){/Straight}{0}
      \FAProp(9.7,10.5)(9.7,5.)(0.,){/Cycles}{0}
      \FALabel(7.5,8.)[r]{${\tiny \tilde g}$}}
    \red{\FALabel(11.5,17.)[r]{${\scriptstyle QCD}$}}
    \FAVert(10.,5.){0}
    \FAVert(10.,10.5){0}
    \FAVert(7.,15.){0}
    \FAVert(13.,15.){0}
  \end{feynartspicture}
  &\parbox{.3cm}{\vspace{-1.5cm}$+$}&
  \begin{feynartspicture}(55,55)(1,1)
    \FADiagram{$\alpha_s^2$}
    \FAProp(0.,15.)(5.,15.)(0.,){/Straight}{1}
    \FAProp(0.,5.)(5.,5.)(0.,){/Straight}{1}
    \FAProp(15.,5.)(20.,5.)(0.,){/ScalarDash}{1}
    \FAProp(15.,15.)(20.,15.)(0.,){/ScalarDash}{1}
    \red{\FAProp(5.,15.)(15.,15.)(0.,){/Straight}{0}
      \FAProp(5.,5.)(15.,5.)(0.,){/Straight}{0}
      \FAProp(15.,15.)(15.,5.)(0.,){/Straight}{0}
      \FALabel(15.,10.)[r]{${\scriptstyle QCD}$}
      \FAProp(5.,15.)(5.,5.)(0.,){/Cycles}{0}
      \FAProp(5.,15.)(5.,5.)(0.,){/Straight}{0}
      \FALabel(2.8,10.)[r]{${\tiny \tilde g}$}}
    \FAVert(5.,15.){0}
    \FAVert(15.,15.){0}
    \FAVert(5.,5.){0}
    \FAVert(15.,5.){0}
  \end{feynartspicture}
  &\parbox{.3cm}{\vspace{-1.5cm}$+$}&
  \begin{feynartspicture}(55,55)(1,1)
    \FADiagram{$\alpha_s^2$}
    \FAProp(0.,15.)(6.,15.)(0.,){/Straight}{1}
    \FAProp(0.,5.)(6.,5.)(0.,){/Straight}{1}
    \FAProp(6.,5.)(13.,10.)(0.,){/ScalarDash}{0}
    \FAProp(6.,15.)(13.,10.)(0.,){/ScalarDash}{0}
    \FAProp(13.,10.)(20.,15.)(0.,){/ScalarDash}{1}
    \FAProp(13.,10.)(20.,5.)(0.,){/ScalarDash}{1}
    \red{\FAProp(6.,15.)(6.,5.)(0.,){/Straight}{0}
      \FAProp(6.,15.)(6.,5.)(0.,){/Cycles}{0}
      \FALabel(5.18,10.)[r]{${\tilde g}$}}
    \FAVert(6.,15.){0}
    \FAVert(6.,5.){0}
    \FAVert(13.,10.){0}
  \end{feynartspicture}
  \parbox{.3cm}{\vspace{-1.5cm}\Bigg) }
\end{tabular}
\caption{Sample of Feynman diagrams to illustrate the virtual
  contributions at $\mathcal{O}(\alphas^2\alpha)$. Three gauge
  invariant subsets of interferences occur at this order. The label of
  perturbative order is attached to each diagram. $EW$ refers to
  electroweakly interacting particles and $QCD$ to strongly
  interacting particles in the loop insertions. The full sets of
  diagrams are shown in Figs.~\ref{fig_EWinsQCD}, \ref{fig_QCDinsEW},
  and~\ref{fig_QCDinsQCD}.
\label{fig_virt}}}


The virtual contributions are given
by the interference of tree-level 
and  one-loop diagrams. In practice three types of
interferences occur at $\mathcal{O}(\alphas^2\alpha)$, as schematically depicted in \figref{fig_virt}. All three
interference terms yield non-vanishing contributions to the cross section.
For each subprocess, the partonic cross section can be written as
\begin{align}
\begin{split}
  d\hat\sigma^{2,\,1}_{\rm virt.} 
=&
  \frac{d\hat t}{16\pi \hat s^2} \overline{\sum} \, 2 \operatorname{Re} \left\lbrace
    \left( \mathcal{M}^{1,\,0} \right)^*
    \mathcal{M}^{1,\,1}_{(\rm EW)} +
    \left( \mathcal{M}^{1,\,0} \right)^*
    \mathcal{M}^{1,\,1}_{(\rm QCD)} \right\rbrace
  \\
  &+  \frac{d\hat t}{16\pi \hat s^2} \overline{\sum}\, 2 \operatorname{Re} \left\lbrace
    \left( \mathcal{M}^{0,\,1}\right)^*
    \mathcal{M}^{2,\,0}\right\rbrace. 
\label{eq_virt}
\end{split}
\end{align}
The first line corresponds to $(a)$ and $(b)$ of \figref{fig_virt} and
is given by the interference of $\mathcal{M}^{1,\,0}$ with
$\mathcal{M}^{1,\,1}$. The amplitude $\mathcal{M}^{1,\,1}$ is split
into two parts, $\mathcal{M}^{1,\,1}_{(\rm EW)}$ and $\mathcal{M}^{1,\,1}_{\rm (QCD)}$,
the first arising from tree-level QCD diagrams with EW insertions
(\figref{fig_virt}a, right), and the latter from tree-level EW
diagrams with QCD insertions (\figref{fig_virt}b, right). 
The second line in \eqref{eq_virt}, corresponding to \figref{fig_virt}c,
is given by the interference of $\mathcal{M}^{0,\,1}$ with the
pure-QCD one loop amplitude $\mathcal{M}^{2,\,0}$.
Care has to be taken with diagrams containing a
four-squark vertex. This vertex includes the electroweak as well as
the strong coupling and the appropriate part has to be selected in each interference 
contribution to match the right order, as indicated in \figref{fig_virt}.
\par

\medskip

The full set of virtual corrections is UV finite after renormalization of the theory 
and the inclusion of the proper set of one-loop counterterms. The renormalization for 
squark--squark production proceeds in close analogy to that for squark--anti-squark 
production described in \cite{Hollik:2008yi} and is sketched here only briefly.
Each of the three interference subsets is
gauge-independent by itself and can be renormalized separately.

In the first group, shown in \figref{fig_virt}a, UV singularities only
arise from gluino-mediated amplitudes with weak insertions
($\mathcal{M}^{1,\,1}_{(\rm EW)}$).  We include the diagrams with
counterterms for the $q\gluino\squarka$ vertex, see
\figref{fig_EWinsQCD}, and evaluate the renormalization constants at
$\Order(\alpha)$. At this order in the perturbative expansion we need
to renormalize quark and squark fields, while the renormalization of
gluino and strong coupling is not required. The regularization of the
divergent amplitudes in this sector is done in dimensional reduction,
and renormalization of quarks and squarks is performed in the on-shell
scheme.
\par

In the second case, \figref{fig_virt}b, neutralino- or
chargino-mediated amplitudes with strong insertions
($\mathcal{M}^{1,\,1}_{(\rm QCD)}$) are considered. To obtain a
UV-finite result, one needs to include diagrams containing
counterterms for the $q\squark\tilde\chi^0$ vertex and, if arising,
for the $q\squark'\tilde\chi^{\pm}$ vertex, see \figref{fig_QCDinsEW}.
The renormalization constants have to be evaluated at
$\Order(\alpha_s)$ and no renormalization of the neutralino or
chargino is required.  Since the gluino does not enter this subset of
one-loop amplitudes, it is thus sufficient to renormalize the quark
and squark sector.  As before, the divergent amplitudes are
regularized in dimensional reduction and on-shell conditions are
imposed to fix the (s)quark renormalization constants.
\par

The third subset, \figref{fig_virt}c, refers to pure-QCD one-loop
amplitudes, \ie~gluino-mediated diagrams with strong insertions
($\mathcal{M}^{2,\,0}$). In this case one has to renormalize the quark
and squark sector as well as the gluino and the strong Yukawa coupling
$\hat g_s$, which appears in the $q \tilde q\tilde g$ vertex. The
renormalization constants in the corresponding amplitudes, see
\figref{fig_QCDinsQCD}, have to be evaluated at
$\mathcal{O}(\alphas)$. The strong scalar coupling $\hat g_s$ is
related to the strong coupling $g_s$ via supersymmetry.  To match the
definition of the strong coupling constant used in the extraction of
the PDFs, $g_s$ has to be given in the $\msbar$ scheme with the
contributions from heavy particles subtracted in the running of
$\alpha_s$.  We thus regularize this part of the virtual corrections
using dimensional regularization. Quarks and squarks are renormalized
on-shell again, in the strong sector the $\overline{\text{MS}}$ scheme
is applied.  Dimensional regularization however induces a finite
difference between $g_s$ and $\hat{g}_s$ at the one-loop level and
violates the supersymmetric relation between the two
couplings~\cite{Hollik:2001cz}. We add the well-known finite shift in
the definition of the renormalization constant for $\hat{g}_s$ in
order to restore SUSY in the physical
amplitudes~\cite{Beenakker:1996ch}.

All counterterms and renormalization constants are explicitly given 
in \cite{Hollik:2008yi}, Appendix B, and need not to be repeated here.

\medskip

The virtual corrections also exhibit photonic and gluonic mass
singularities of infrared (IR) and collinear origin. In
$\mathcal{M}^{1,\,1}_{\rm (EW)}$, mass singularities arise if two
external particles exchange a low-energetic massless photon while
collinear singularities appear if one of the massless initial-state
quarks splits collinearly into a quark and a photon. In order to
obtain an IR finite result, real photon radiation at
$\Order(\alpha_s^2 \alpha)$ has to be added.  In contrast in
$\mathcal{M}^{1,\,1}_{\rm (QCD)}$, massless gluons running in the
loops give rise to mass singularities in the soft and collinear
limit. Similarly, the diagrams contributing to $\mathcal{M}^{2,\,0}$
suffer from gluonic IR and collinear singularities. Hence we have to
include real gluon bremsstrahlung at $\Order(\alpha_s^2 \alpha)$ in
order to cancel the IR singularities.
We regularize the photonic singularities by means of mass regularization. Owing to the photon-like appearance of the gluon in the
respective diagrams, it is also possible to regularize these IR singularities by 
a fictitious gluon mass.

\subsection{Real corrections}
\label{subsec_real}

\FIGURE[t]{
\begin{tabular}{m{.35\textwidth}cm{.3cm}c}
  \parbox{4.0cm}{\vspace{-1.5cm}
    {photon bremsstrahlung}} &
  \parbox{4pt}{ \vspace*{-1.55cm}$\bigg|$}
  \begin{feynartspicture}(55,55)(1,1)
    \FADiagram{$\alpha_s\sqrt{\alpha}$}
    \FAProp(0.,16.)(5.,16.)(0.,){/Straight}{1}
    \FAProp(5.,16.)(10.,16.)(0.,){/Straight}{1}
    \FAProp(10.,16.)(20.,16.)(0.,){/ScalarDash}{1}
    \FAProp(5.,16.)(15.,20.)(0.,){/Sine}{0}
    \red{\FAProp(10.,16.)(10.,6.)(0.,){/Straight}{0}
      \FAProp(9.7,16.)(9.7,6.)(0.,){/Cycles}{0}
      \FALabel(8.18,11.)[r]{${\tiny \tilde g}$}}
    \FAProp(0.,6.)(10.,6.)(0.,){/Straight}{0}
    \FAProp(10.,6.)(20.,6.)(0.,){/ScalarDash}{1}
    \FAVert(10.,16.){0}
    \FAVert(10.,6.){0}
    \FAVert(5.,16.){0}
  \end{feynartspicture}
  \parbox{4pt}{ \vspace*{-1.55cm}$\bigg|^2$}
  &\\
  \parbox{4.0cm}{\vspace{-1.5cm}
    {gluon bremsstrahlung}} &
  \begin{feynartspicture}(55,55)(1,1)
    \FADiagram{$\sqrt{\alpha_s}^3$}
    \FAProp(0.,16.)(5.,16.)(0.,){/Straight}{1}
    \FAProp(5.,16.)(10.,16.)(0.,){/Straight}{1}
    \FAProp(10.,16.)(20.,16.)(0.,){/ScalarDash}{1}
    \FAProp(5.,16.)(15.,20.)(0.,){/Cycles}{0}
    \red{\FAProp(10.,16.)(10.,6.)(0.,){/Straight}{0}
      \FAProp(9.7,16.)(9.7,6.)(0.,){/Cycles}{0}
      \FALabel(8.18,11.)[r]{${\tiny \tilde g}$}}
    \FAProp(0.,6.)(10.,6.)(0.,){/Straight}{0}
    \FAProp(10.,6.)(20.,6.)(0.,){/ScalarDash}{1}
    \FAVert(10.,16.){0}
    \FAVert(10.,6.){0}
    \FAVert(5.,16.){0}
  \end{feynartspicture}
  &\parbox{.3cm}{\vspace{-1.5cm}$\times$}&
  \begin{feynartspicture}(55,55)(1,1)
    \FADiagram{$\sqrt{\alpha_s}\alpha$}
    \FAProp(0.,16.)(5.,16.)(0.,){/Straight}{1}
    \FAProp(5.,16.)(10.,16.)(0.,){/Straight}{1}
    \FAProp(10.,16.)(20.,16.)(0.,){/ScalarDash}{1}
    \FAProp(5.,16.)(15.,20.)(0.,){/Cycles}{0}
    \blue{\FAProp(10.,16.)(10.,6.)(0.,){/Straight}{0}
      \FAProp(10.,16.)(10.,6.)(0.,){/Sine}{0}
      \FALabel(9.18,11.)[r]{${\tiny \tilde \chi}$}}
    \FAProp(0.,6.)(10.,6.)(0.,){/Straight}{0}
    \FAProp(10.,6.)(20.,6.)(0.,){/ScalarDash}{1}
    \FAVert(10.,16.){0}
    \FAVert(10.,6.){0}
    \FAVert(5.,16.){0}
  \end{feynartspicture}
  \\
  \parbox{4.0cm}{\vspace{-1.5cm}
    {real quark radiation}} &
  \begin{feynartspicture}(55,55)(1,1)
    \FADiagram{$\sqrt{\alpha_s}^3$}
    \FAProp(0.,16.)(5.,16.)(0.,){/Cycles}{0}
    \FAProp(5.,16.)(10.,16.)(0.,){/Straight}{1}
    \FAProp(10.,16.)(20.,16.)(0.,){/ScalarDash}{1}
    \FAProp(5.,16.)(15.,20.)(0.,){/Straight}{-1}
    \red{\FAProp(10.,16.)(10.,6.)(0.,){/Straight}{0}
      \FAProp(9.7,16.)(9.7,6.)(0.,){/Cycles}{0}
      \FALabel(8.18,11.)[r]{${\tiny \tilde g}$}}
    \FAProp(0.,6.)(10.,6.)(0.,){/Straight}{0}
    \FAProp(10.,6.)(20.,6.)(0.,){/ScalarDash}{1}
    \FAVert(10.,16.){0}
    \FAVert(10.,6.){0}
    \FAVert(5.,16.){0}
  \end{feynartspicture}
  &\parbox{.3cm}{\vspace{-1.5cm}$\times$}&
  \begin{feynartspicture}(55,55)(1,1)
    \FADiagram{$\sqrt{\alpha_s}\alpha$}
    \FAProp(0.,16.)(5.,16.)(0.,){/Cycles}{0}
    \FAProp(5.,16.)(10.,16.)(0.,){/Straight}{1}
    \FAProp(10.,16.)(20.,16.)(0.,){/ScalarDash}{1}
    \FAProp(5.,16.)(15.,20.)(0.,){/Straight}{-1}
    \blue{\FAProp(10.,16.)(10.,6.)(0.,){/Straight}{0}
      \FAProp(10.,16.)(10.,6.)(0.,){/Sine}{0}
      \FALabel(9.18,11.)[r]{${\tiny \tilde \chi}$}}
    \FAProp(0.,6.)(10.,6.)(0.,){/Straight}{0}
    \FAProp(10.,6.)(20.,6.)(0.,){/ScalarDash}{1}
    \FAVert(10.,16.){0}
    \FAVert(10.,6.){0}
    \FAVert(5.,16.){0}
  \end{feynartspicture}
\end{tabular}
\caption{Sample of Feynman diagrams for the three subsets of real
  emission contributions at $\mathcal{O}(\alphas^2\alpha)$. The order
  in  the perturbative expansion is specified for each
  diagram. The full sets of diagrams are given in the Appendix,
  \cf~Figures~\ref{fig_realphoton}, \ref{fig_realgluon}, and
  \ref{fig_realquark}.
\label{fig_real}}}


Three independent bremsstrahlung processes contribute at
$\Order(\alpha_s^2 \alpha)$, as depicted in \figref{fig_real}. Real
photon and real gluon radiation processes have to be combined with the
corresponding subset of virtual corrections to obtain an IR finite
result. Also real quark radiation gives nonzero contributions from the
interference of QCD and EW mediated diagrams and has to be included in
the cross section at $\Order(\alpha_s^2 \alpha)$.

\subsubsection{Real photon emission}
\label{subsec_realphoton}

The real photon emission at $\mathcal{O}(\alphas^2\alpha)$,
\bee
  q(p_1)\, q^{\prime}(p_2)\, \rightarrow \,\squarka(p_3)\, \squarkb'(p_4)\, \gamma(k),
\label{eq_NLOmom}
\eee
is given by the squared matrix element of a QCD tree-level diagram
with an external photon attached (see \figref{fig_real}, top and
\figref{fig_realphoton} for the full set of diagrams). The integration
over the photon phase space is IR divergent in the soft-photon region,
\ie~for $k^0 \rightarrow 0$. Further singularities arise in the
collinear region if $p_i \cdot k \rightarrow 0$ for $i=\{1,2\}$. We use phase space
slicing and apply a cut
on the photon energy, $k^0 > \delta_s \sqrt{\hat s}/2$,
and on the angle $\theta$ between the photon and incoming partons, 
 $|\cos(\theta)| < 1- \delta_{\theta}$, to split off the singular regions. 
In the hard, non-collinear region the integration is convergent and is
performed numerically. 
The cross sections in the soft- and collinear region can be approximated analytically 
and are given in Appendix \ref{subsec_Asoftphoton}.

\par 
By combining the real photon emission with the virtual EW-type
corrections (\figref{fig_virt}a), the soft singularities cancel.
Remaining initial-state collinear singularities are universal and have
to be absorbed via factorization in the PDFs, see
section~\ref{subsec_fact}.  In the following, we will refer to this
UV-, IR- and collinear-finite combination as the EW-type corrections.

\subsubsection{Real gluon emission}
\label{subsec_realguon}

Real gluon bremsstrahlung at $\Order(\alpha_s^2 \alpha)$ 
proceeds via the partonic process
\bee
  q(p_1) \,q^{\prime}(p_2)\, \rightarrow\, \squarka(p_3)\, \squarkb'(p_3)\, g(k).
\eee
It is given by the interference term of a QCD- and an
EW tree-level diagram, both with an external gluon attached on
(\figref{fig_real}, center and \figref{fig_realgluon}).
In the considered processes the gluon is Abelian like and we can
treat soft and collinear singularities by  mass regularization in close analogy 
to the photonic case. However the eikonal current has to be modified in order to take
color correlations into account. 
Different to real photon emission, 
collinear singularities only arise for diagonal $\squarka\squarka$, 
$\Sup_L \Sdown_L$, and $\Scharm_L \Sstrange_L$ production. 
This can be seen by noticing
that in the collinear cone the cross section becomes proportional
to the corresponding $2 \rightarrow 2$ process, which in this case
would be squark--squark production at $\mathcal{O}(\alphas\alpha)$;
\ie~the interference of tree-level QCD and EW diagrams must be
non-vanishing. 
Explicit expressions for the cross
sections in the soft- and collinear regions are given in Appendix
\ref{subsec_Asoftgluon}.
\par

By combining real gluon emission and the two virtual QCD-type
corrections (\figref{fig_virt}b,c), the IR singularities
cancel. Remaining collinear singularities are again absorbed into the
PDFs, as described in Section~\ref{subsec_fact}.  In the following we
will refer to this UV-, IR- and collinear-finite combination as the
QCD-type corrections.

\subsubsection{Real quark emission}
\label{subsec_realquark}

Finally, also real quark radiation contributes at $\mathcal{O}(\alphas^2\alpha)$,
\begin{align}
\begin{split}
  g(p_1)\,q(p_2)\, &\rightarrow\, \squarka(p_3) \,\squarkb'(p_4) \,\bar q'(k),
\\
\hspace*{-2cm}\text{and if~} q \neq q'\qquad
  g(p_1)\,q^{\prime}(p_2)\, &\rightarrow\, \squarka(p_3) \,\squarkb'(p_4) \,\bar q(k),
\end{split}
\end{align}
via the interference of a QCD-type diagram with an EW-type diagram, as
shown in \figref{fig_real}, bottom (see \figref{fig_realquark} for the
complete listing of diagrams). This process can be regarded as
completely independent to the virtual corrections, since it is IR
finite by itself. However it has to be taken into account in a
consistent analysis of electroweak corrections up to
$\mathcal{O}(\alphas^2\alpha)$.  Initial-state collinear singularities are present for
processes with non-vanishing tree-level interferences. The cross
section in the collinear region is given in Appendix
\ref{subsec_Acollquark}. As before, these singularities are absorbed
via factorization into the PDFs, see Section~\ref{subsec_fact}.
\par 

Different to photon and gluon bremsstrahlung, the internal gluino,
neutral\-ino or charg\-ino can go on-shell in specific SUSY scenarios,
if heavier than one of the external squarks. In these cases, we
include a Breit-Wigner width for the resonant particle in the
corresponding propagators to regularize the poles.  Note that physical
resonances do not occur.  This is different to the case of real quark
radiation in \eg~gluino--squark production
processes~\cite{Hollik:2008vm}, where internal squarks can go on-shell
in both the EW- and the QCD-mediated diagrams.

\subsection{Factorization of initial-state collinear singularities}
\label{subsec_fact}

The remaining collinear singularities have to  be absorbed by redefining the PDFs. 
At $\Order(\alpha_s^2 \alpha)$ this can be achieved by the replacement~\cite{Baur:1998kt, Hollik:2007sq}
\begin{align}
\begin{split}
  f_q(x,\mu_F)\, \to \,& \, f_q(x,\mu_F) \left( 1-\frac{\alpha e_q^2+\alphas
      C_F}{\pi} \kappa_{v+s} -\frac{1}{4}\frac{\alpha
    e_q^2}{\pi}f_{v+s}\right)
\\
  &-\int_{x}^{1-\delta_s} \frac{dz}{z}
  f_q\Big(\frac{x}{z},\mu_F\Big) \left( \frac{\alpha e_q^2+\alphas C_F}{2\pi}
  \kappa_c(z) - \frac{\alpha e_q^2}{2\pi} f_c(z)\right) 
\\
  &-\int_x^1 \frac{dz}{z} f_g\Big(\frac{x}{z},\mu_F\Big)\, \frac{\alphas
    C_F}{2\pi} P_{qg}(z) \ln\left( \frac{\mu_F^2}{m_q^2} \right),
\label{eq_PDFredef}
\end{split}
\end{align}
where $e_q$ denotes the electric charge of quark $q$, $C_F=4/3$, and
\begin{align}
\begin{split}
  \kappa_{v+s} &= 1 -\ln\delta_s -\ln^2\delta_s
  +\left(\ln\delta_s+\frac{3}{4}\right)
  \ln\left(\frac{\mu_F^2}{m_{q}^2}\right),
\\
  \kappa_c(z) &= P_{qq}(z)\, \ln\left( \frac{\mu_F^2}{m_q^2}
  \frac{1}{(1-z)^2} -1 \right).
\end{split}
\end{align}
The factorization-scheme dependent functions are 
\begin{align}
\begin{split}
  f_{v+s} &= 9 +\frac{2\pi^2}{3} +3\ln\delta_s -2\ln^2\delta_s,
\\
  f_c(z) &= P_{qq}(z) \,\ln\left( \frac{1-z}{z}\right) -\frac{3}{2}
  \frac{1}{1-z} + 2z +3,
\end{split}
\end{align}
with the splitting functions
\bee
  P_{qq}(z) &= \frac{1+z^2}{1-z}, \qquad 
  P_{qg}(z) &= z^2+(1-z^2).
\eee
The factorization is done in the $\overline{\text{MS}}$ scheme at NLO
QCD and in the physical DIS scheme at NLO EW.
The replacement of the PDFs in \eqref{eq_Lumi} gives
further contributions of $\mathcal{O}(\alphas^2\alpha)$ to the total
cross section. The first and
second line in \eqref{eq_PDFredef} cancel the remaining singularities
in the EW-type and QCD-type corrections. The third line in \eqref{eq_PDFredef} cancels
the collinear singularities in the real quark radiation.

\newcommand {\Born} {\sigma^{\rm Born}} 
\newcommand {\TreeEW} {\Delta\sigma^{\rm tree~EW}} 
\newcommand {\NLOEW} {\Delta\sigma^{\rm NLO~EW}} 
\newcommand {\EW} {\Delta\sigma^{\rm EW}}
\newcommand {\NLO} {\sigma^{\rm NLO}}
\newcommand {\dTreeEW}{\delta^{\rm tree~EW}}
\newcommand {\dNLOEW}{\delta^{\rm NLO~EW}}
\newcommand {\dNLO}{\delta^{\rm EW}}
\newcommand {\qL}{\tilde{q}_{L}}
\newcommand {\qR}{\tilde{q}_{R}}
\newcommand {\uL}{\tilde{u}_{L}}
\newcommand {\uR}{\tilde{u}_{R}}
\newcommand {\dL}{\tilde{d}_{L}}
\newcommand {\dR}{\tilde{d}_{R}}
\newcommand {\cL}{\tilde{c}_{L}}
\newcommand {\cR}{\tilde{c}_{R}}
\newcommand {\sL}{\tilde{s}_{L}}
\newcommand {\sR}{\tilde{s}_{R}}
\newcommand {\SPA} {SPS1a${}^\prime$}
\section{Numerical results}
\label{sect_results}
In the following we illustrate the impact of the EW contributions on
the production cross section. Since we have 36 processes contributing
to squark--squark production and the same amount for
anti-squark--anti-squark production, we present (at least partly)
inclusive results. We refer to four different combinations of
(anti-)squarks in the final state, which differ with respect to the
chirality of the produced particles:
\begin{itemize}
\item  $ \qL \qL^\prime$ or ``LL'' refers to the inclusive production
  of two left-handed squarks  and two left-handed anti-squarks.
\item  $\qL \qR^\prime$ or ``LR'' refers to the inclusive production
  of one left-handed and one right-handed squark and the charge conjugated process.
\item  $\qR \qR^\prime$ or ``RR'' refers to the inclusive production
  of two right-handed squarks and two right-handed anti-squarks.
\item  $\squark \squark^\prime$ or ``incl.'' refers to the inclusive production
  of all (anti-)squarks. It is given by the sum of the three cases above,
  taking all 72 subprocesses of squark--squark and
  anti-squark--anti-squark final states into account.
\end{itemize}
  We focus here on these chirality-based classes since  
squarks of different chiralities are, in principle,
experimentally distinguishable by their decay chains, 
see \eg~Section 5.1.2 of~\cite{Weiglein:2004hn}.

\medskip
In the discussion we refer to the following quantities, based on the cross section 
definitions in Section~\ref{sect_tree}. The leading order cross section is denoted by
$\Born = \sigma^{2,\,0}$. 
The tree-level EW 
and the NLO EW contributions to the cross section
are labeled by 
\bee
\TreeEW = (\sigma^{1,\,1}+\sigma^{0,\,2}), \qquad
\NLOEW = \sigma^{2,\,1},
\eee
 respectively, and
$\EW = \TreeEW + \NLOEW$ will be referred to as the EW contribution.
The total sum of the LO cross section with the EW contributions is denoted by
$\NLO=\Born+\EW$. Relative EW contributions are defined by
\begin{align}
\dTreeEW = \TreeEW / \Born, \qquad
\dNLOEW = \NLOEW / \Born, \qquad
\dNLO = \EW /\Born.
\end{align}
In distributions $\delta$ denotes the relative EW contribution
  defined as $\delta = (\mathcal{O}_{\rm NLO}-\mathcal{O}_{\rm Born}) /
  \mathcal{O}_{\rm Born}$, where $\mathcal{O}$ is a generic observable
  and $\mathcal{O}_{\rm NLO}$ is the sum of the Born and the EW
  contribution.

\subsection{Input Parameters}

The Standard Model input parameters are chosen in correspondence with 
\cite{AguilarSaavedra:2005pw, CDFtopmass},
\begin{align}
\begin{split}
M_Z & = 91.1876~{\GeV}, \qquad\quad~  M_W = 80.4247685~\GeV,\\
\alpha^{-1} & = 137.036, \qquad\qquad \alphas(M_Z) = 0.119,\\
m_t & = 170.9~{\GeV}, \qquad\quad\quad\, m_b^{\rm OS} = 4.7~\GeV.
\end{split}
\end{align}
The strong coupling constant $\alphas$ has been defined in the $\msbar$ scheme 
using the two-loop renormalization group equation with five light flavors.

For the SUSY parameters, we refer to three benchmark mSUGRA scenarios,
the \SPA{} scenario, the SPS2, and SPS5 scenario
\cite{AguilarSaavedra:2005pw,Allanach:2002nj}. The \SPA{} scenario can
be considered as a ``typical'' mSUGRA scenario. It has been proposed
by the SPA convention and should be used for comparisons with other
calculations. The SPS2 scenario features relatively heavy squarks with
light charginos and neutralinos and a gluino lighter than the
squarks. The SPS5 scenario leads to a very light $\Stop_1$ with
moderate light-flavor squark masses.
In each scenario, the particle spectrum is determined by universal GUT
scale parameters, \cf~Table~\ref{tab_SPSinput}, which act as boundary
conditions for the renormalization group running of the soft-breaking
parameters. We use the program Softsusy~\cite{Allanach:2001kg} to
evolve the soft-breaking parameters down to the SUSY scale $M_{\rm SUSY}$.
We choose a common SUSY scale $M_{\rm SUSY} = 1$\,TeV for
all scenarios, in reference to the SPA convention. At $M_{\rm SUSY}$ a
consistent translation of the squark masses into the on-shell scheme
is performed.  The left-handed down-type squark is treated as a
dependent mass parameter, fixed by SU(2) invariance. It is set to its
corresponding on-shell value obtained at one-loop accuracy according
to \cite{Hollik:2003jj}.  The on-shell mass parameters for the light
flavor squarks together
with the masses of the gluino and the lightest neutralino/chargino 
are summarized in Table~\ref{tab_SPSval}.

\TABULAR[t]{c|c|c|c|c|c}{
  \hline\trule
  & $\boldsymbol{m_0}$
  & $\boldsymbol{m_{1/2}}$
  & $\boldsymbol{A_0}$
  & $\boldsymbol{\tan\beta}$
  & $\boldsymbol{{\rm sign}(\mu)}$\\
  \hline
  \SPA{} & 70~\GeV & 250~\GeV & $-$300~\GeV & 10 & $+$\\
  \hline
  SPS2 & 1450~\GeV & 300~\GeV & 0 & 9.66 & +\\
  \hline
  SPS5 & 150~\GeV & 300~\GeV & $-$1000~\GeV & 4.82 & $+$
  \\ \hline
}{High energy input parameters for the different SUSY scenarios
  considered. The mass parameters $m_0$, $m_{1/2}$ and $A_0$ are
  given at the GUT scale, $\tan\beta$ is evaluated at
  $M_{\rm SUSY}=1$\,TeV. \label{tab_SPSinput}}
\TABULAR[t]{c|c|c|c|c|c|c|c}{
  \hline\trule
  & $\boldsymbol{\uL}$
  & $\boldsymbol{\uR}$
  & $\boldsymbol{\dL}$
  & $\boldsymbol{\dR}$
  & $\boldsymbol{\gluino}$
  & $\boldsymbol{\neu_1}$
  & $\boldsymbol{\cha_1}$\\
  \hline
  \SPA{} & 561 & 543 & 566 & 539 & 609 & 101 & 180\\
  \hline
  SPS2 & 1559 & 1554 & 1561 & 1555 & 785 & 120 & 199\\
  \hline
  SPS5 & 677 & 655 & 681 & 654 & 724 & 123 & 225 \\
  \hline
}{On-shell masses of the squarks, the gluino, and the lightest neutralino and 
chargino within the different SUSY scenarios
  considered. All masses are given in \GeV.\label{tab_SPSval}}

\medskip
The technical cuts needed for the regularization of soft and
collinear singularities are set to $\delta_s=10^{-3}\sqrt{\hat s}$
and $\delta_{\theta}=10^{-4}$. We checked numerically that these values
are sufficiently small to justify the eikonal approximation.
\FIGURE[t]{
  \hspace*{2cm}\includegraphics[width=.5\textwidth,]{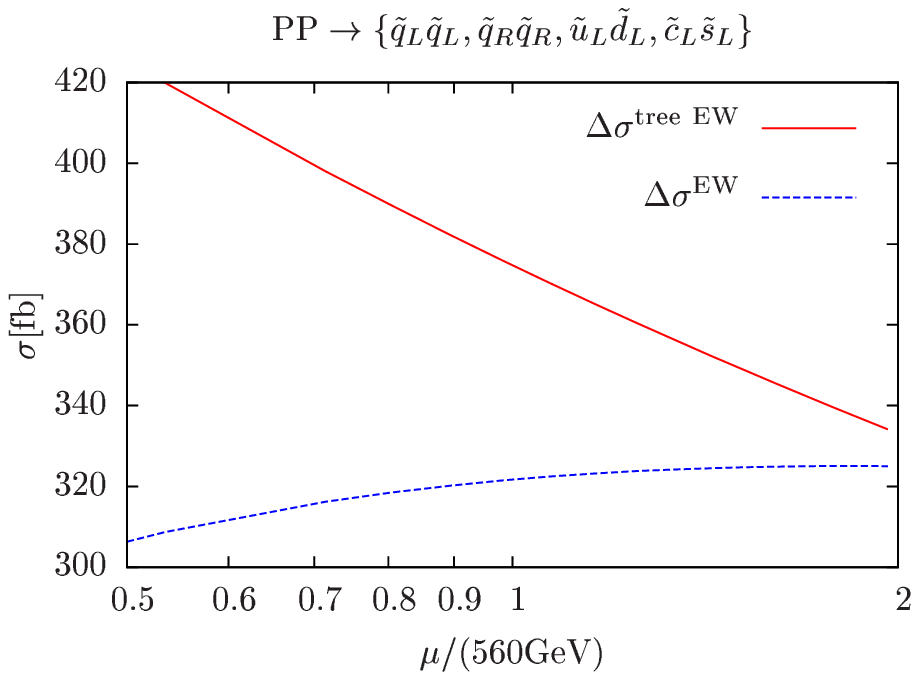}\hspace*{2cm}
  \caption{Dependence of the hadronic cross section $\sigma$ on the
    renormalization and factorization scale $\mu_R$ and $\mu_F$ for
    the \SPA{} scenario. Both scales are set to a common value
    $\mu_{R}=\mu_{F}=\mu$ with $\mu$ varied by a factor two around
    $\overline{m}_{\squark}=560~\GeV$. Only processes that contribute
    at $\mathcal{O}(\alphas\alpha)$ are considered. The red line shows
    the EW tree-level contribution to the cross section, while the
    green curve shows the EW contribution up to NLO EW. For the latter
    case the scale dependence is reduced.}
  \label{fig_scale}
}
For the calculation of hadronic observables we use the MRST2004QED
parton distribution functions \cite{Martin:2004dh}.
The factorization and renormalization scales are set to a common value,
$\mu = \mu_{R}=\mu_{F}=\overline{m}_{\squark}$ with
$\overline{m}_{\squark}$ being the average mass of the light-flavor
squarks.
\figref{fig_scale} shows the scale dependence for the \SPA{}
scenario. Here we are inclusive with respect to processes that
contribute at $\mathcal{O}(\alphas\alpha)$. We compare the tree-level
EW contribution with the cross section including the EW NLO
contributions at $\mathcal{O}(\alphas^2\alpha)$, \ie~one order higher
in the strong coupling. One finds that the scale dependence is
considerably reduced.


\subsection{Total hadronic cross sections}
\TABULAR[t]{c|c|c|c||c|c|c}{
  \hline\trule
    \multirow{2}{*}{\bf \SPA{}}
    & {$\boldsymbol{\Born}$}
    & {$\boldsymbol{\TreeEW}$}
    & {$\boldsymbol{\NLOEW}$}
    & {\multirow{2}{*}{$\boldsymbol{\dTreeEW}$}}
    & {\multirow{2}{*}{$\boldsymbol{\dNLOEW}$}}
    & {\multirow{2}{*}{$\boldsymbol{\dNLO}$}}\\
    & {$\mathcal{O}(\alphas^2)$}
    & {$\mathcal{O}(\alphas\alpha+\alpha^2)$}
    & {$\mathcal{O}(\alphas^2\alpha)$}
    &
    &
    &\\
    \hline\trule    
    ${\boldsymbol{\qL \qL^{\prime}}}$ & 1717.6(8) & 378.9(1) & ~$-74.8(6)$ & 22.1 \% & $-$4.4 \% & 17.7 \%\\
    \hline\trule
    ${\boldsymbol{\qR \qR^{\prime}}}$ & 1981.9(7) & 31.81(2) & ~$-1.60(9)$ & ~1.6 \% & $-$0.1 \% & ~ 1.5 \%\\
    \hline\trule
    ${\boldsymbol{\qL \qR^{\prime}}}$ & 1743.8(4) & 2.538(1) & $-70.71(4)$ & ~0.1 \% & $-$4.1 \% & $-3.9 \%$ \\    
    \hline
    \hline\trule             
    ${\boldsymbol{\squark \squark^{\prime}}}$ & ~~5443(1) & 413.3(1) & $-147.1(6)$ & ~7.6 \% & $-$2.7 \% & ~ 4.9 \%\\
    \hline
  }{Hadronic cross sections for squark--squark production at the LHC
    within the SPS1a${}^\prime$ scenario. Shown are the LO cross
    section, the tree-level EW as well as NLO EW contributions and the
    relative corrections as defined in the text.  Anti-particles are
    included. The numbers in brackets refer to the integration
    uncertainty in the last digit. All cross sections are given in
    femtobarn (fb).  \label{tab_SPA}}

\TABULAR[t]{c|c|c|c||c|c|c}{
  \hline\trule
    \multirow{2}{*}{\bf SPS2}
    & {$\boldsymbol{\Born}$}
    & {$\boldsymbol{\TreeEW}$}
    & {$\boldsymbol{\NLOEW}$}
    & {\multirow{2}{*}{$\boldsymbol{\dTreeEW}$}}
    & {\multirow{2}{*}{$\boldsymbol{\dNLOEW}$}}
    & {\multirow{2}{*}{$\boldsymbol{\dNLO}$}}\\
    & {$\mathcal{O}(\alphas^2)$}
    & {$\mathcal{O}(\alphas\alpha+\alpha^2)$}
    & {$\mathcal{O}(\alphas^2\alpha)$}
    &
    &
    &\\
    \hline\trule
    ${\boldsymbol{\qL \qL^{\prime}}}$ & ~7.359(1) & 1.0326(2) & $-$0.5776(7) & 14.0 \% & $-$7.8 \% & ~ 6.2 \%\\ 
  \hline\trule
    ${\boldsymbol{\qR \qR^{\prime}}}$ & ~7.529(1) & 0.1005(1) & $-$0.0052(1) & ~1.3 \% & $-$0.1 \% & ~ 1.3 \%\\  
      \hline\trule
    ${\boldsymbol{\qL \qR^{\prime}}}$ & 14.651(1) & 0.0136(1) & $-$0.8676(2) & ~0.1 \% & $-$5.9 \% & $-$5.8 \%\\
    \hline
    \hline\trule
    ${\boldsymbol{\squark \squark^{\prime}}}$ & 29.539(2) & 1.1468(2) & $-$1.4506(7) & ~3.9 \% & $-$4.9 \% & $-$1.0 \%\\
  \hline
}{Same as Table~\ref{tab_SPA} but for the SPS2 scenario.
  \label{tab_SPS2}}

\TABULAR[t]{c|c|c|c||c|c|c}{
  \hline\trule
    \multirow{2}{*}{\bf SPS5}
    & {$\boldsymbol{\Born}$}
    & {$\boldsymbol{\TreeEW}$}
    & {$\boldsymbol{\NLOEW}$}
    & {\multirow{2}{*}{$\boldsymbol{\dTreeEW}$}}
    & {\multirow{2}{*}{$\boldsymbol{\dNLOEW}$}}
    & {\multirow{2}{*}{$\boldsymbol{\dNLO}$}}\\
    & {$\mathcal{O}(\alphas^2)$}
    & {$\mathcal{O}(\alphas\alpha+\alpha^2)$}
    & {$\mathcal{O}(\alphas^2\alpha)$}
    &
    &
    &\\
    \hline\trule
    ${\boldsymbol{\qL \qL^{\prime}}}$  & ~774.7(1) & 185.71(4) & ~$-$35.9(1) & 24.0 \% & $-$4.6 \% & ~19.3 \%\\
    \hline\trule
    ${\boldsymbol{\qR \qR^{\prime}}}$  & ~888.0(1) & 16.332(5) & ~$-$0.69(2) & 1.8 \% & $-$0.1 \% & ~ 1.8 \%\\
    \hline\trule
    ${\boldsymbol{\qL \qR^{\prime}}}$  & 758.00(9) & 1.1559(3) & $-$33.68(1) & 0.2 \% & $-$4.4 \% & $-$4.3 \%\\
    \hline
    \hline\trule
    ${\boldsymbol{\squark \squark^{\prime}}}$  & 2420.7(3) & 203.20(4) & ~$-$70.3(1) & 8.4 \% & $-$2.9 \% & ~ 5.5 \%
    \\   \hline
}{Same as Table~\ref{tab_SPA} but for the SPS5 scenario.
  \label{tab_SPS5}}

Tables \ref{tab_SPA}--\ref{tab_SPS5} give the results for the
hadronic cross sections for squark--squark production at the LHC
 within the \SPA{}, SPS2, and SPS5 scenario, respectively. 
We refer to the production of squarks of different chiralities separately. 
 Renormalization and factorization
scales are set to $\mu=560~\GeV$ (\SPA{}), $\mu=1560~\GeV$ (SPS2)
and $\mu=666~\GeV$ (SPS5). 

The Born cross section is QCD mediated and does not depend on the
chirality of the produced squarks. Indeed the cross sections for the
diagonal production of two squarks, LL or RR, become equal for
degenerate masses.
The Born cross section for non-diagonal LR production, however, is
different in general since the final-state particles are distinguishable.
In the \SPA{} and SPS5~scenario it happens to be of similar size as
the LL and RR production cross sections. In the SPS2~scenario,
however, the LR cross section is enhanced and accounts for $50\%$ of
the total cross section for inclusive squark--squark production. As we
will see below, \cf~\figref{fig_scan3}\,(b), the relative yield of the
LR production cross section is determined by the ratio of squark and
gluino masses, becoming more important if the exchanged gluino is
lighter than the final-state squarks.

Due to the nature of the electroweak interaction, the EW cross section
contributions depend strongly on the chirality of the produced
squarks.
The impact of tree-level EW contributions is largest in case of LL
production ($15-25\%$ in the considered scenarios) and around one
order of magnitude smaller for RR production. It is even further
suppressed for LR production, where the $\Order(\alpha_s \alpha)$
tree-level interference contributions are completely absent.
The situation is different for the NLO EW contributions. These are
equally important in the case of LL and LR production (reducing the LO
prediction by about $4-8\%$ in the considered scenarios), but
negligible for RR production. In all three cases the NLO EW
corrections are negative and partially compensate the EW tree-level
contributions.
Summing over all processes, 
the EW contributions to the total cross section 
for inclusive squark--squark production decrease
 from about $8\%$ to about $5\%$ in
\SPA{} and SPS5 after the inclusion of NLO EW corrections.
In the SPS2 scenario, where LR production is the dominant
production mechanism, the NLO EW corrections even
overcompensate the EW tree-level contributions and the 
result turns negative.

\medskip


In order to further investigate the dependence of the EW contributions
on squark and gluino masses, we perform a parameter scan on those 
quantities. The independent squark masses are chosen equal to a common
value $m_{\squark}$, with the dependent fourth squark mass set to its
corresponding on-shell value.  All other parameters are fixed to their
\SPA{} values.
The renormalization and factorization scale are identified with the
common squark mass $\mu_{R}=\mu_{F}=m_{\squark}$.

%
\FIGURE[t]{
    \includegraphics[width=.49\textwidth]{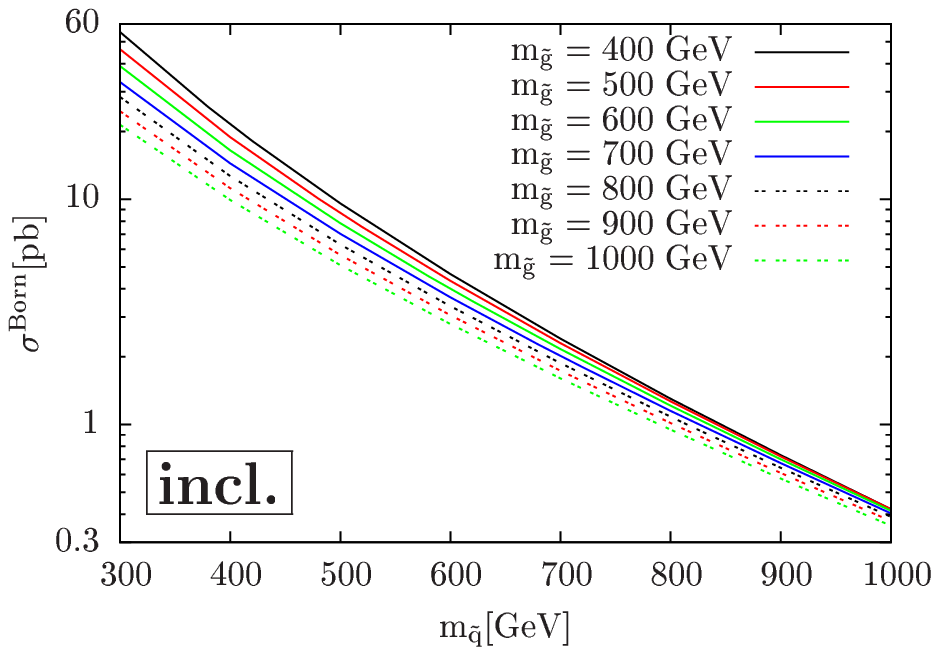}
    \includegraphics[width=.49\textwidth]{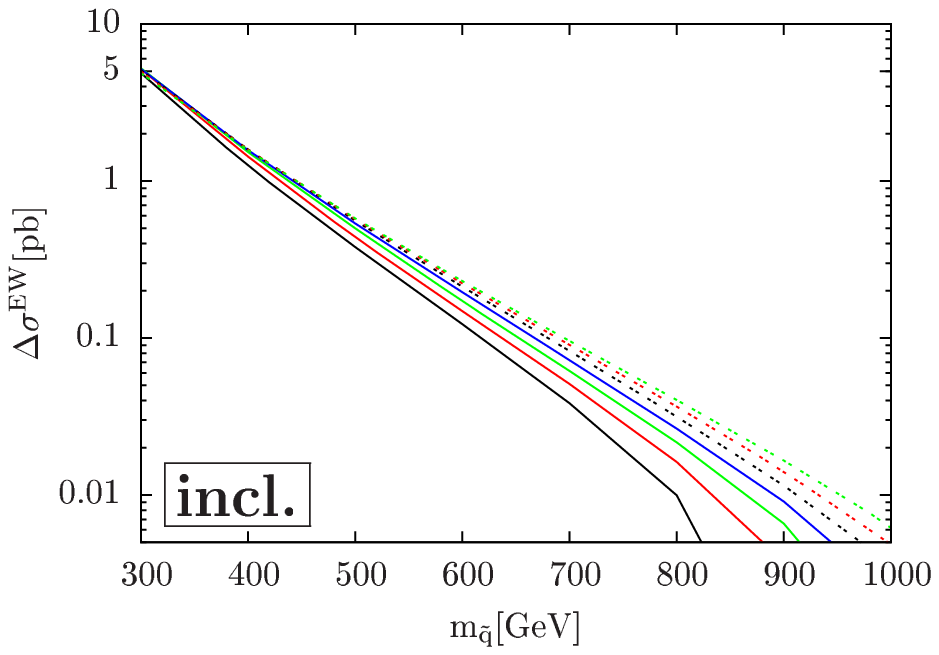}
  \caption{Hadronic Born cross section (left) and EW contributions
    (right) for inclusive squark--squark production as a function of a
    common squark mass $m_{\squark}$. Different gluino masses
    $m_{\gluino}$ are considered, all other parameters are
    set to their \SPA{} values.
    \label{fig_scan1}}
}
To start with, we show in \figref{fig_scan1} the Born cross
section (left panel) and the EW cross section contributions (right panel) 
for inclusive squark--squark 
production as a function of the common squark mass and
 for different values of the gluino mass.
Both the Born and the EW contribution strongly decrease for growing
squark masses. While the Born cross section is quite sensitive to the
gluino mass for low squark masses, the EW contribution is almost
independent in this regime. For high squark masses the behavior is vice versa.


\FIGURE[t]{
  \includegraphics[width=.49\textwidth]{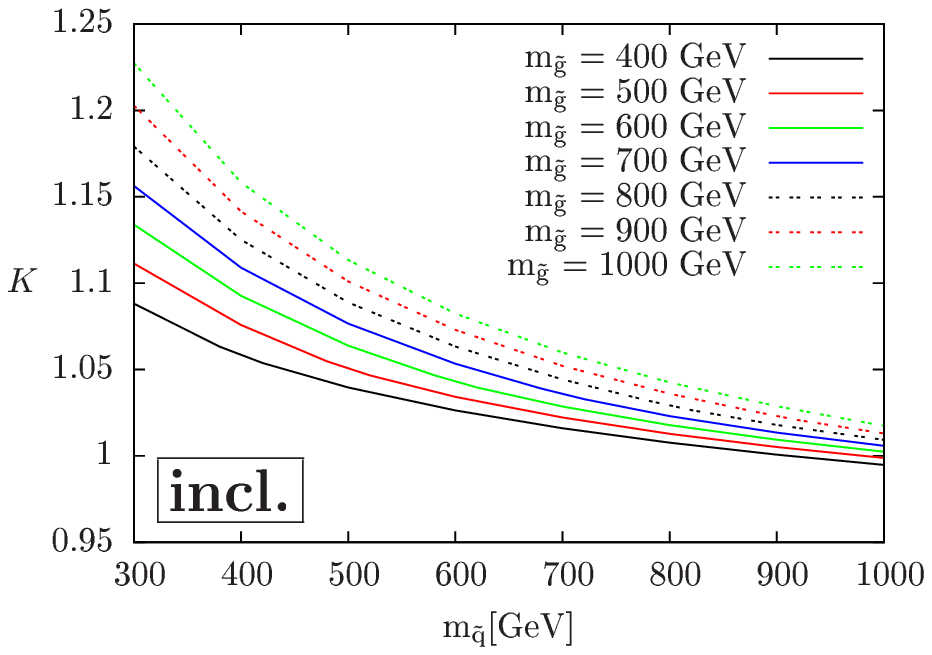}
  \includegraphics[width=.49\textwidth]{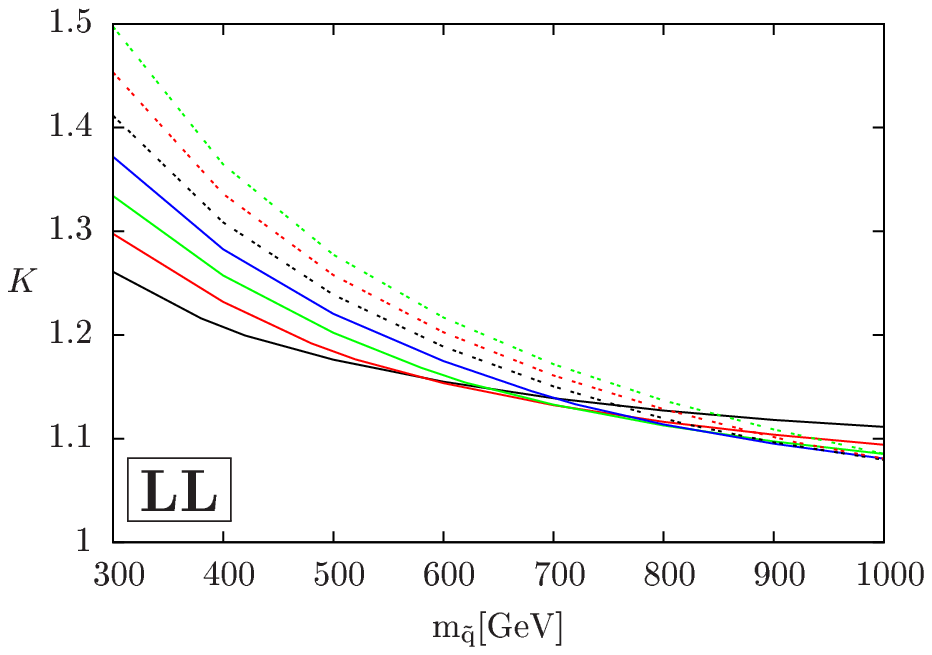}\\[15pt]
  \includegraphics[width=.49\textwidth]{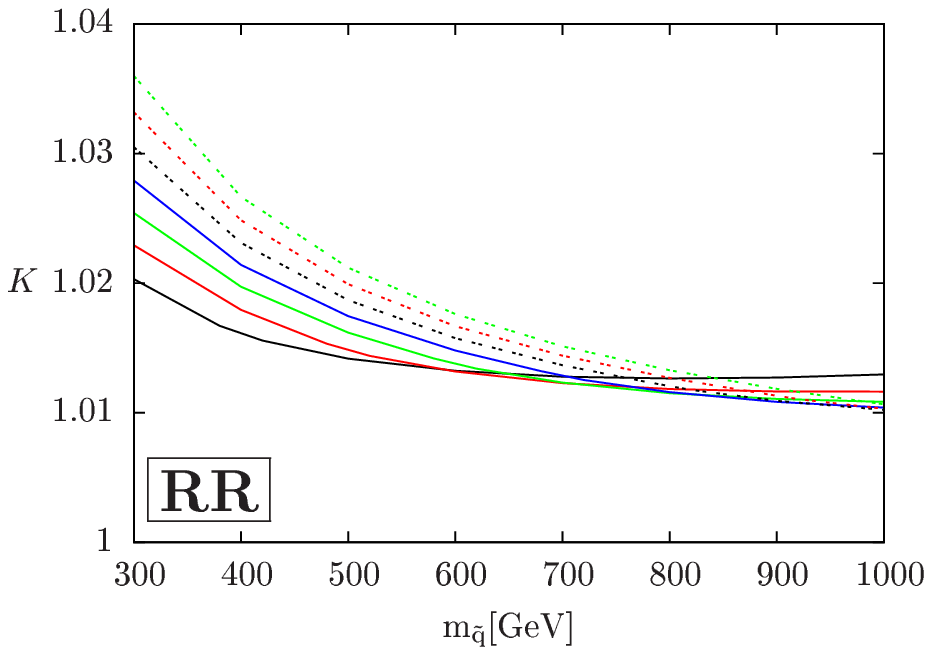}
  \includegraphics[width=.49\textwidth]{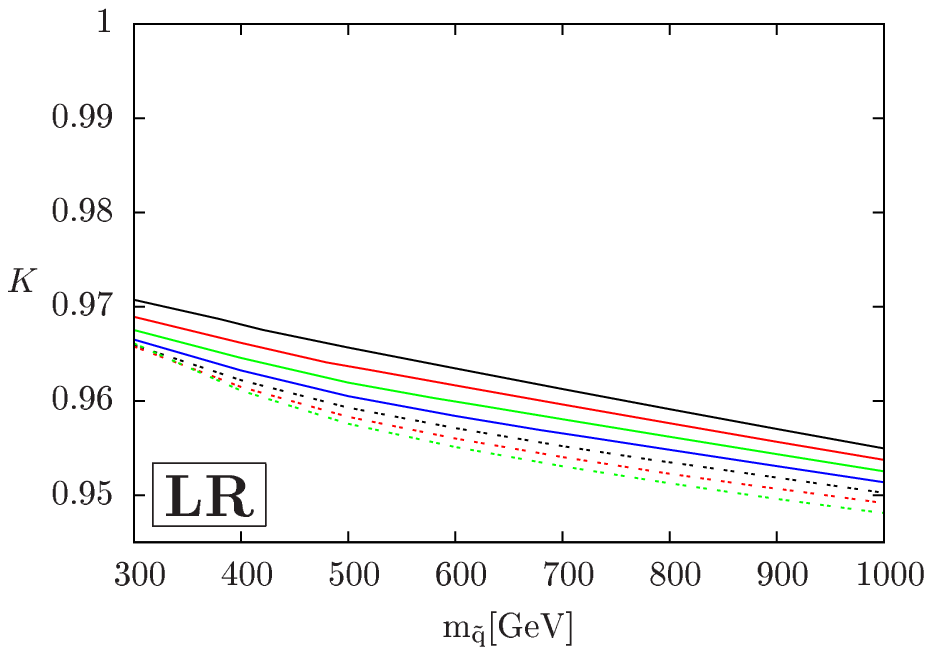}
  \caption{$K$ factor defined as $K=\NLO/\Born$, as a function of a
    common squark mass.  Different gluino masses $m_{\gluino}$ are
    considered, all other parameters are set to their \SPA{} values.
    The labels ``incl.'' and ``LL'', ``RR'', ``LR'' refer to inclusive
    squark--squark production and chirality-grouped subprocesses as
    explained in the text.}
  \label{fig_scan2}
}
\figref{fig_scan2} shows the $K$~factor, defined as $K=\NLO/\Born$,
for the same parameter range as considered in \figref{fig_scan1}. The three different combinations
of chiralities of the final-state squarks, as well as the
inclusive case are considered separately. 
The $K$~factor is largest for two left-handed squarks in the final
state. Here, the EW contributions alter the LO cross section prediction between
$10-50\%$, being most important in case of light squarks and a heavy gluino. The
EW contribution is enhanced by the large tree-level EW
contribution.
In the case of RR production, the EW contributions are below
$3\%$ in most parts of the parameter space.
For LR production the EW contributions
are mainly given by the NLO EW corrections, leading to a
 $K$~factor smaller than unity. The LO cross section is reduced by $-3\%$ to
$-5\%$, most strongly in a scenario with heavy squarks and a heavy gluino. 
Altogether one finds for inclusive squark--squark production 
 EW contributions that range from
$9\%$ for $m_{\gluino}=400~\GeV$ up to $22\%$ for
$m_{\gluino}=1000~\GeV$ for light squarks. For heavy squarks, the EW
contributions are only at the percent level due to the interplay of
positive EW corrections in the LL and RR case and negative EW
corrections in the LR case, suppressing the EW contributions by one
order in magnitude.


\FIGURE[t]{
  \includegraphics[width=.49\textwidth]{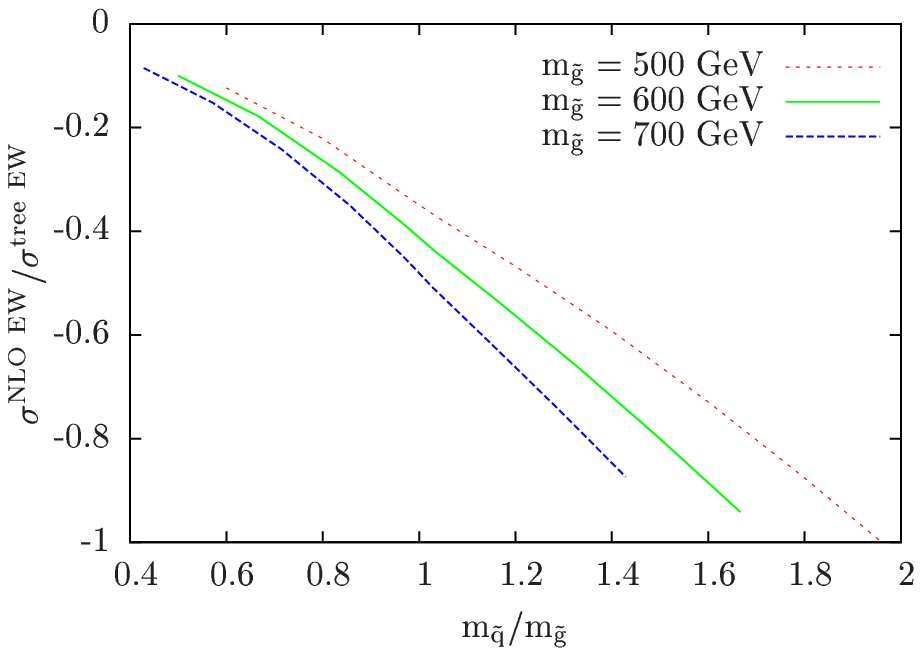}
  \includegraphics[width=.49\textwidth]{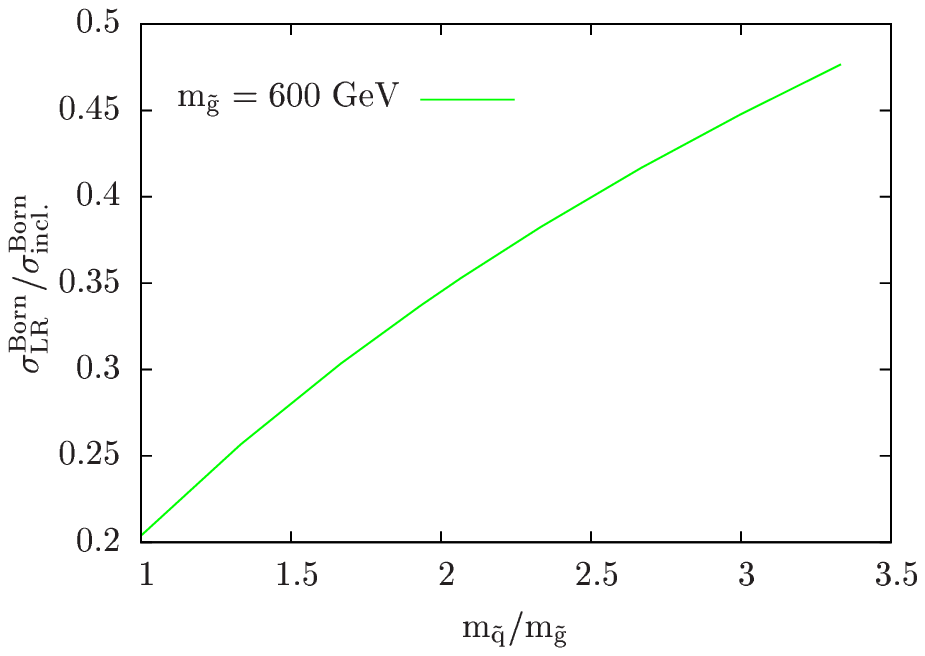}
  {\small \hspace*{1.25cm} (a) \hspace*{7.cm} (b) }
  \caption{(a) Ratio of NLO EW to tree-level EW
    contributions for inclusive squark--squark production.
 (b) Relative contribution of LR final states to the
    inclusive Born cross section for a fixed gluino mass
    $m_{\gluino}=600~\GeV$. }
  \label{fig_scan3}
  }

We can understand the smallness of the EW contribution for high squark
and low gluino masses 
by having a closer look at the interplay of the tree-level EW and NLO EW
contributions. The ratio $\NLOEW/\TreeEW$  for inclusive 
squark--squark production is shown in \figref{fig_scan3}\,(a).
As one can see, the NLO EW corrections become more important for
larger ratios $m_{\squark}/m_{\gluino}$ and reach the same size as the
tree-level EW contributions for about $m_{\squark}/m_{\gluino}\gtrsim
1.5$, depending on the precise value of the gluino mass. This is due
to the fact that the LR contribution becomes more relevant for
increasing $m_{\squark}/m_{\gluino}$, see \figref{fig_scan3}\,(b). Owing
to the suppressed tree-level contributions, the EW contributions to LR
production are negative and partially compensate the positive yield
from LL and RR production.
\figref{fig_scan3}\,(a) also confirms our observation from the SPS2
scenario that the NLO EW corrections compensate the tree-level EW
contributions in the inclusive cross section,
\cf~Table~\ref{tab_SPS2}, which seems to be a generic feature in
scenarios with the squark heavier than the gluino.

\medskip

\subsection{Differential distributions}

Here, we illustrate the results for the \SPA{} scenario. In
Figures~\ref{fig_ptdist}, \ref{fig_minvdist}, and \ref{fig_etadist} we
consider the differential distributions of the EW contributions with
respect to various kinematical variables.  In the left panels, the tree-level
EW contributions and the three gauge-invariant subsets of NLO EW
contributions (EW-type corrections, QCD-type corrections, real quark
radiation), as well as the summed EW contributions are shown. In the
right panels, the impact of the EW contributions relative to the Born
cross section, $\delta$, is given.

%
\FIGURE{
    \includegraphics[width=.49\textwidth]{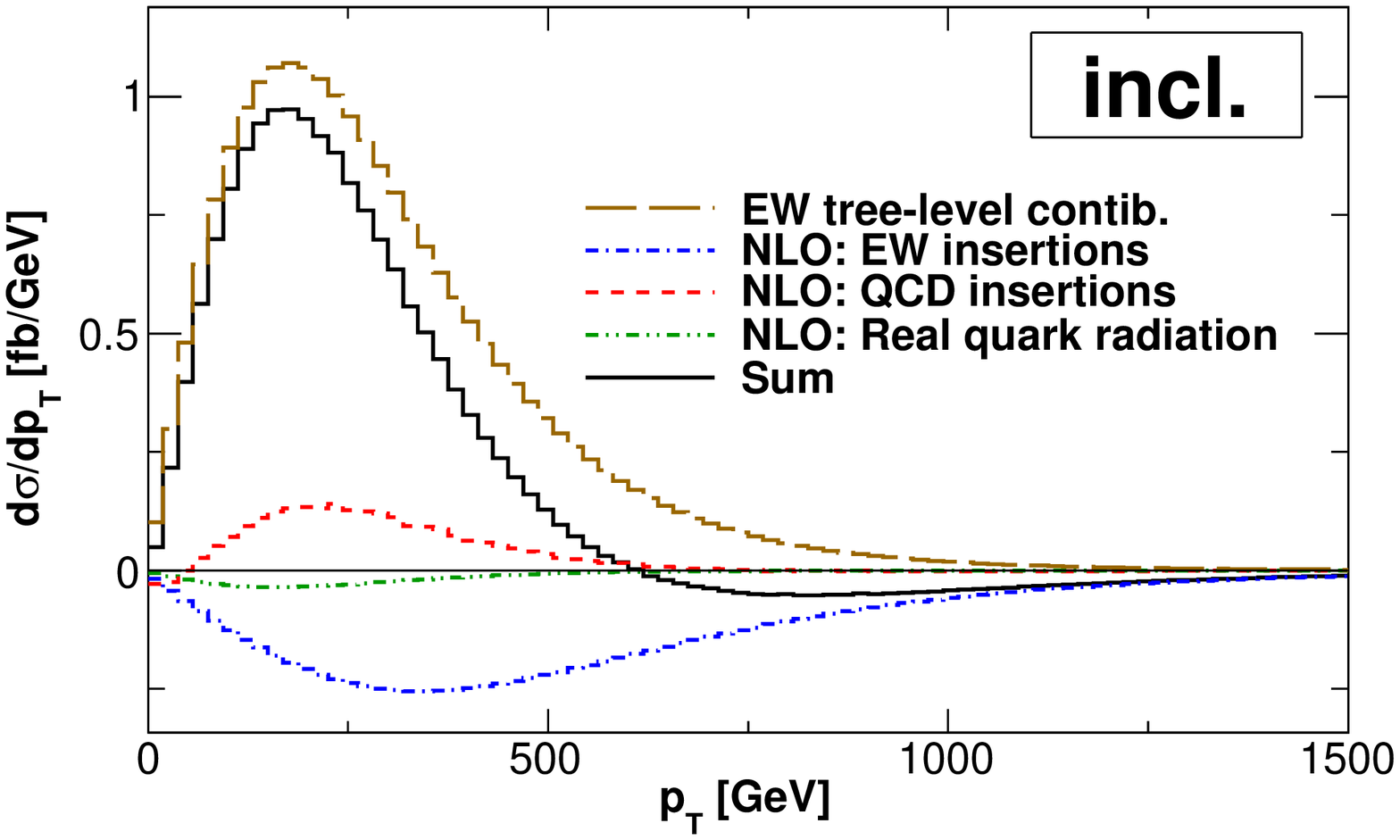}
    \includegraphics[width=.49\textwidth]{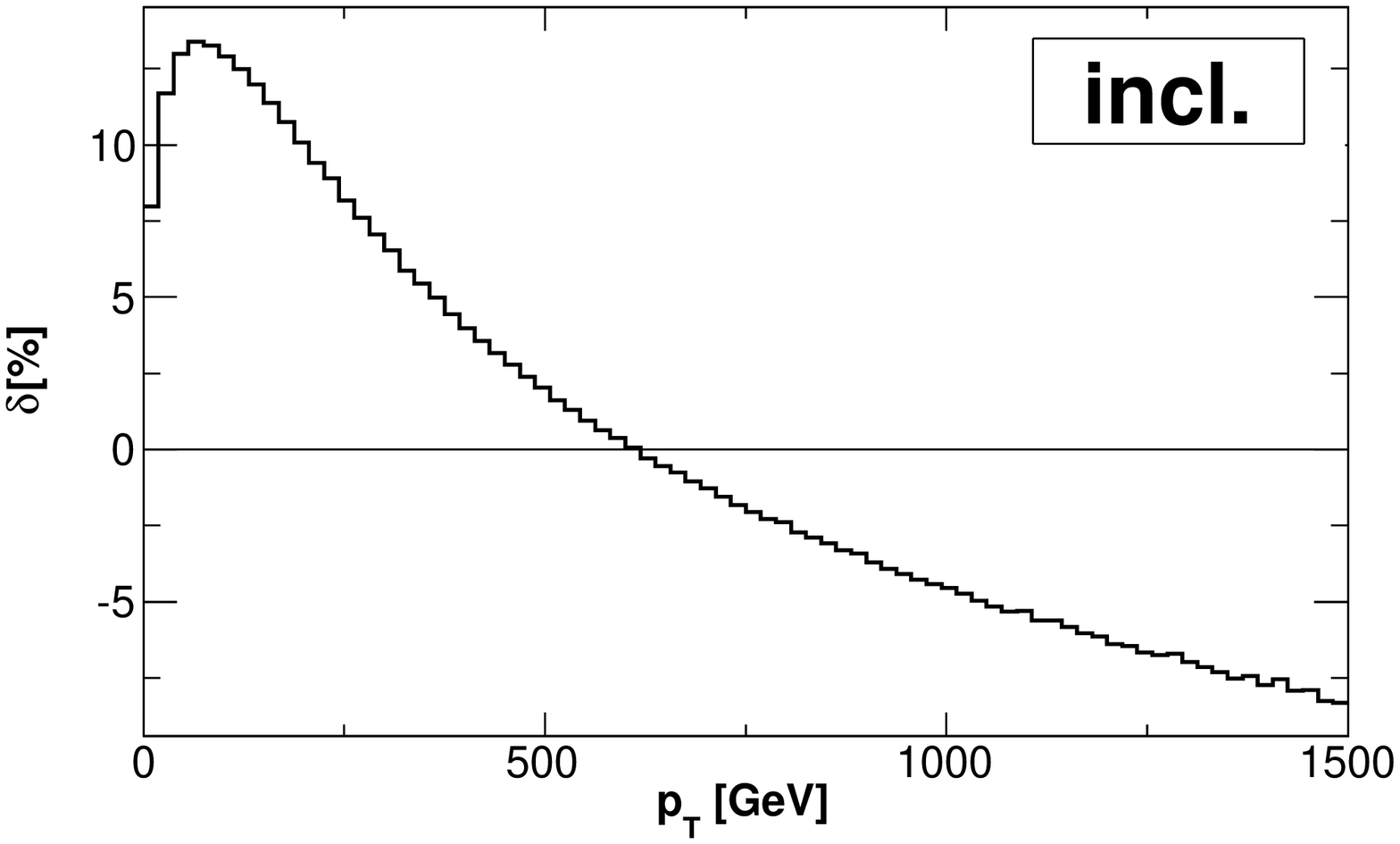}
    \includegraphics[width=.49\textwidth]{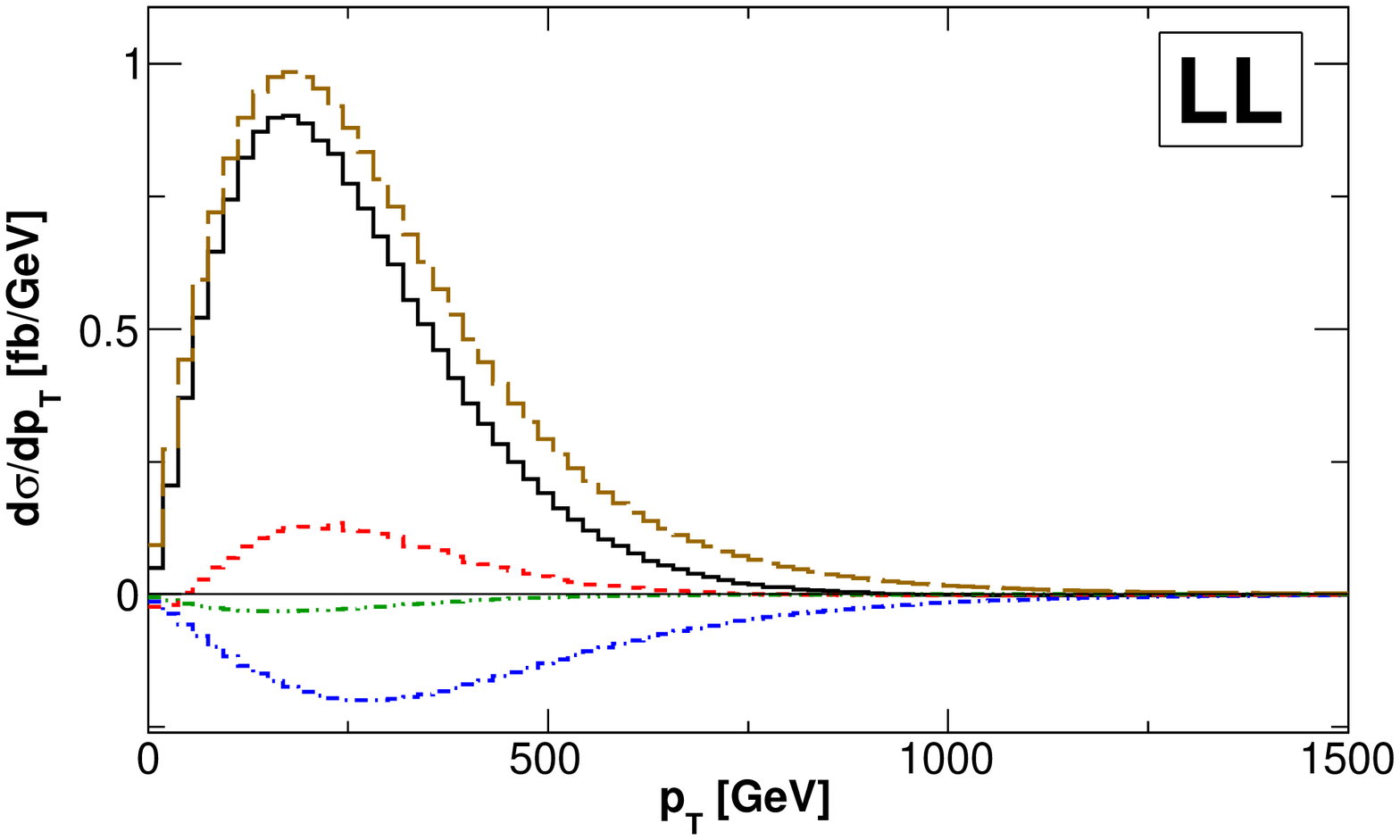}
    \includegraphics[width=.49\textwidth]{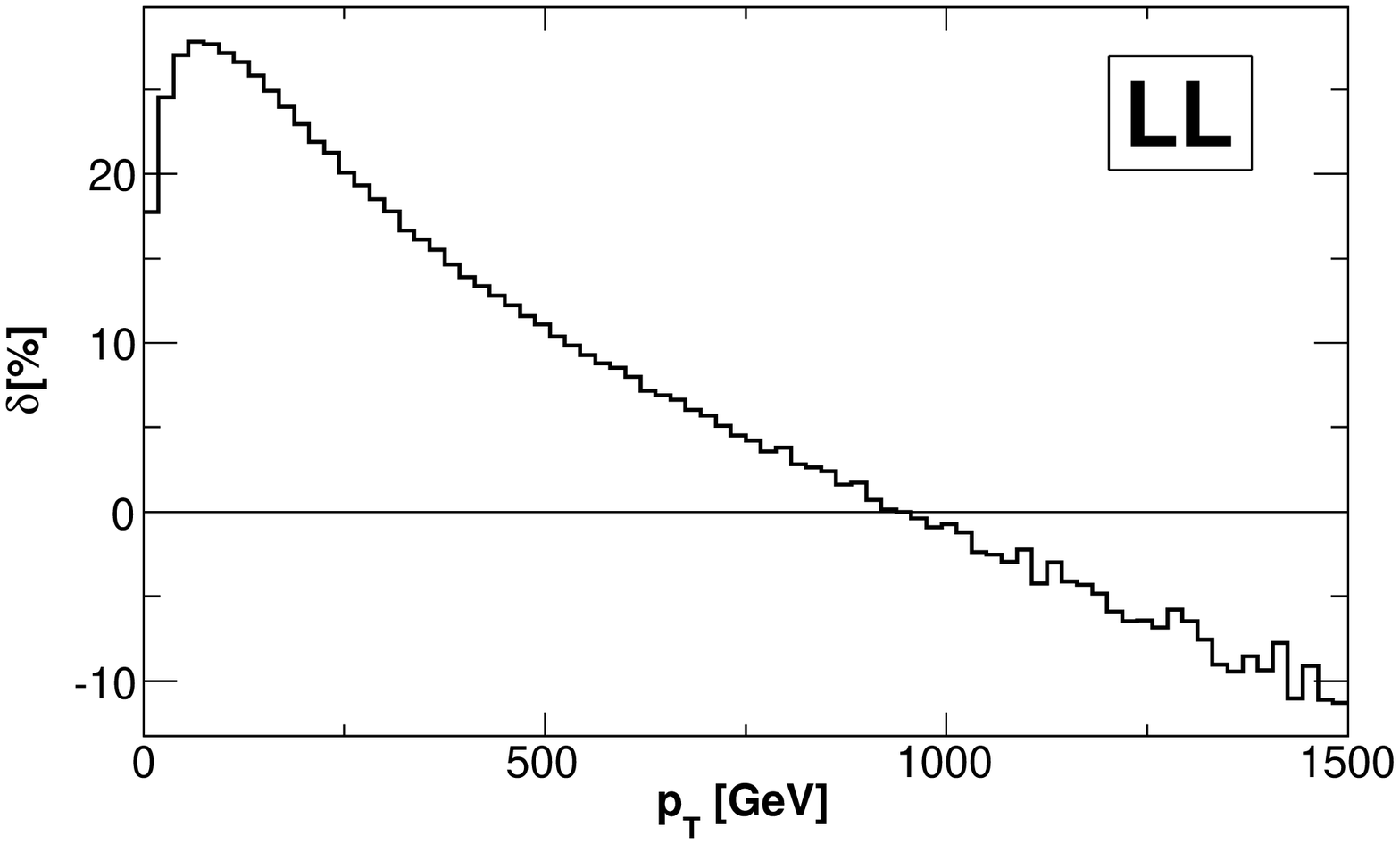}
    \includegraphics[width=.49\textwidth]{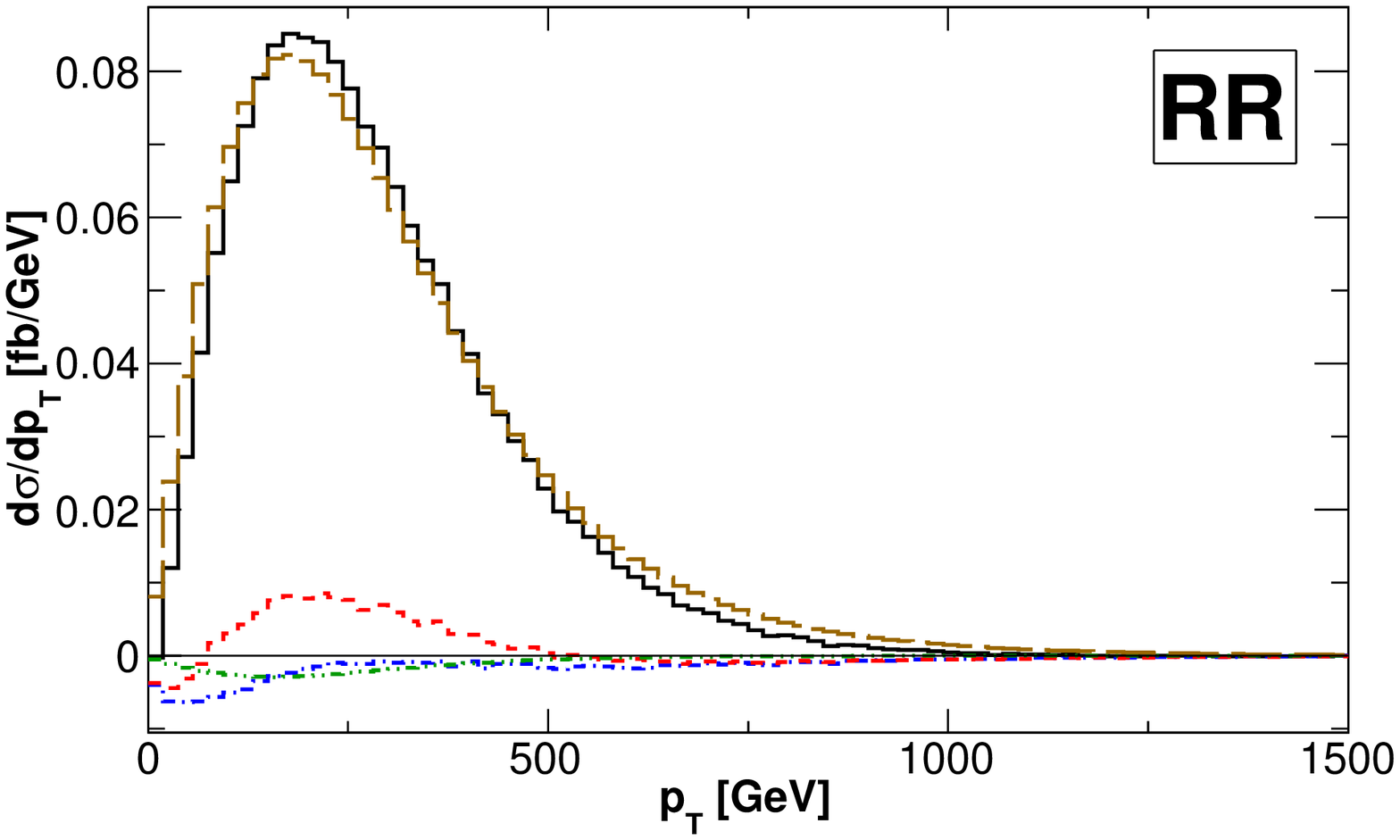}
    \includegraphics[width=.49\textwidth]{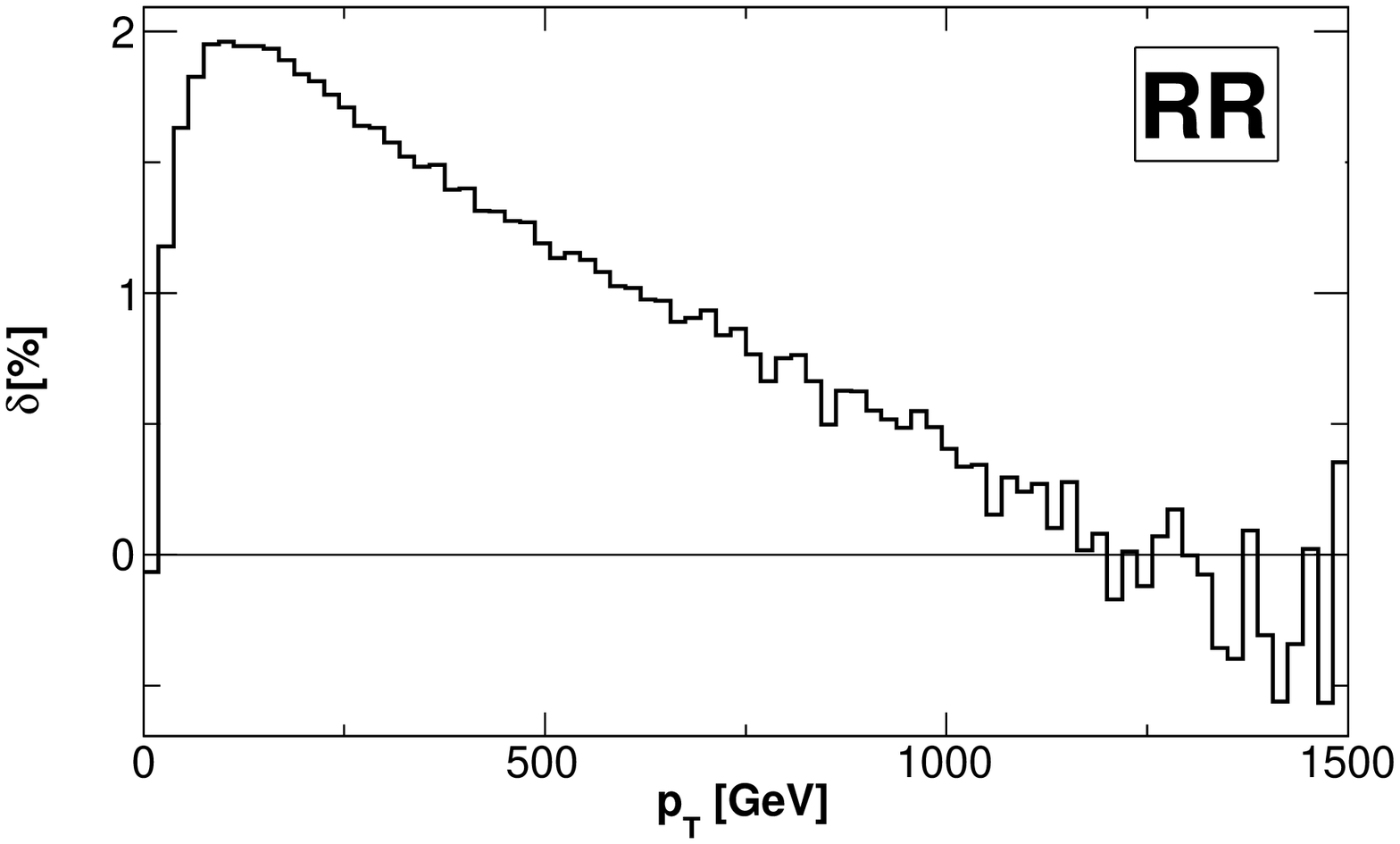}
    \includegraphics[width=.49\textwidth]{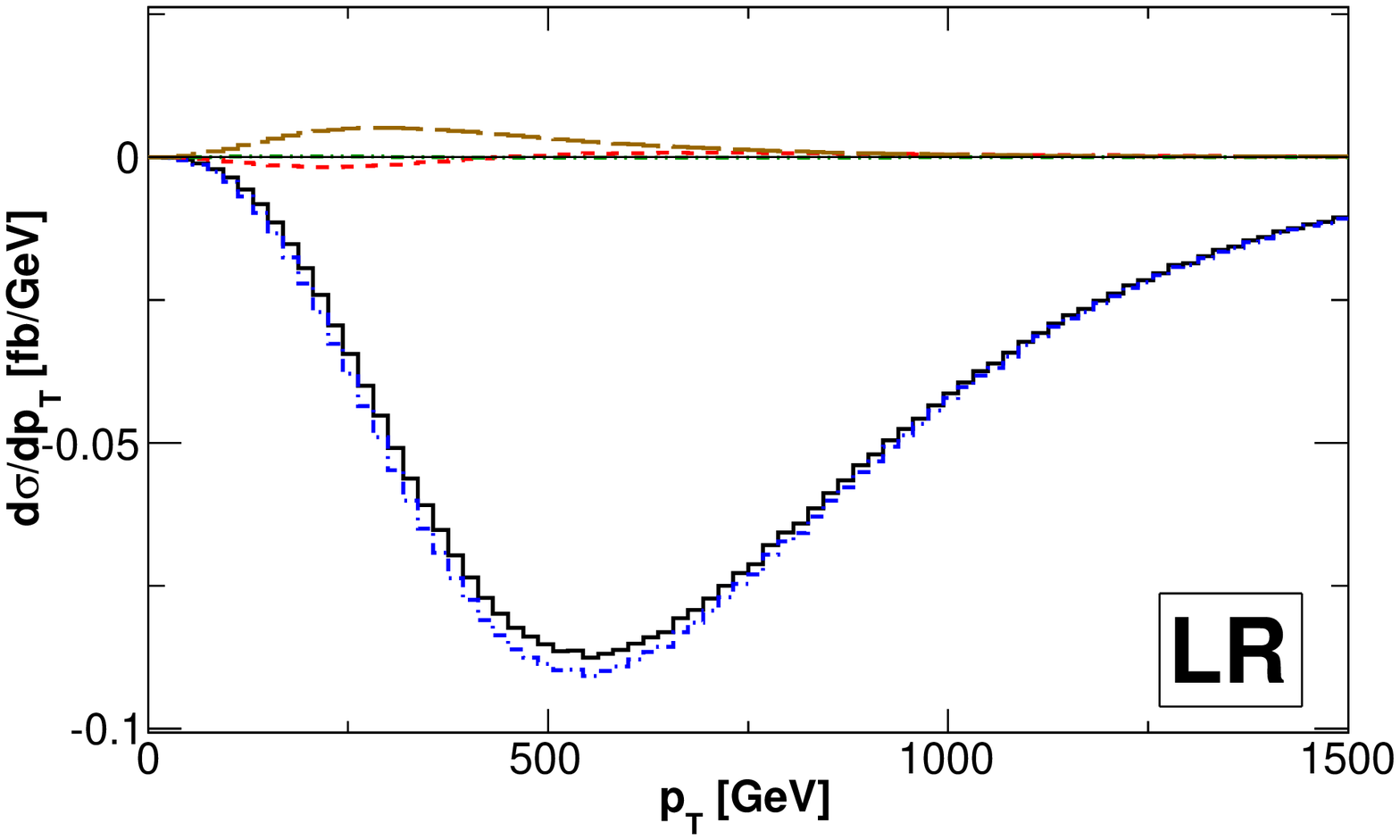}
    \includegraphics[width=.49\textwidth]{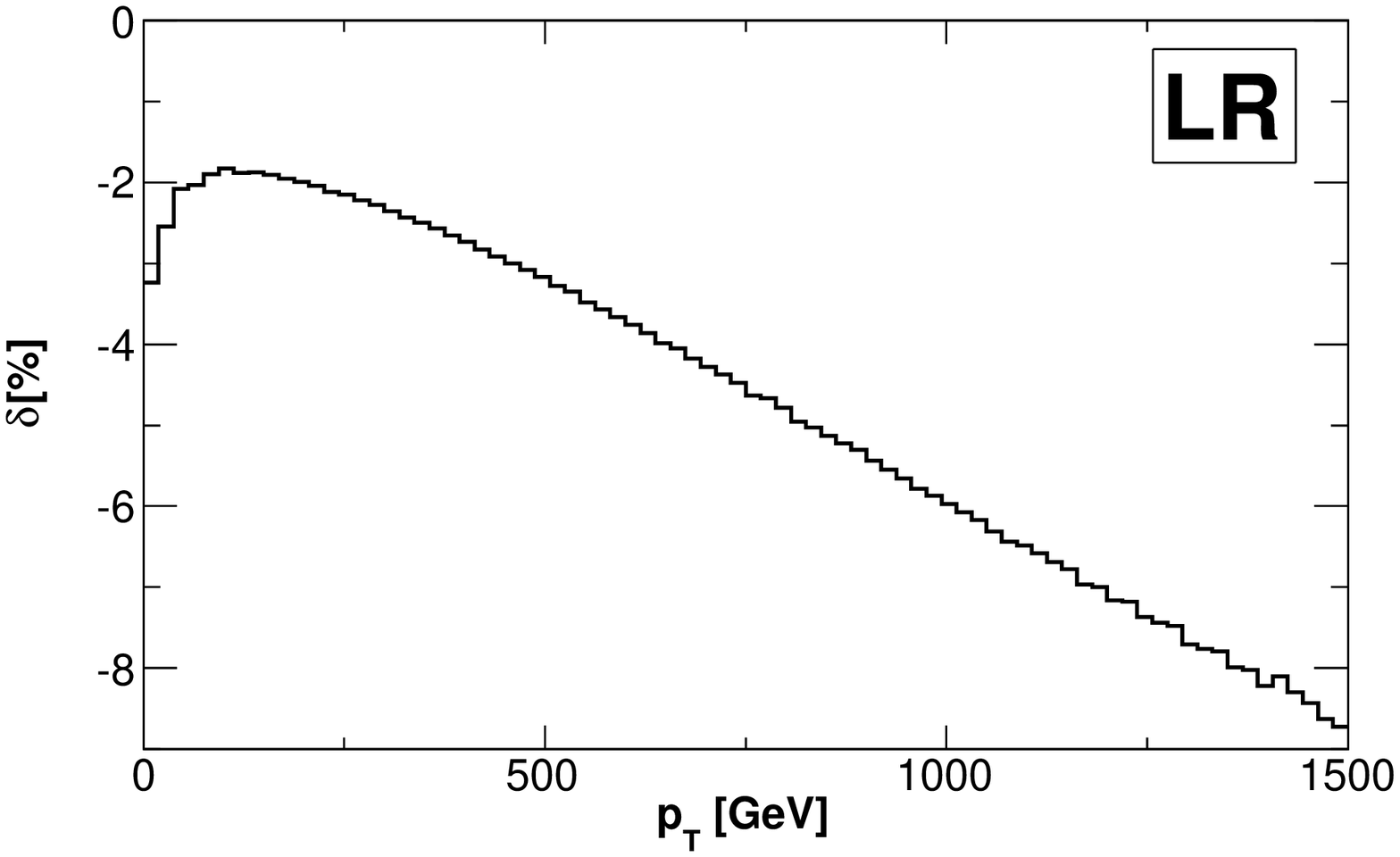}
  \caption{Transverse momentum distributions for squark--squark
    production at the LHC within the SPS1a' scenario. Shown are the
    tree-level and NLO EW cross section contributions (left) and the
    impact of EW contributions relative to the QCD Born cross section
    (right) for inclusive $\squark\squark'$ production (top),
    production of two left-handed squarks $\squark_L \squark'_L$
    (second), production of two right-handed squarks $\squark_R
    \squark'_R$ (third), and non-diagonal $\squark_L \squark'_R$
    production (bottom). Charge conjugated processes are included.}
\label{fig_ptdist}
}

\figref{fig_ptdist} refers to the distribution with respect to the
transverse momentum $\pT$ of the squark with highest $\pT$.  The
tree-level EW contributions are always positive with a maximum at
about 250 GeV and dominate the sum over a wide range of the
phase-space for LL, RR, and inclusive squark--squark
production. Again, they are suppressed for LR production.
The interplay of the NLO EW contributions is more complicated. For all
processes, the real quark radiation is small and mostly negative.
For LL production, large cancellations among the EW- and QCD-type
corrections occur.  As a result, the relative yield is dominated from
the tree-level contributions in the small-$\pT$ region where it is
large and positive (up to $25\%$).  For higher values of $\pT$, the
relative corrections turn negative and grow up to $-10\%$.  In case of
RR production the EW-type corrections are suppressed from the
chirality and the QCD-type corrections are more important. However the
relative EW contributions in total do not exceed a few percent.
Finally in the LR case, the QCD-type corrections are negligible since
they are related to QCD--EW interferences. The dominant
contribution arises here from the EW-type corrections.  The relative
contributions are always negative, between $-2\%$ for small values of
$\pT$ and up to $-10\%$ in the high-$\pT$ region.  It is important to
note that even though the relative NLO~EW contributions to the integrated
cross section are comparable for LL and LR production,
\cf~Table~\ref{tab_SPA}, they originate from distinct sources and the
differential distributions differ strongly.

%
%
\FIGURE{
    \includegraphics[width=.49\textwidth]{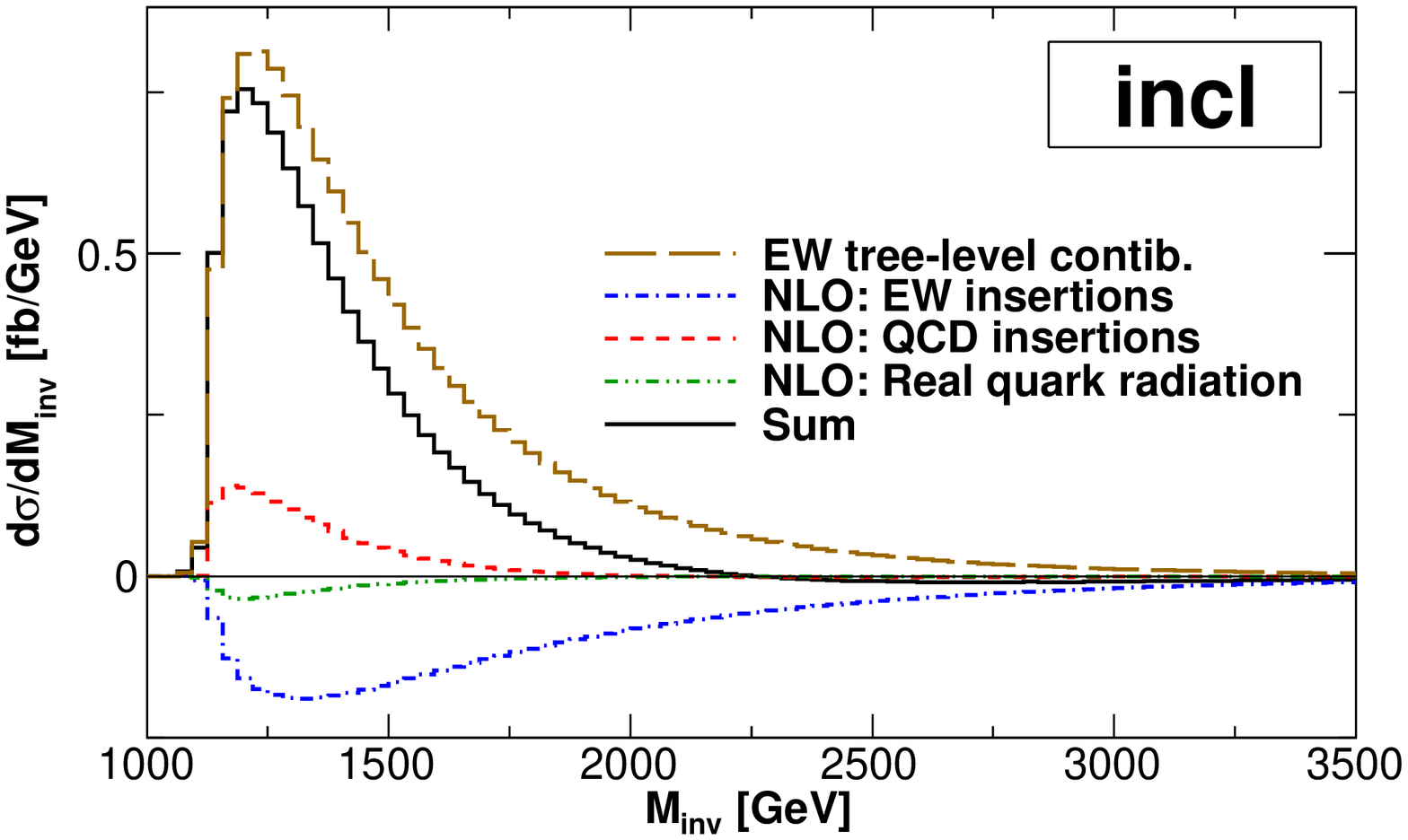}
    \includegraphics[width=.49\textwidth]{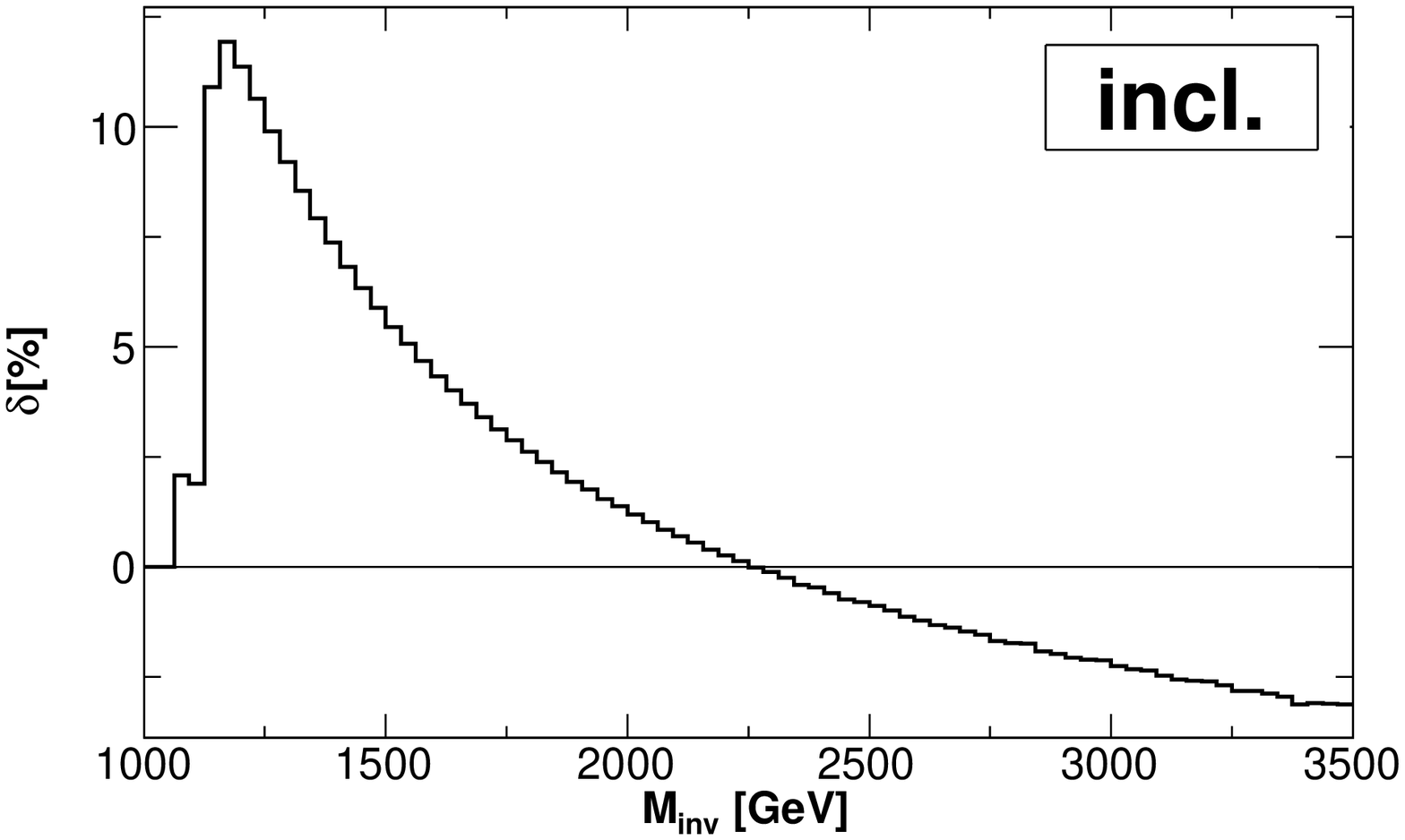}
    \includegraphics[width=.49\textwidth]{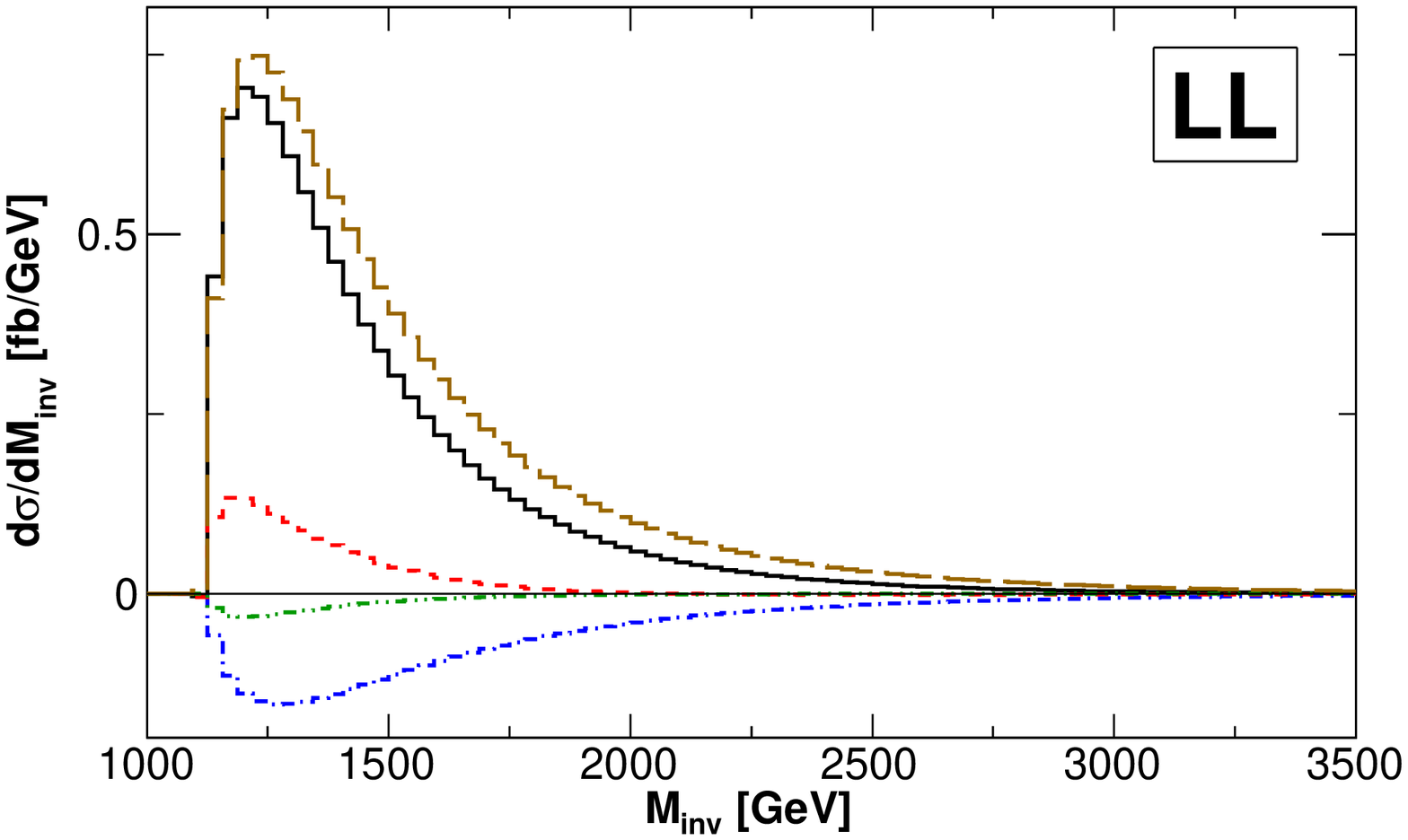}
    \includegraphics[width=.49\textwidth]{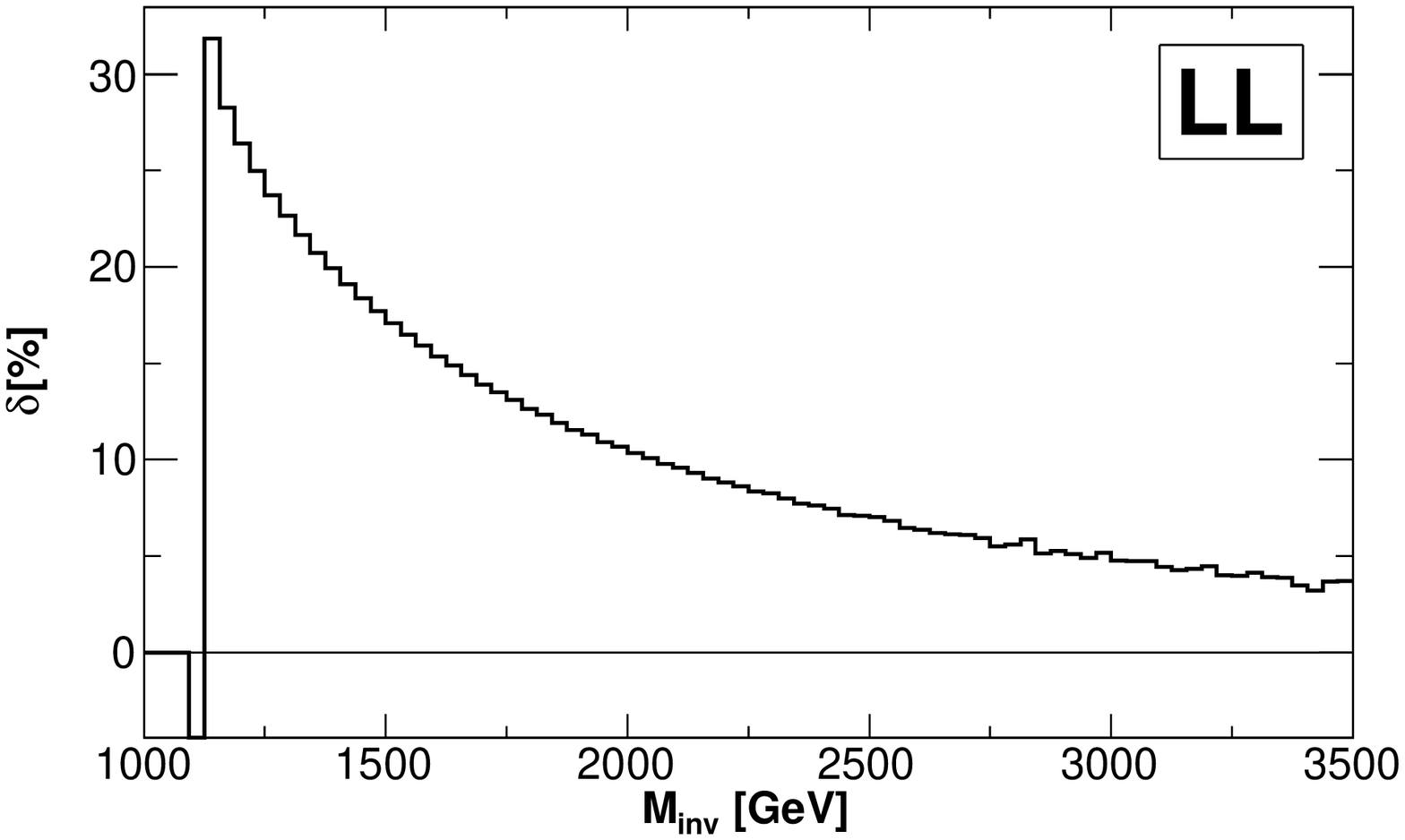}
    \includegraphics[width=.49\textwidth]{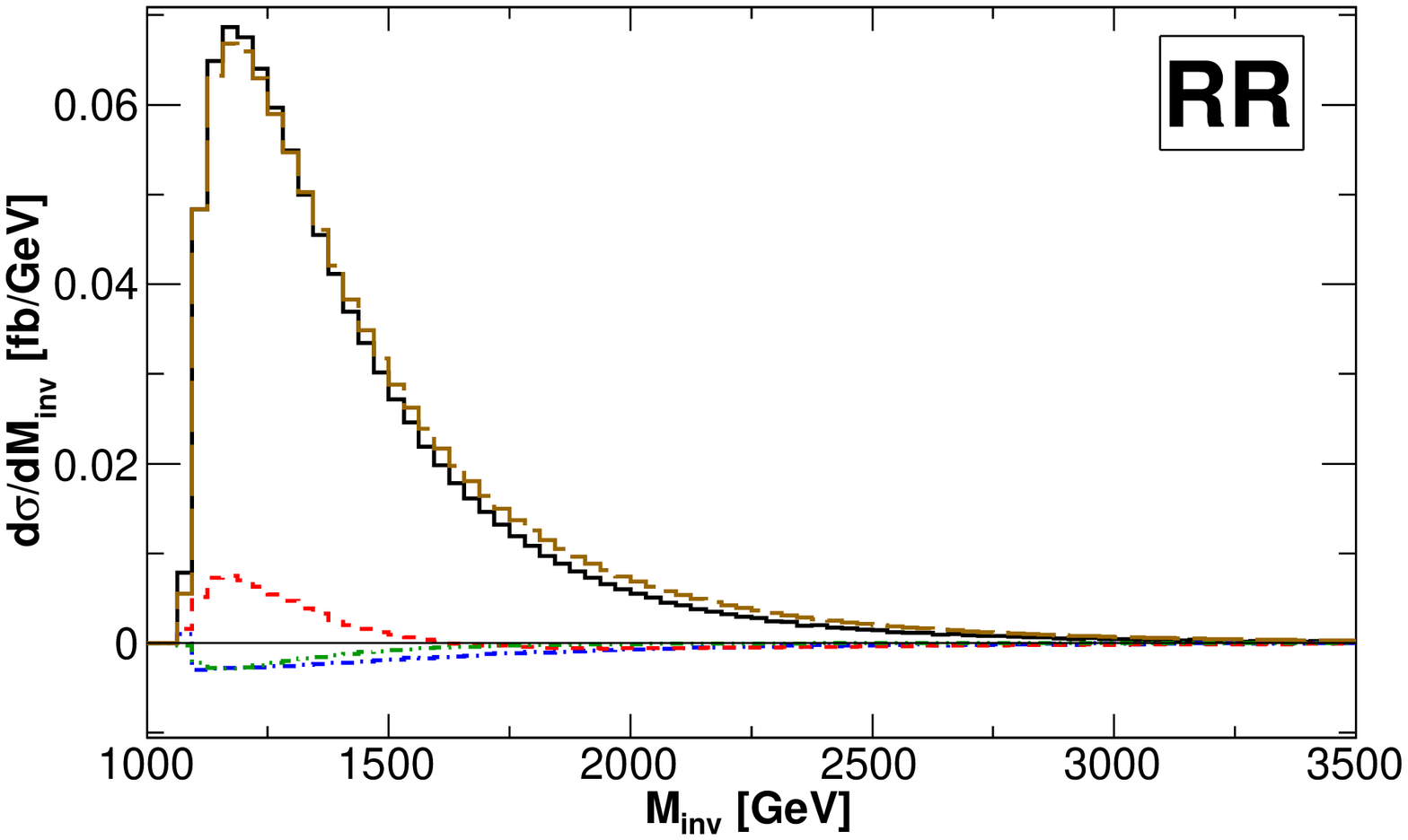}
    \includegraphics[width=.49\textwidth]{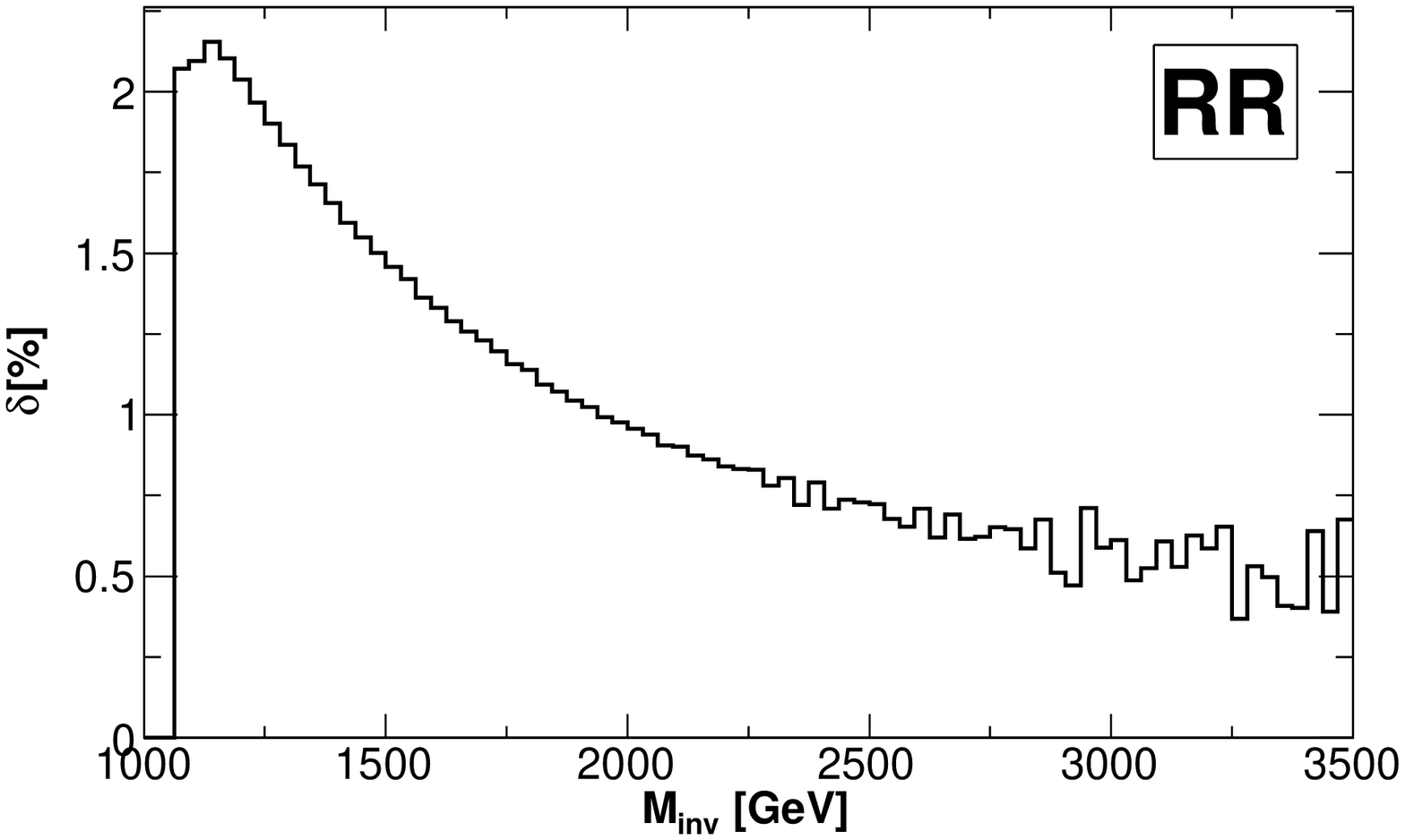}
    \includegraphics[width=.49\textwidth]{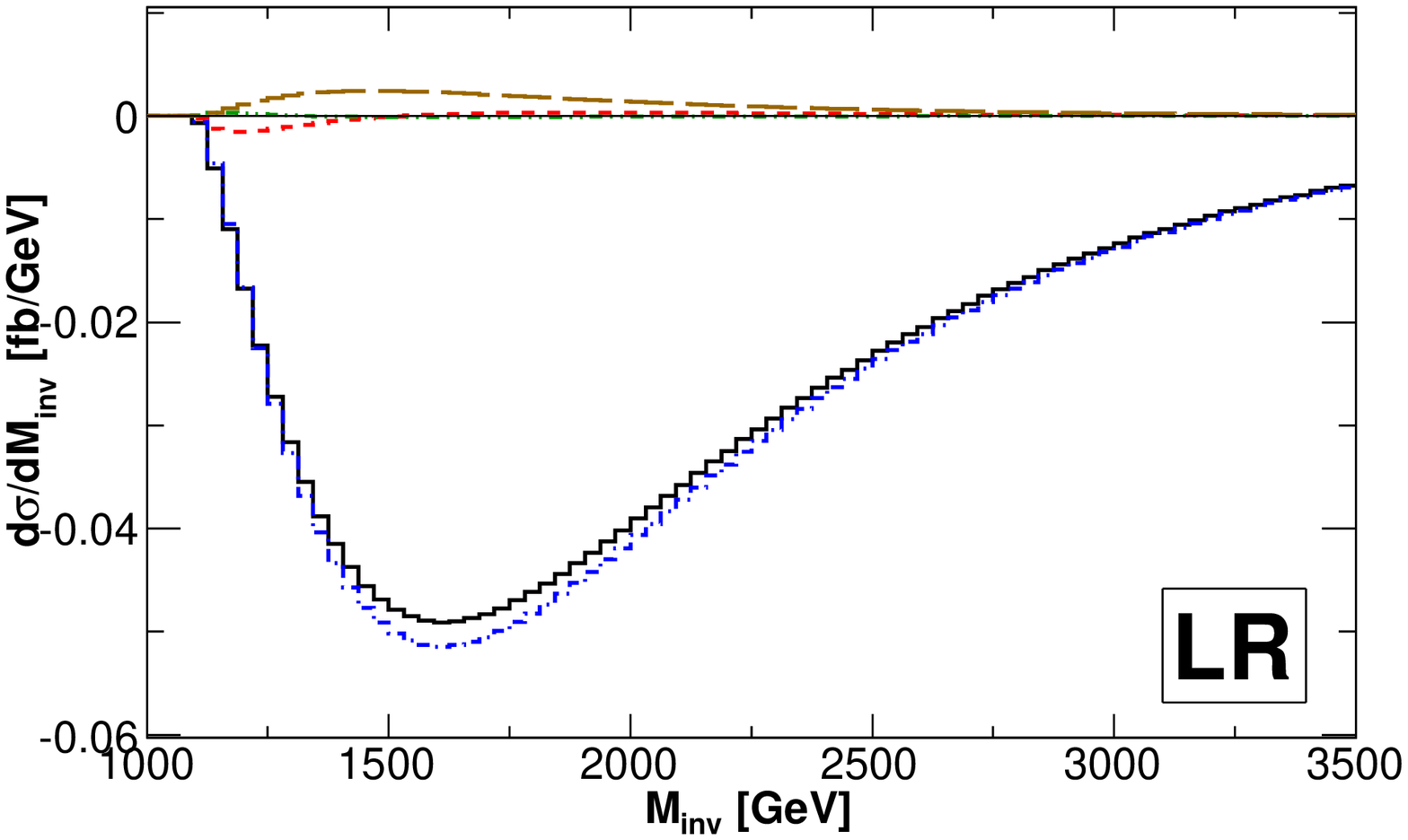}
    \includegraphics[width=.49\textwidth]{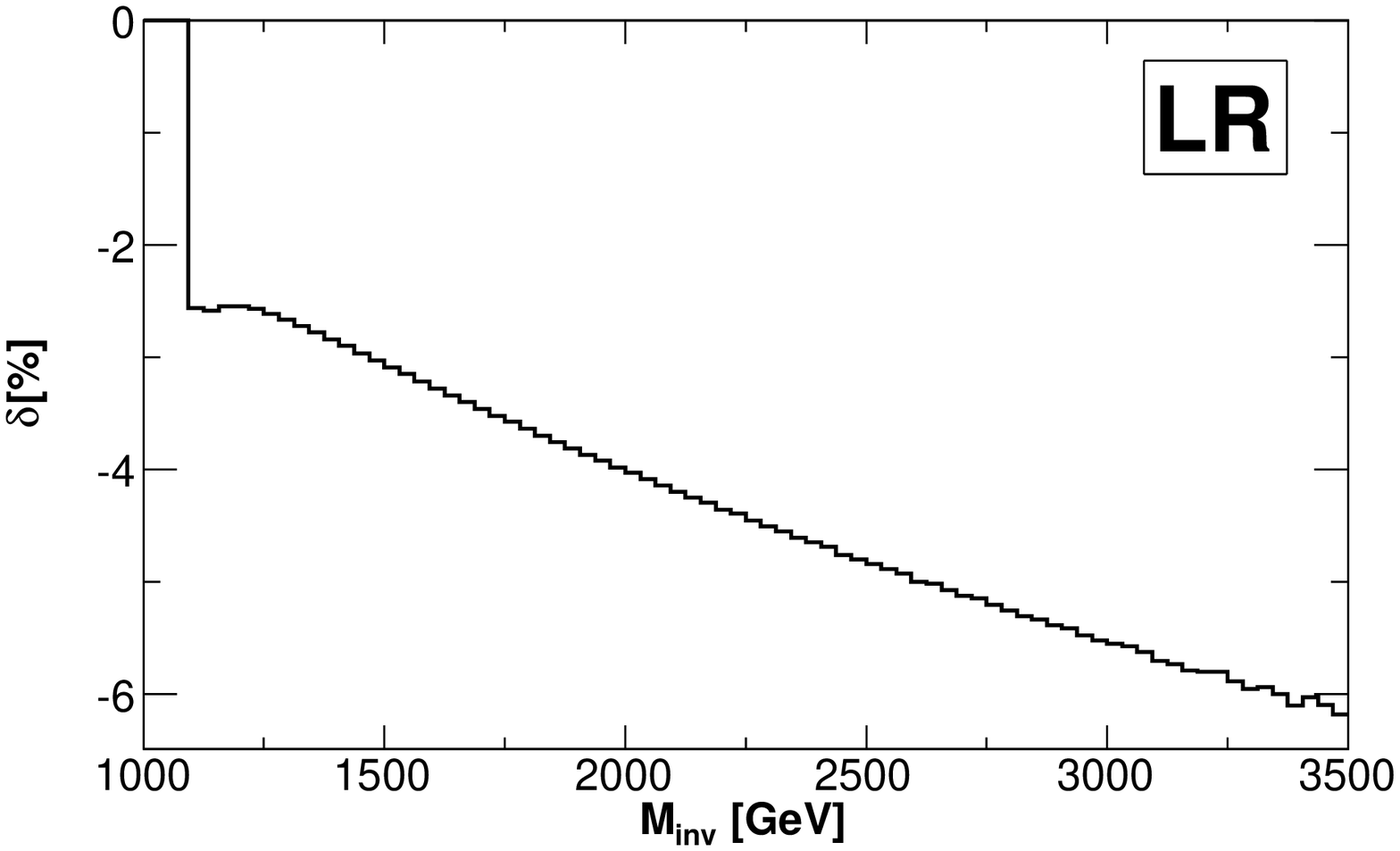}
  \caption{Invariant mass distributions for squark--squark
    production at the LHC within the SPS1a' scenario. Shown are the
    tree-level and NLO EW cross section contributions (left) and the
    impact of EW contributions relative to the QCD Born cross section
    (right) for inclusive $\squark\squark'$ production (top),
    production of two left-handed squarks $\squark_L \squark'_L$
    (second), production of two right-handed squarks $\squark_R
    \squark'_R$ (third), and non-diagonal $\squark_L \squark'_R$
    production (bottom). Charge conjugated processes are included.}
\label{fig_minvdist}
}

In \figref{fig_minvdist} the distributions with respect to the invariant mass of 
the squark pair, $M_{\rm inv} = (p_3 + p_4)^2$, are shown. The interplay 
of the various subsets of EW contributions is similar as for the $\pT$~distributions. 
For LL and RR production, the EW tree-level contributions are dominant and peak at around $M_{\rm inv} \approx 1200$\,GeV. They shift the relative EW corrections  
to positive values, up to $30\%$ in the low-$M_{\rm inv}$ region for LL production.
In case of non-diagonal LR production, where the EW tree-level contributions are suppressed, the relative corrections are negative and grow up to $-5\%$ for the intermediate and high-energy region. Finally, in the inclusive case, we find a 
strong energy dependence of the relative EW corrections, ranging from $+10\%$ for 
$M_{\rm inv} \approx 1200$\,GeV to  $-5\%$ for \mbox{$M_{\rm inv} > 3500$\,GeV}.

%
%
\FIGURE{
    \includegraphics[width=.49\textwidth]{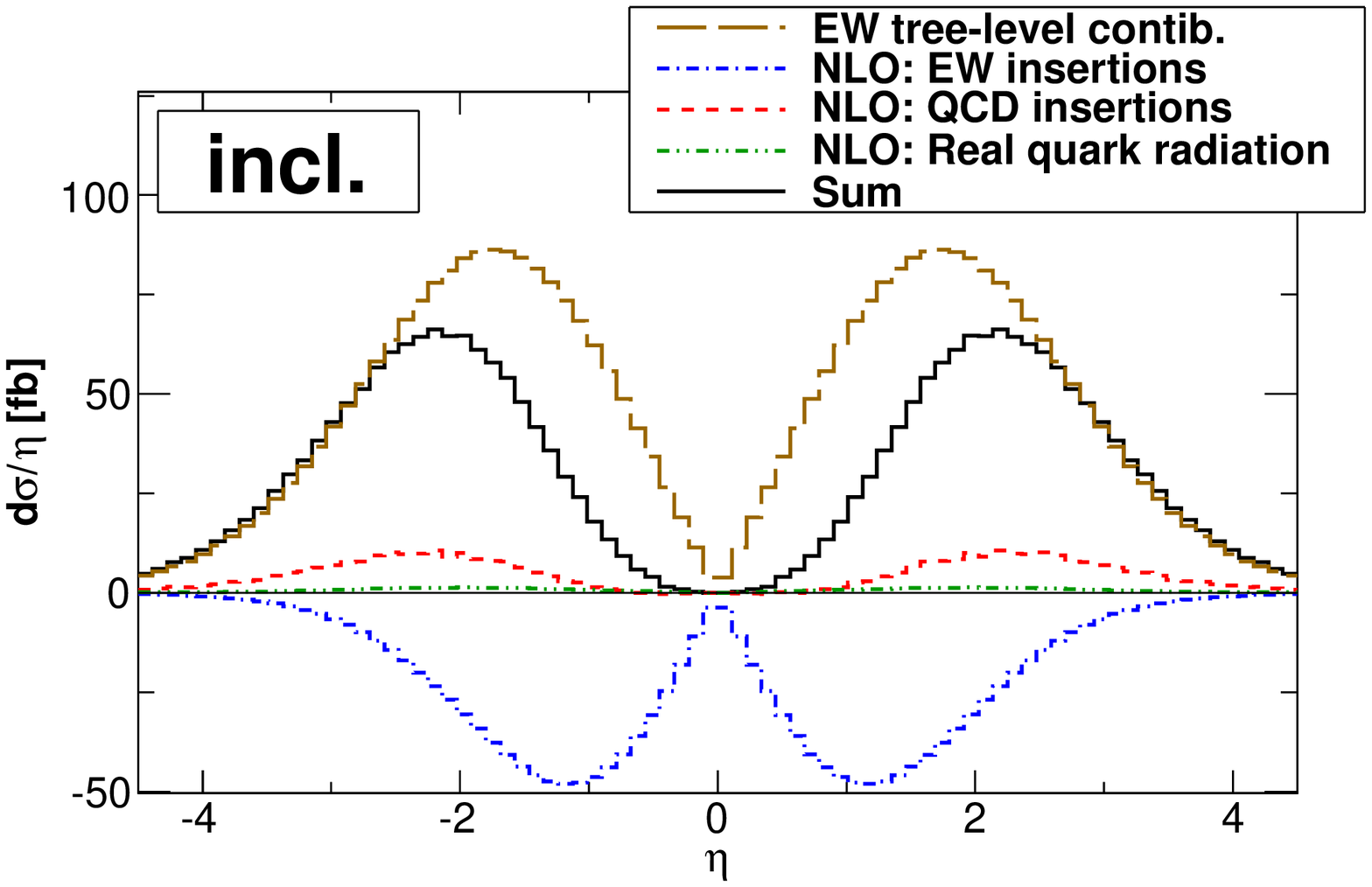}
    \includegraphics[width=.49\textwidth]{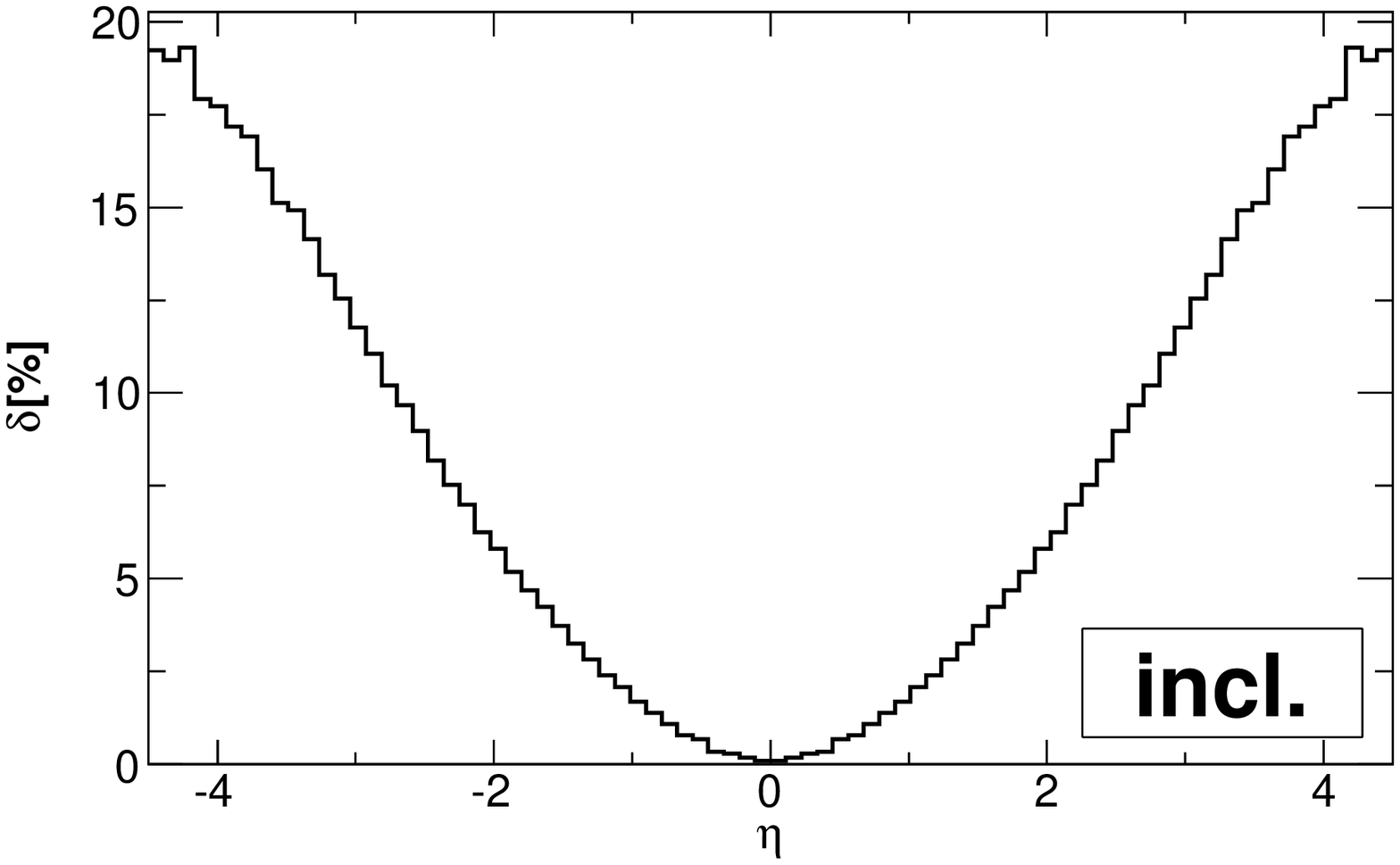}
    \includegraphics[width=.49\textwidth]{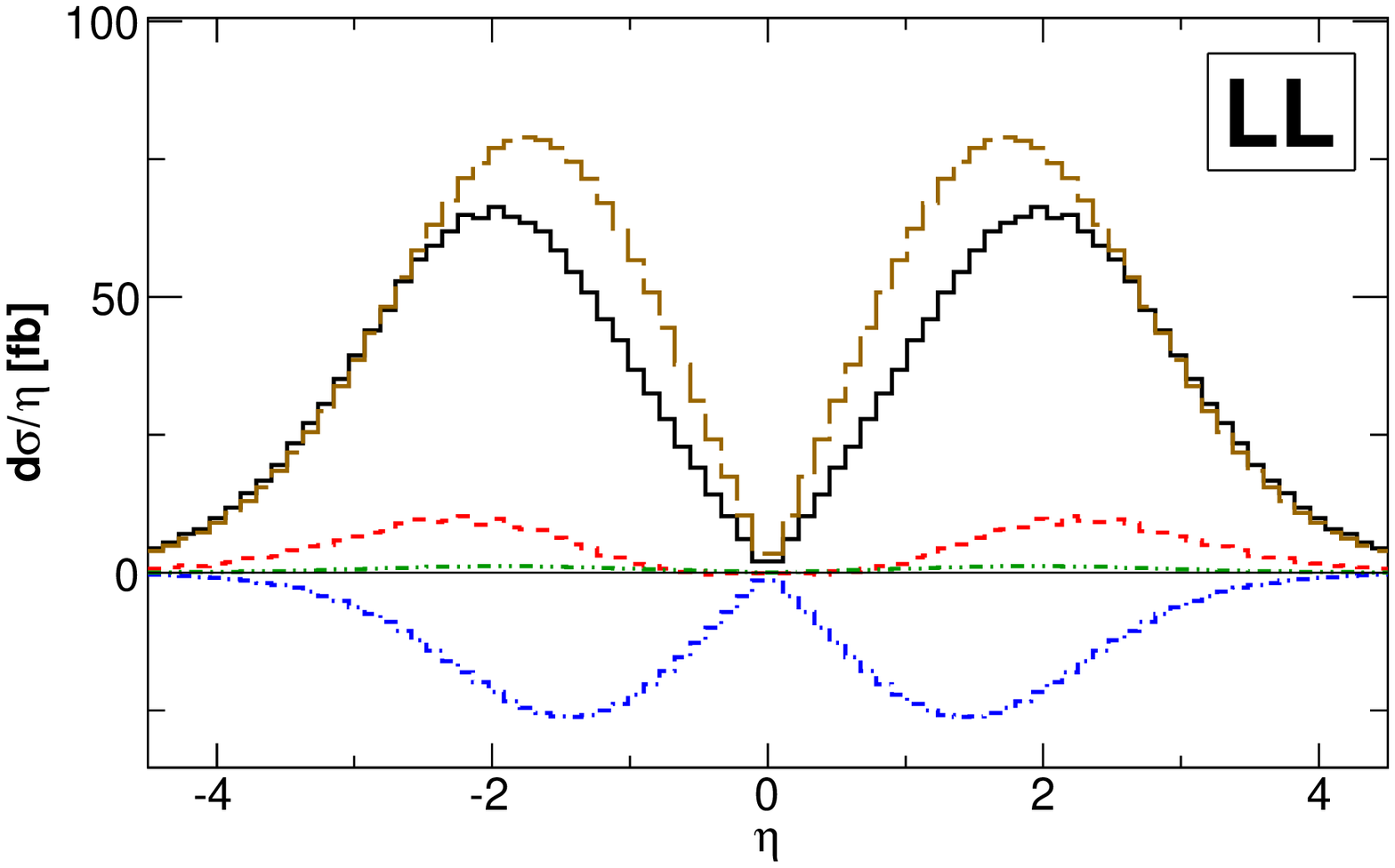}
    \includegraphics[width=.49\textwidth]{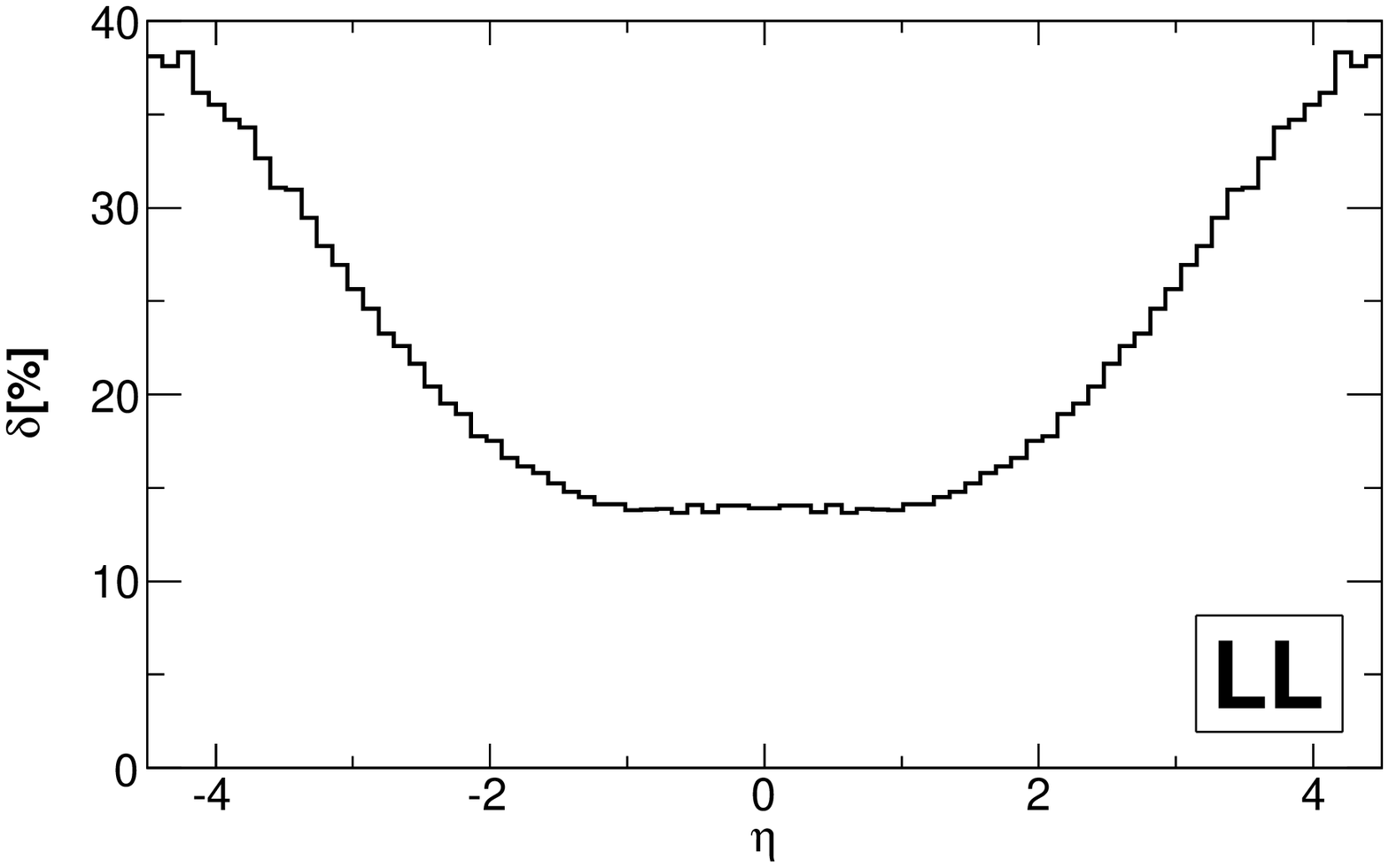}
    \includegraphics[width=.49\textwidth]{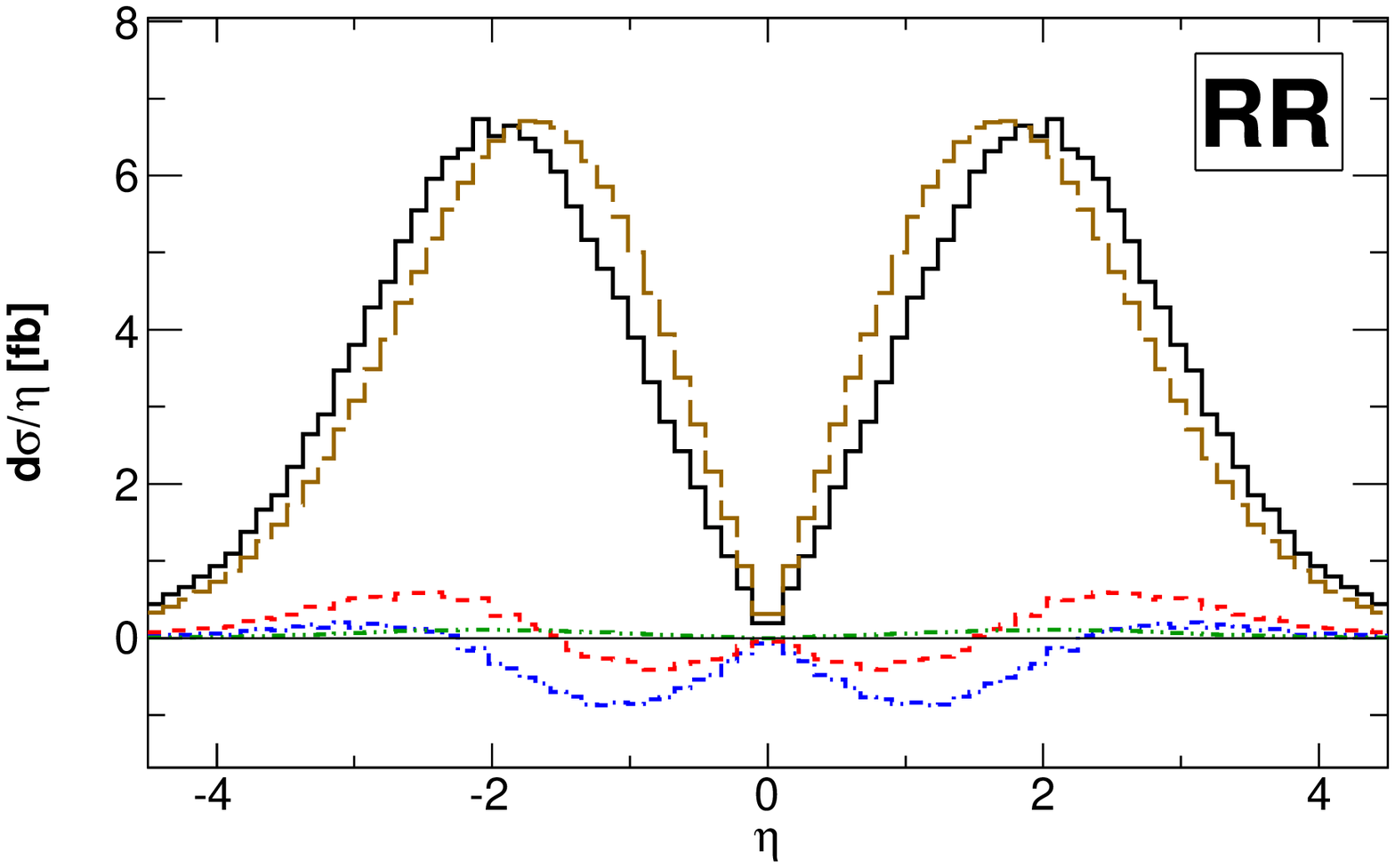}
    \includegraphics[width=.49\textwidth]{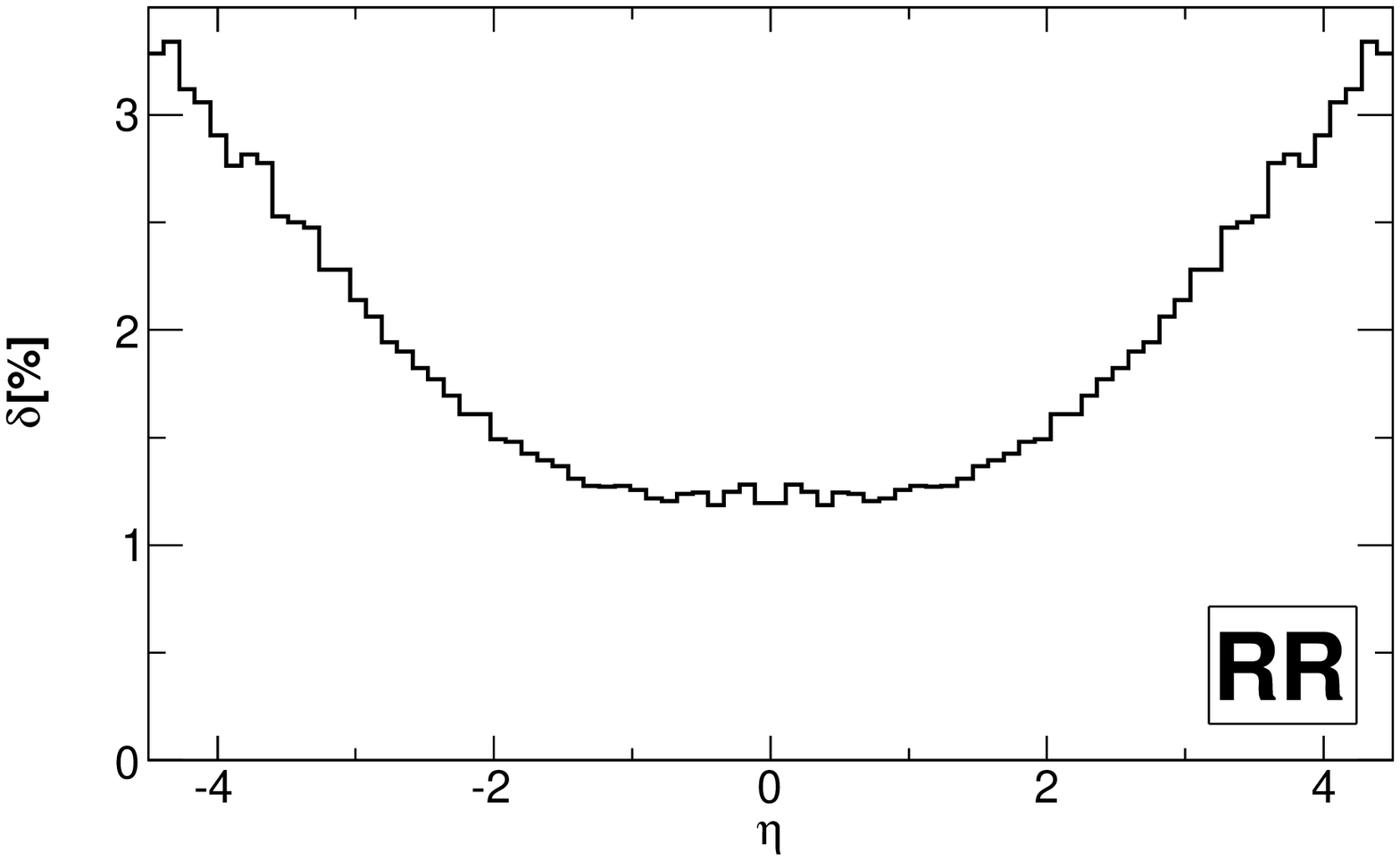}
    \includegraphics[width=.49\textwidth]{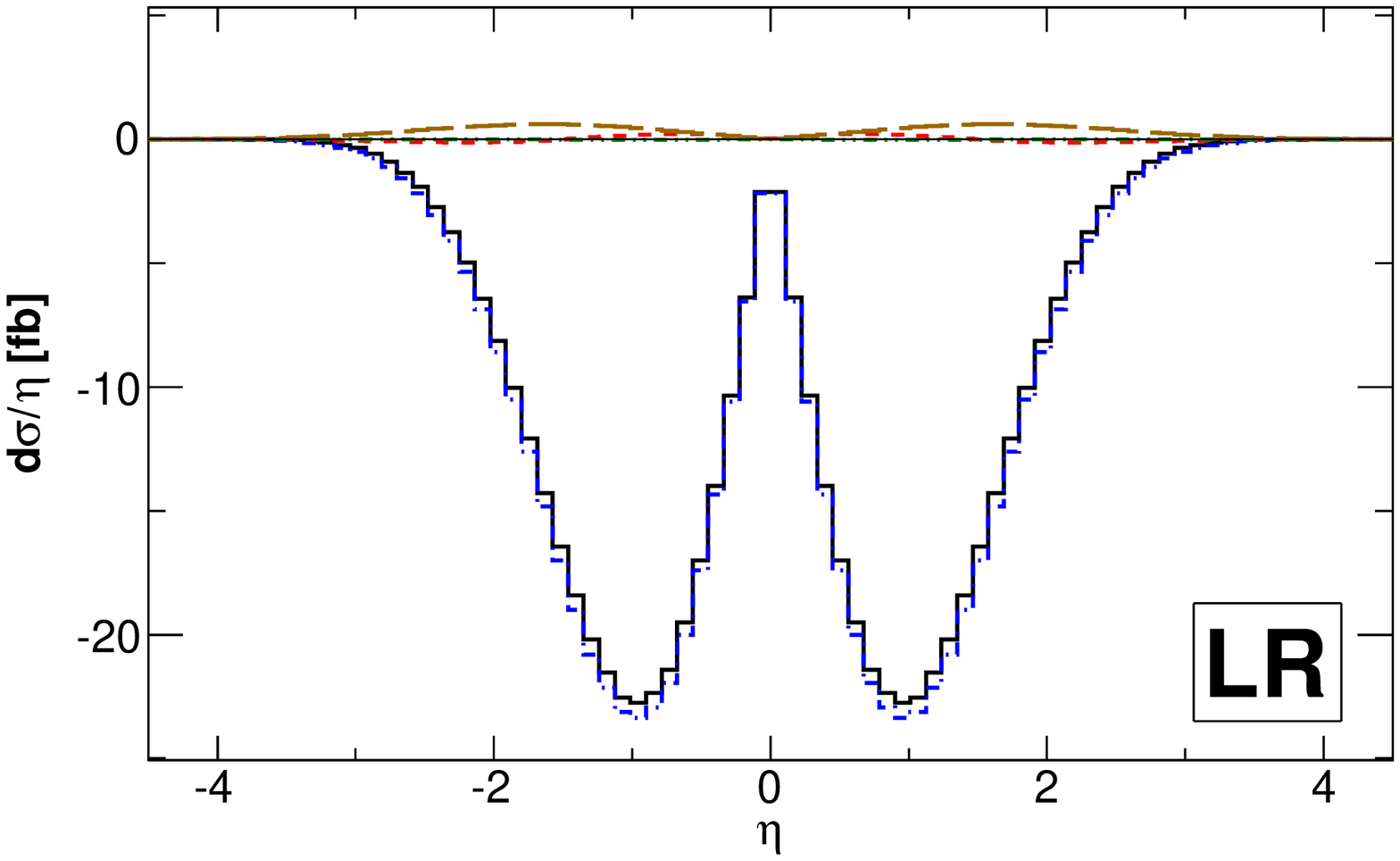}
    \includegraphics[width=.49\textwidth]{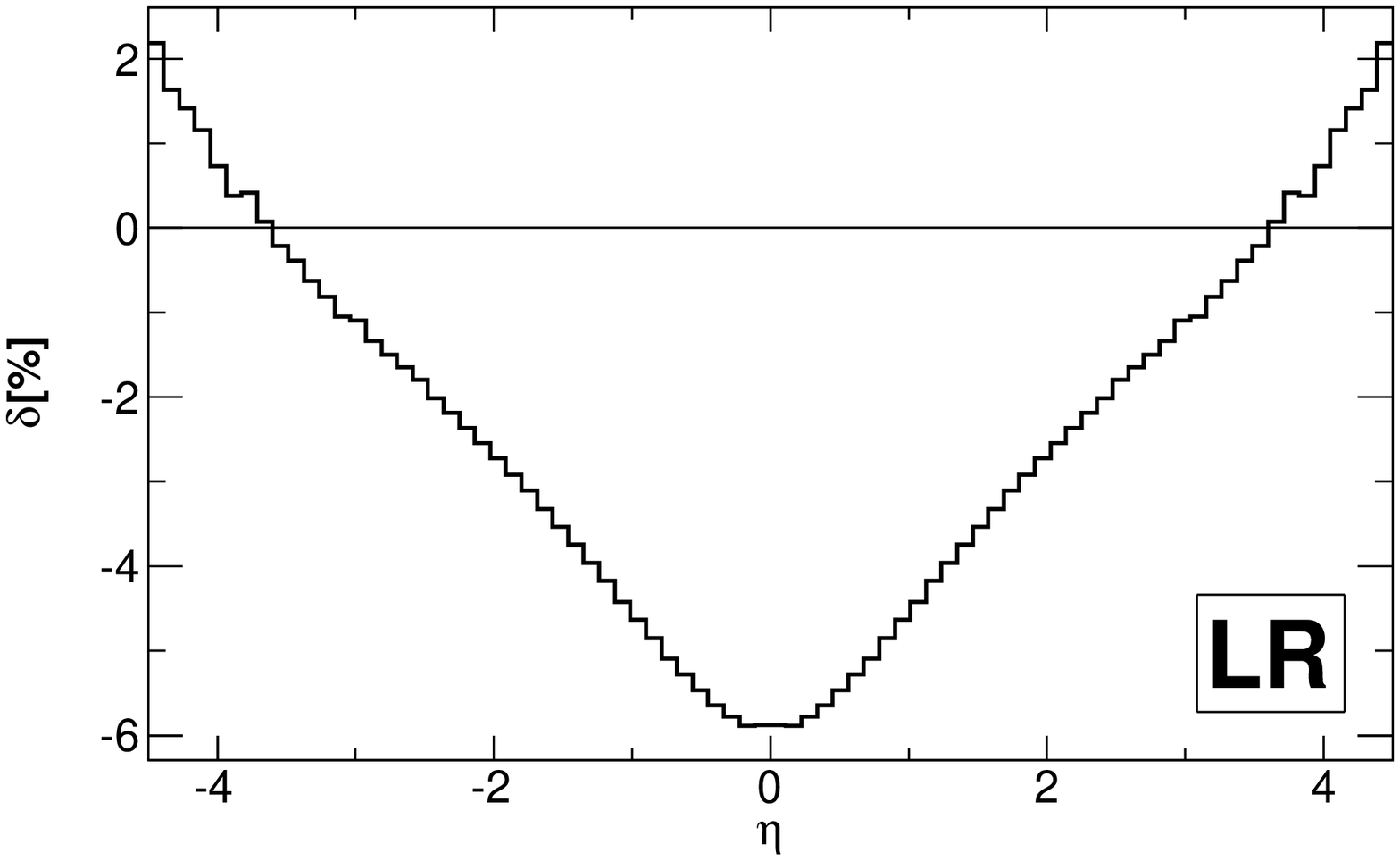}
  \caption{Pseudo-rapidity distributions for squark--squark
    production at the LHC within the SPS1a' scenario. Shown are the
    tree-level and NLO EW cross section contributions (left) and the
    impact of EW contributions relative to the QCD Born cross section
    (right) for inclusive $\squark\squark'$ production (top),
    production of two left-handed squarks $\squark_L \squark'_L$
    (second), production of two right-handed squarks $\squark_R
    \squark'_R$ (third), and non-diagonal $\squark_L \squark'_R$
    production (bottom). Charge conjugated processes are included.}
\label{fig_etadist}
}

In \figref{fig_etadist} we present the pseudo-rapidity distributions,
where always the squark with the higher absolute value of the
pseudo-rapidity $\eta$ (in the laboratory frame) is considered. All EW
contributions are vanishing in the central $\eta\approx 0$ region.
The characteristics of the rapidity gap in the distributions depend on
the precise quantity considered and is enhanced by our choice of
referring to the larger $\eta$. The tree-level contributions peak at
around $|\eta|=2$ and dominate the total result, if present. The
distributions for EW-type and QCD-type NLO corrections differ in sign
and shape from each other, leading to large cancellations over a wide
phase-space range.  In total, the EW contributions alter the LO
distributions by up to $20-40\%$ for LL and up to $10-20\%$ for
inclusive squark--squark production in the strong forward region for
$|\eta|>2$.

\clearpage


\medskip

Up to now, our discussion has only been for inclusive combinations of
final-state squarks for given chiralities. To get further insight on
the cross section, we show in Table~\ref{tab_SPA_detailed} the cross
section divided up into the various subprocesses for squark--squark
production within the \SPA{} scenario.  Again, anti-particles are
included. Owing to the degenerate masses of first- and second-generation
squarks, we do not distinguish between final states that result from
exchanging both squarks with their first or second generation
counterpart, i.e. $\uL \uL$ production also includes $\cL \cL$
production, etc.. This reduces the number of distinct subprocesses
from 36 down to 22.
The contributions to $\TreeEW$ are always positive and are largest for
$\uL \dL$ production due to the interference of gluino and chargino
exchange diagrams and constitute $57\%$ of the inclusive tree-level EW
contribution, see also Table~\ref{tab_SPA}. One even finds that the
inclusive tree-level EW contribution is given to $98\%$ by only five
processes, namely $\uL\uL$, $\uR\uR$, $\dL\dL$, $\uL\dL$ and
$\uL\sL$. The contributions to $\NLOEW$ are mostly negative, reducing
the importance of EW contributions. In contrast to the tree-level EW
case, many processes contribute with a significant amount to the
inclusive NLO EW contribution of the cross section. Especially for
processes with  squarks of different generations,
$\EW$ is mostly dominated by NLO EW contributions. The size of the NLO
EW contributions is often reduced due to the interplay of QCD-type and
EW-type corrections as shown in \figref{fig_uLdL} in the case of
$\uL\dL$ production. The different types of NLO EW corrections
partially cancel. Furthermore, the sum contains corrections of positive
and negative sign, leading to an integrated result $\NLOEW$ that
is considerably smaller than the corrections affecting the LO result in
various phase-space regions.

\TABULAR[t]{c|c|c|c||c|c|c}{
    \hline\trule\trule
    \multirow{2}{*}{\hspace{-7pt}\bf \SPA{} \hspace{-5pt}}
    & {$\boldsymbol{\Born}$}
    & {$\boldsymbol{\TreeEW}$}
    & {$\boldsymbol{\NLOEW}$}
    & {\multirow{2}{*}{$\boldsymbol{\dTreeEW}$}}
    & {\multirow{2}{*}{$\boldsymbol{\dNLOEW}$}}
    & {\multirow{2}{*}{$\boldsymbol{\dNLO}$}}\\
    & {$\mathcal{O}(\alphas^2)$}
    & {$\mathcal{O}(\alphas\alpha+\alpha^2)$}
    & {$\mathcal{O}(\alphas^2\alpha)$}
    &
    &
    &\\
    \hline
    $\boldsymbol{\uL \uL}$ & ~486.8(3) & 93.78(5) & ~$-$30.5(2) & 19.27 \% & $-$6.26 \% &   13.00 \% \\ 
    $\boldsymbol{\dL \dL}$ & 143.83(8) & 29.18(2) & ~$-$9.85(6) & 20.29 \% & $-$6.85 \% &   13.44 \% \\  
    $\boldsymbol{\uL \dL}$ & ~692.6(7) & 234.8(2) & ~~$-$9.5(6) & 33.90 \% & $-$1.38 \% &   32.52 \% \\ 
    $\boldsymbol{\uL \sL}$ & ~211.3(2) & 17.95(3) & ~$-$8.53(1) & ~8.50 \% & $-$4.04 \% &   ~4.46 \% \\  
    $\boldsymbol{\uL \cL}$ & 102.96(8) & 1.864(2) & $-$8.885(7) & ~1.81 \% & $-$8.63 \% & $-$6.82 \% \\  
    $\boldsymbol{\dL \sL}$ & ~80.19(6) & 1.390(2) & $-$7.526(4) & ~1.73 \% & $-$9.39 \% & $-$7.65 \% \\   
    \hline
    $\boldsymbol{\uR \uR}$ & ~537.1(4) & ~28.58(2) & ~$-$4.44(8) &  5.32 \% & $-$0.83 \% &   ~4.49 \% \\  
    $\boldsymbol{\dR \dR}$ & ~173.1(1) & ~2.414(2) & $-$0.318(7) &  1.39 \% & $-$0.18 \% &   ~1.21 \% \\  
    $\boldsymbol{\uR \dR}$ & ~799.1(6) & 0.4458(8) &  ~ ~~3.41(3) &  0.06 \% &   ~0.43 \% &   ~0.48 \% \\  
    $\boldsymbol{\uR \sR}$ & ~253.0(2) & 0.1276(2) &  ~ ~~1.36(1) &  0.05 \% &   ~0.54 \% &   ~0.59 \% \\  
    $\boldsymbol{\uR \cR}$ & 118.95(9) & 0.2365(4) & $-$1.337(8) &  0.20 \% & $-$1.12 \% & $-$0.93 \% \\  
    $\boldsymbol{\dR \sR}$ & 100.65(8) & 0.0126(1) & $-$0.281(2) &  0.01 \% & $-$0.28 \% & $-$0.27 \% \\  
    \hline                        						           			    
    $\boldsymbol{\uL \uR}$ & ~629.7(4) & ~1.288(1) & ~$-$26.41(4) & 0.20 \% & $-$4.19 \% & $-$3.99 \% \\ 
    $\boldsymbol{\dL \dR}$ & 165.49(9) & 0.0792(1) & ~$-$7.027(4) & 0.05 \% & $-$4.25 \% & $-$4.20 \% \\  
    $\boldsymbol{\uL \dR}$ & ~328.5(2) & 0.1720(1) & ~$-$12.30(1) & 0.05 \% & $-$3.75 \% & $-$3.69 \% \\ 
    $\boldsymbol{\uR \dL}$ & ~321.4(2) & 0.6026(6) & ~$-$13.81(2) & 0.19 \% & $-$4.30 \% & $-$4.11 \% \\ 
    $\boldsymbol{\uL \sR}$ & ~82.26(4) & 0.0450(1) & ~$-$2.809(3) & 0.05 \% & $-$3.42 \% & $-$3.36 \% \\   
    $\boldsymbol{\uR \sL}$ & ~79.90(4) & 0.1556(1) & ~$-$3.167(4) & 0.19 \% & $-$3.96 \% & $-$3.77 \% \\   
    $\boldsymbol{\uL \cR}$ & ~38.08(2) & 0.0832(1) & ~$-$1.388(2) & 0.22 \% & $-$3.65 \% & $-$3.43 \% \\   
    $\boldsymbol{\uR \cL}$ & ~38.08(2) & 0.0832(1) & ~$-$1.388(2) & 0.22 \% & $-$3.65 \% & $-$3.44 \% \\   
    $\boldsymbol{\dL \sR}$ & ~30.24(2) & 0.0149(1) & $-$1.2015(9) & 0.05 \% & $-$3.97 \% & $-$3.92 \% \\   
    $\boldsymbol{\dR \sL}$ & ~30.24(2) & 0.0149(1) & $-$1.2015(9) & 0.05 \% & $-$3.97 \% & $-$3.92 \% \\
    \hline
}{Hadronic cross section for squark--squark production at the LHC 
    within the \SPA{} scenario. Charge conjugated processes are included. 
    $ \Sup \Sup$ final states include $ \Scharm \Scharm$, $\Sdown \Sdown$
    include $\Sstrange \Sstrange$, and $\Sup \Sstrange$ includes
    $\Scharm \Sdown$. All cross sections are given in femtobarn (fb).
    \label{tab_SPA_detailed}
  }

%
\FIGURE[t]{
    \includegraphics[width=.49\textwidth]{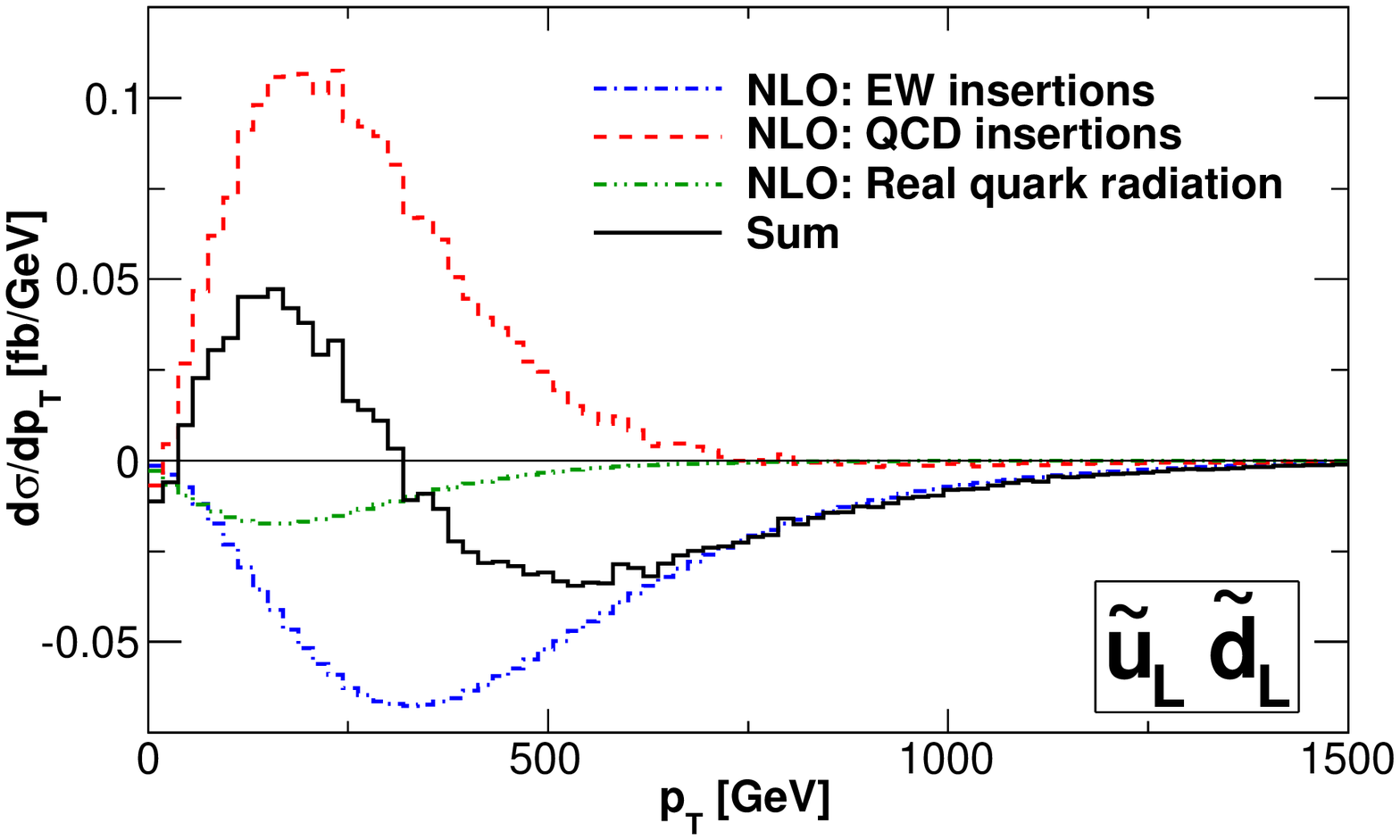}
    \includegraphics[width=.49\textwidth]{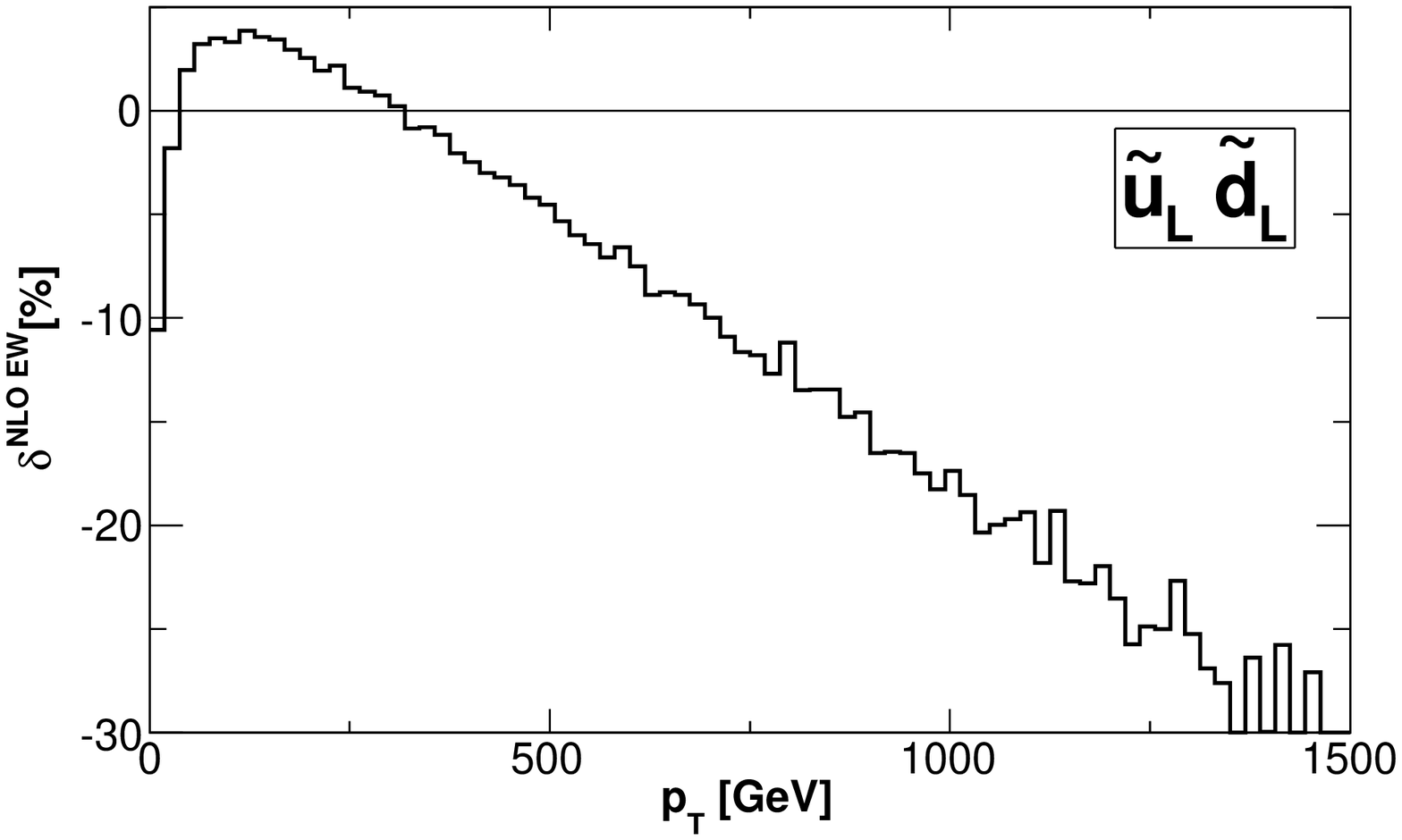}
    {\small \hspace*{1.25cm} (a) \hspace*{7.cm} (b) }
    \caption{(a) Transverse momentum distribution of the hardest squark
      for $\uL \dL$ production within the \SPA{} scenario. Strong
      cancellations occur between the different contributions at NLO
      EW. (b) Relative NLO EWcontribution,
	defined as the ratio of
      $d \sigma^{\rm NLO~EW}/d p_T$ and $d \sigma^{\rm Born}/d p_T$.
    }
    \label{fig_uLdL}
}

\section{Conclusions}

We have studied the hadronic production of two squarks or two
anti-squarks within the MSSM, including tree-level EW and NLO EW cross
section contributions of $\Order(\alpha_s~\alpha~+~\alpha^2)$ and
$\Order(\alpha_s^2\alpha)$, respectively.  In contrast to pure QCD
computations, one has to treat processes with squarks of different
flavor or chirality separately.

At $\Order(\alpha_s^2\alpha)$, numerous QCD--EW interference terms
occur. Virtual corrections arise from the interference contributions
of tree-level QCD amplitudes and mixed EW--QCD one-loop diagrams, as
well as from the interference of tree-level EW and pure-QCD one-loop
amplitudes.  Bremsstrahlung corrections comprise real photon, real
gluon, and real quark radiation processes.

We have performed a detailed numerical analysis for two
left-handed squarks ($\squark_L\squark_L$), two right-handed squarks
($\squark_R\squark_R$), and one left-handed and one right-handed squark
($\squark_L\squark_R$), as well as for inclusive squark--squark
($\squark\squark$) production at the LHC. The tree-level EW
contributions are largest for $\squark_L\squark_L$ production, where
they are enhanced by $\Order(\alpha_s\alpha)$ interference terms and
can easily reach the $20\%$ level.  The interference contributions are
suppressed for $\squark_R\squark_R$ production from the chiral
couplings and vanish for $\squark_L\squark_R$ production in the limit
of no L-R mixing.  At NLO, partial cancellations occur among the
various EW contributions. As a result, the integrated cross section is
reduced by the NLO EW contributions by a few percent for most
subprocesses. The interplay of tree-level and NLO EW contributions is
not universal and depends sensitively on the ratio of squark and
gluino masses as well as on their absolute values. The full EW
contributions affect the integrated cross section for inclusive
squark--squark production at the percent level (about $5\%$ in \SPA{}
and SPS5, $-1\%$ in SPS2).  In the distributions, however, the EW
contributions range from $-10\%$ to $20\%$ and even larger values for
exclusive $\squark_L\squark_L$ production.


\begin{acknowledgments}
E.\,M. and M.\,T. thank their former home institution MPI for Physics for
hospitality during the final stages of the work.
M.\,T. is supported by the DOE grant DE-FG02-95ER40896.\\
This work was supported in part by the European Community's Marie-Curie Research
Training Network under contract MRTN-CT-2006-035505
`Tools and Precision Calculations for Physics Discoveries at Colliders'
(HEPTOOLS).

\end{acknowledgments}

\appendix

\section{Tree-level cross sections}
\label{sect_App_treelevel}

In this Appendix, we give explicit expressions for the tree-level 
differential cross sections 
\eqref{eq_treeQCD}--\eqref{eq_treeEW} 
for all squark--squark production subprocesses. 
We closely follow  \cite{Bornhauser:2007bf} and 
express the color- and spin-averaged squared $t$- and $u$-channel matrix elements 
and their interference in terms of the following functions,
\bee
\Phi({\tilde \xi}_1,{\tilde \xi}_2,\squark_{\alpha},\squark_{\beta}^\prime) &=& \frac{1}{4}
c_\Phi({\tilde \xi}_1,{\tilde \xi}_2) \,\frac{1}{\hat t - m_{{\tilde \xi}_1}^2} \frac{1}{\hat t - m_{{\tilde \xi}_2}^2}\,
\Big[ A({\tilde \xi}_1,{\tilde \xi}_2,\squark_{\alpha},\squark_{\beta}^\prime) 
\label{eq_tree1}
\\
&& \times \left( \hat t \hat u -
m_{\squark_{\alpha}}^2 m_{\squark_{\beta}^\prime}^2 \right) +
B({\tilde \xi}_1,{\tilde \xi}_2,\squark_{\alpha},\squark_{\beta}^\prime) m_{{\tilde \xi}_1} m_{{\tilde \xi}_2} \hat s \Big],
\nonumber\\
\Theta({\tilde \xi}_1,{\tilde \xi}_2,\squark_{\alpha},\squark_{\beta}^\prime) &=& \frac{1}{4}
c_{\Theta}({\tilde \xi}_1,{\tilde \xi}_2)\, \frac{1}{\hat u - m_{{\tilde \xi}_1}^2} \frac{1}{\hat u - m_{{\tilde \xi}_2}^2}\,
\Big[ C({\tilde \xi}_1,{\tilde \xi}_2,\squark_{\alpha},\squark_{\beta}^\prime) 
\label{eq_tree2}
\\
&& \times \left( \hat t \hat u - m_{\squark_{\alpha}}^2
  m_{\squark_{\beta}^\prime}^2 \right) +
  D({\tilde \xi}_1,{\tilde \xi}_2,\squark_{\alpha},\squark_{\beta}^\prime) m_{{\tilde \xi}_1} m_{{\tilde \xi}_2} \hat s \Big],
\nonumber\\
\Psi({\tilde \xi}_1,{\tilde \xi}_2,\squark_{\alpha},\squark_{\beta}^\prime) &=& \frac{1}{4}
c_\Psi({\tilde \xi}_1,{\tilde \xi}_2)\, \frac{1}{\hat t - m_{{\tilde \xi}_1}^2} \frac{1}{\hat u - m_{{\tilde \xi}_2}^2}\,
F({\tilde \xi}_1,{\tilde \xi}_2,\squark_{\alpha},\squark_{\beta}^\prime) m_{{\tilde \xi}_1} m_{{\tilde \xi}_2} \hat s .
\label{eq_tree3}
\eee
Here,  ${\tilde \xi}_{1,2}\in \{\gluino,\,\neu,\,\cha\}$ denote the exchanged particle in the 
$t$- or $u$-channel, respectively, and 
$c_{\{\Phi, \Theta, \Psi\}}$ are color matrices which are summarized in Table~\ref{tab_color}.
Note that the color factors $c_\Psi(\tilde{\chi},\tilde \chi)$ differ from 
\cite{Bornhauser:2007bf}. 
\TABULAR[b]{c|c|c|c}{
    \hline\trule
    \hspace{10pt} $\boldsymbol{\tilde \xi_1,\tilde \xi_2}$ \hspace{10pt} &
    $\boldsymbol{c_\Phi(\tilde \xi_1,\tilde \xi_2)}$ &
    $\boldsymbol{c_\Theta(\tilde \xi_1,\tilde \xi_2)}$ &
    \hspace{10pt} $\boldsymbol{c_\Psi(\tilde \xi_1,\tilde \xi_2)}$ \hspace{10pt}
\\ \hline
    $\gluino,\gluino$ & 2/9 & 2/9 & $-2/27$ 
\\ \hline
    $\tilde{\chi}, \tilde{\chi}$ & 1 & 1 & 1/3 
\\ \hline
    $\gluino, \tilde{\chi} $ & 0 & 0 & 4/9
\\    \hline
}{
  Color factors in Eqs.~\ref{eq_tree1}--\ref{eq_tree3}, with
  $\tilde{\chi}$ denoting any neutralino $\neu_k$ or chargino
  $\cha_k$.
\label{tab_color}
}

The coupling constants are collected
in the abbreviations $A,B,C,D,F$, as given by
\begin{subequations}
\bee
\begin{split}
A({\tilde \xi}_1,{\tilde \xi}_2,\squark_{\alpha},\squark_{\beta}^\prime) &=&
    a_{{\tilde \xi}_1,\squark_\alpha} 
    a_{{\tilde \xi}_2,\squark_\alpha}^\ast
    b_{{\tilde \xi}_1,\squark_\beta^\prime}^\prime
    b_{{\tilde \xi}_2,\squark_\beta^\prime}^{\prime\ast} +
    a_{{\tilde \xi}_1,\squark_\beta^\prime}^\prime
    a_{{\tilde \xi}_2,\squark_\beta^\prime}^{\prime\ast}
    b_{{\tilde \xi}_1,\squark_\alpha} 
    b_{{\tilde \xi}_2,\squark_\alpha}^{\ast},
\\
B({\tilde \xi}_1,{\tilde \xi}_2,\squark_{\alpha},\squark_{\beta}^\prime) &=&
    a_{{\tilde \xi}_1,\squark_\alpha} 
    a_{{\tilde \xi}_2,\squark_\alpha}^\ast
    a_{{\tilde \xi}_1,\squark_\beta^\prime}^\prime
    a_{{\tilde \xi}_2,\squark_\beta^\prime}^{\prime\ast} +
    b_{{\tilde \xi}_1,\squark_\alpha} 
    b_{{\tilde \xi}_2,\squark_\alpha}^{\ast}
    b_{{\tilde \xi}_1,\squark_\beta^\prime}^\prime
    b_{{\tilde \xi}_2,\squark_\beta^\prime}^{\prime\ast},
\\
C({\tilde \xi}_1,{\tilde \xi}_2,\squark_{\alpha},\squark_{\beta}^\prime) &=&
    c_{{\tilde \xi}_1,\squark_\beta} 
    c_{{\tilde \xi}_2,\squark_\beta^\prime}^\ast
    d_{{\tilde \xi}_1,\squark_\alpha}^\prime
    d_{{\tilde \xi}_2,\squark_\alpha}^{\prime\ast} +
    c_{{\tilde \xi}_1,\squark_\alpha}^\prime
    c_{{\tilde \xi}_2,\squark_\alpha}^{\prime\ast}
    d_{{\tilde \xi}_1,\squark_\beta^\prime} 
    d_{{\tilde \xi}_2,\squark_\beta^\prime}^{\ast}, 
\label{eq_App_const}
\\
D({\tilde \xi}_1,{\tilde \xi}_2,\squark_{\alpha},\squark_{\beta}^\prime) &=&
    c_{{\tilde \xi}_1,\squark_\beta^\prime} 
    c_{{\tilde \xi}_2,\squark_\beta^\prime}^\ast
    c_{{\tilde \xi}_1,\squark_\alpha}^\prime
    c_{{\tilde \xi}_2,\squark_\alpha}^{\prime\ast} +
    d_{{\tilde \xi}_1,\squark_\beta^\prime} 
    d_{{\tilde \xi}_2,\squark_\beta^\prime}^{\ast}
    d_{{\tilde \xi}_1,\squark_\alpha}^\prime
    d_{{\tilde \xi}_2,\squark_\alpha}^{\prime\ast},
\\
F({\tilde \xi}_1,{\tilde \xi}_2,\squark_{\alpha},\squark_{\beta}^\prime) &=&
    a_{{\tilde \xi}_1,\squark_\alpha}
    c_{{\tilde \xi}_2,\squark_\beta^\prime}^\ast
    a_{{\tilde \xi}_1,\squark_\beta^\prime}^\prime
    c_{{\tilde \xi}_2,\squark_\alpha}^{\prime\ast} +
    b_{{\tilde \xi}_1,\squark_\alpha}
    d_{{\tilde \xi}_2,\squark_\beta^\prime}^{\ast}
    b_{{\tilde \xi}_1,\squark_\beta^\prime}^\prime
    d_{{\tilde \xi}_2,\squark_\alpha}^{\prime\ast},
\end{split}
\eee
\end{subequations}
where the notation refers to the labels as listed in \figref{fig_labels}.
\FIGURE[t]{
\qquad
\begin{feynartspicture}(80,80)(1,1)
  \FADiagram{}
  \FAProp(0.,15.)(10.,15.)(0.,){/Straight}{0}
  \FAProp(-0.,5.)(10.,5.)(0.,){/Straight}{0}
  \FAProp(10.,15.)(10.,5.)(0.,){/Straight}{0}
  \FALabel(9.18,10.)[r]{$\tilde \xi_{1,2}$}
  \FAProp(10.,15.)(20.,15.)(0.,){/Straight}{0}
  \FALabel(20.,14.18)[t]{$\squark_\beta^\prime$}
  \FALabel(10.,15.82)[b]{$a^\prime,b^\prime$}
  \FAProp(10.,5.)(20.,5.)(0.,){/Straight}{0}
  \FALabel(20.,4.18)[t]{$\squark_\alpha$}
  \FALabel(10.,4.18)[t]{$a,b$}
  \FALabel(-3,10.)[r]{$t$-channel:}
  \FAVert(10.,15.){0}
  \FAVert(10.,5.){0}
\end{feynartspicture}
\qquad\qquad\qquad\qquad
\begin{feynartspicture}(80,80)(1,1)
  \FADiagram{}
  \FAProp(0.,15.)(10.,15.)(0.,){/Straight}{0}
  \FAProp(-0.,5.)(10.,5.)(0.,){/Straight}{0}
  \FAProp(10.,15.)(10.,5.)(0.,){/Straight}{0}
  \FALabel(9.18,10.)[r]{${\tilde \xi}_{1,2}$}
  \FAProp(10.,15.)(20.,5.)(0.,){/Straight}{0}
  \FALabel(20.,14.18)[t]{$\squark_\beta^\prime$}
  \FALabel(10.,15.82)[b]{$c^\prime,d^\prime$}
  \FAProp(10.,5.)(20.,15.)(0.,){/Straight}{0}
  \FALabel(20.,4.18)[t]{$\squark_\alpha$}
  \FALabel(10.,4.18)[t]{$c,d$}
  \FALabel(-3,10.)[r]{$u$-channel:}
  \FAVert(10.,15.){0}
  \FAVert(10.,5.){0}
\end{feynartspicture}\qquad
\caption{Notations for the couplings in the
  tree-level formulas in Appendix~\ref{sect_App_treelevel}. Labels
  $a$,$c$ refer to the couplings to left-handed squarks, $b$ and $d$
  to that of right-handed squarks. Couplings at the upper and lower
  vertex, respectively, are denoted by distinct labels for convenience.
  \vspace*{1ex} }
\label{fig_labels}
}
Finally the explicit coupling constants $a_{\tilde \xi_i, \squarka},\,
b_{\tilde \xi_i, \squarka},\dots$
are given in Table~\ref{tab_couplings}.

\TABULAR[t]{c|c|c}{
\hline\trule
$\boldsymbol{\tilde{\xi}_i, \squarka}$
& $\boldsymbol{a, a', c, c'}$ & $\boldsymbol{b,b',d,d'}$ 
\\[.5ex] \hline
$\neu_k,\Sup_\alpha $ & $ -\frac{ie}{\sqrt{2} s_w}
      \Big( \frac{1}{3} \frac{s_w}{c_w} N^\ast_{k1} + N^\ast_{k2} \Big) \,\delta_{L\alpha}$
      & $ \frac{4ie}{3\sqrt{2} c_w}\,N_{k1}\, \delta_{R\alpha}$
\\[.5ex] \hline
$\neu_k,\Sdown_\alpha$ & $ -\frac{ie}{\sqrt{2} s_w}
      \Big( \frac{1}{3} \frac{s_w}{c_w}  N^\ast_{k1} - N^\ast_{k2}\Big)\, \delta_{L\alpha}$
      & $ -\frac{2ie }{3\sqrt{2}c_w}\,N_{k1}\,\delta_{R\alpha}$
\\[.5ex] \hline\trule
$\cha_k,\Sup_\alpha^{}$ & $ -\frac{ie}{s_w}  V^\ast_{k1}\,\delta_{L\alpha}$
    &$ 0$
\\[.5ex] \hline\trule
$\cha_k,\Sdown_\alpha^{}$ & $ -\frac{ie}{s_w} U^\ast_{k1} \,\delta_{L\alpha} $
    & $0$
\\[.5ex] \hline \trule
$\gluino,\Sup_\alpha^{}$ & $ -\sqrt{2} i \hat g_s\,\delta_{L\alpha}$
    &$\sqrt{2} i \hat g_s\,\delta_{R\alpha}$
\\[.5ex] \hline\trule
$\gluino,\Sdown_\alpha^{}$ & $ -\sqrt{2} i \hat g_s\,\delta_{L\alpha}$
    &$ \sqrt{2} i \hat g_s\,\delta_{R\alpha}$
\\[.5ex]
\hline
}{ Coupling constants $a_{\tilde \xi_i, \squarka}, b_{\tilde \xi_i,
    \squarka},\dots$ for exchange particle $\tilde \xi_i$ and produced
  light-flavor squark $\squarka$, following the conventions of
  \cite{Hahn:2006qw}. L-R mixing of the squark mass eigenstates is
  neglected.  $N,U,V$ are the unitary matrices diagonalizing the
  neutralino and chargino mass matrix, respectively and $s_w =
  \sin\theta_w$ and $c_w = \cos\theta_w$.
\label{tab_couplings}
}

\medskip

For the differential cross sections, we refer to the three classes 
of subprocesses introduced in \eqref{eq_threeclasses}. 
$\alpha,\beta=\{\rm L,R\}$ label the chirality of the squarks, 
$k,l$ label the four (two) mass eigenstates of neutralinos (charginos).

\begin{itemize}
\item $\boldsymbol{PP \rightarrow \Squarka \Squarkb}$
  \textbf{(two squarks of the same flavor)}\\
  The partonic process for this class of processes 
  is $qq \rightarrow \squark_\alpha \squark_\beta$, \ie~all 
  quarks and squarks are of the same flavor. The differential 
  cross sections at $\Order(\alpha_s^2),\, 
    \Order(\alpha^2),\, \Order(\alpha_s \alpha)$ read, according to the
    notation introduced in Section~\ref{sect_tree},
 \begin{align}
\begin{split}
  d\hat\sigma^{2,\,0} = & \,
  \bigg\lbrace \Phi(\gluino,\gluino,\squark_\alpha,\squark_\beta) +
  \Theta(\gluino,\gluino,\squark_\alpha,\squark_\beta) +
  2 \operatorname{Re}
  \left\{ \Psi(\gluino,\gluino,\squark_\alpha,\squark_\beta) \right\}
  \bigg\rbrace \,\frac{d\hat t}{16\pi\hat s^2},\\
  d\hat\sigma^{0,\,2} = &\,
  \sum_{k,l=1}^4\bigg\lbrace
  \Phi(\neu_k,\neu_l,\squark_\alpha,\squark_\beta) +
  \Theta(\neu_k,\neu_l,\squark_\alpha,\squark_\beta)\\
  &\qquad~ + 2 \operatorname{Re}
  \left\{ \Psi(\neu_k,\neu_l,\squark_\alpha,\squark_\beta)
  \right\} \bigg\rbrace \,\frac{d\hat t}{16\pi\hat s^2},\\
   d\hat\sigma^{1,\,1} =&\, 
   \sum_{k=1}^4 2 \operatorname{Re}
  \left\{ \Psi(\gluino,\neu_l,\squark_\alpha,\squark_\beta)
  \right\} \, \frac{d\hat t}{16\pi\hat s^2}.
\end{split}
\end{align}
As can be seen from the couplings in Table~\ref{tab_couplings}, the
interference terms $\Psi$ and thus in particular the interference
contribution $d\hat\sigma^{1,\,1}$ are only present for diagonal
squark--squark production (\ie~$\alpha=\beta$).  This is a result from
the absence of L-R mixing for the light-flavor squarks.
%
\item $\boldsymbol{PP \to \Squarka \Squarkb'}$ 
\textbf{(two squarks of different flavor in the same SU(2) doublet)}\\
The only contributing partonic process is $qq' \rightarrow \Squarka
\Squarkb'$, with $q'$ being the SU(2) partner of $q$.  The tree-level
contributions to the cross section read:
\begin{align}
\begin{split}
  d\hat\sigma^{2,\,0} =&\,
  \Phi(\gluino,\gluino,\Squarka,\Squarkb')\,
  \frac{d\hat t}{16\pi\hat s^2},\\
  d\hat\sigma^{0,\,2}=&\,
  \bigg\lbrace
  \sum_{k,l=1}^4 \Phi(\neu_k,\neu_l,\Squarka,\Squarkb') +
  \sum_{k,l=1}^2 \Theta(\cha_k,\cha_l,\Squarka,\Squarkb')\\
  &+\sum_{k=1}^4\sum_{l=1}^2 2 \Re \{\Psi(\neu_l,\cha_k,\Squarka,\Squarkb')\}
  \bigg\rbrace\, \frac{d\hat t}{16\pi\hat s^2},\\
  d\hat\sigma^{1,\,1} =&\,
   \sum_{k=1}^2 2 \operatorname{Re}
  \left\{ \Psi(\gluino,\cha_l,\Squarka,\Squarkb')
  \right\} \,\frac{d\hat t}{16\pi\hat s^2}. 
\end{split}
\end{align}
In this case, the interference terms are related to chargino-mediated diagrams 
and thus the interference contribution 
    $d\hat\sigma^{1,\,1}$ is only non-zero for the production of two left-handed 
squarks (\ie~$\alpha=\beta=L$). 
%
\item $\boldsymbol{PP \rightarrow \squark_\alpha \squark_\beta^\prime}$ 
    \textbf{(two squarks in different doublets)} \\
    This class describes the production of two squarks of different
    flavor and of different generation, arising from the partonic
    process $qq' \to \squarka \squarkb'$, with $q \neq q'$. The
    tree-level cross sections read as follows,
\begin{subequations}
\begin{align}
\begin{split}
  d\hat\sigma^{2,\,0} = & \,
  \Phi(\gluino,\gluino,\squark_\alpha,\squark_\beta^\prime)\,
  \frac{d\hat t}{16\pi\hat s^2},\\
\end{split}
\end{align}
\begin{align}
\begin{split}
  d\hat\sigma^{0,\,2} =& \,
  \bigg\lbrace \sum_{k,l=1}^4
  \Phi(\neu_k,\neu_l,\squark_{\alpha},\squark'_{\beta}) 
  + \delta_{q u} \delta_{q' s} \,\delta_{\squark \Sdown} \,\delta_{\squark' \Scharm} 
  \sum_{k,l=1}^2
  \Phi(\cha_k,\cha_l,\squark_{\alpha},\squark'_{\beta}) 
\\
  &+ \delta_{q c} \delta_{q' d}\, \delta_{\squark \Sup} \,\delta_{\squark' \Sstrange} 
  \sum_{k,l=1}^2
  \Phi(\cha_k,\cha_l,\squark_{\alpha},\squark'_{\beta}) 
  \bigg\rbrace\,
  \frac{d\hat t}{16\pi\hat s^2},\\
  d\hat\sigma^{1,\,1} =&\, 0 .
\end{split}
\end{align}
\end{subequations}
  Here, two additional chargino-mediated partonic processes
  ($us\to\tilde{d}_L\tilde{c}_L$ and $cd\to\tilde{u}_L\tilde{s}_L$)
  can give an $\Order(\alpha^2)$ contribution.  The
  $\mathcal{O}(\alphas\alpha)$ interference contribution vanishes for
  this class of processes.
\end{itemize}

\section{Bremsstrahlung integrals}
\label{sect_App_IRformulae}

Here we give the soft and collinear singular integrals appearing in
the phase space integration of the real radiation processes, with
either a photon, gluon or (anti-)quark radiated.

\subsection{Soft and collinear photon bremsstrahlung}
 \label{subsec_Asoftphoton}

The cross section for real photon emission factorizes in the soft limit 
from the original cross section without photon emission,
\bee
  d\hat\sigma_{12\rightarrow 34\gamma}^{2,\,1}\Big|_{\text{soft}} &=&-\frac{\alpha}{2\pi} \Big\{ 
  \sum_{i,j=1;i<j}^4 e_i e_j \sigma_i \sigma_j\, \mathcal{I}_{ij}
  \Big\} \, d\hat\sigma_{12\rightarrow 34}^{2,\,0},
\eee
where now the considered process is given as a subscript for clarification and 
particles are labeled by $i=1\ldots4$ according to the definition of 
 momenta $p_i$ in \eqsref{eq_LOmom}{eq_NLOmom}.
$e_i$~is the charge of the $i$th~particle and
$\sigma_i=\pm 1$ depending on whether the particle is incoming or
outgoing, respectively. The phase-space integrals $\mathcal{I}_{ij}= \mathcal{I}_{ji}$ are universal and 
well-known~\cite{Denner:1991kt}.  In the limit of vanishing
initial-state masses (and two massive final-state particles), they are
given by
\begin{align}
\begin{split}
  \mathcal{I}_{ii} &= \ln\left( \frac{4(\Delta E)^2}{\lambda^2} \right) +
  \ln\left(\frac{m_i^2}{\hat s_{12}}\right)\qquad\qquad\qquad\qquad\qquad\qquad\qquad~~ \text{for }i=\{1,2\},
\\
  \mathcal{I}_{ii} &= \ln\left( \frac{4(\Delta E)^2}{\lambda^2} \right) +
  \frac{1}{\beta_i} \ln\left(\frac{1-\beta_i}{1+\beta_i}\right) \qquad\qquad\qquad\qquad\qquad\qquad\,
  \text{for } i=\{3,4\},
\\
  \mathcal{I}_{12} &= \sum_{i=1,2}\left[ \ln\left(\frac{\hat s_{12}}{m_i^2}\right)
  \ln\left(\frac{4(\Delta E)^2}{\lambda^2}\right) - \frac{1}{2}
  \ln^2\left(\frac{\hat s_{12}}{m_i^2}\right) -\frac{\pi^2}{3} \right],
\\
  \mathcal{I}_{34} &= \frac{1}{v_{34}} \sum_{i=3,4} \left[
    \ln\left(\frac{1+\beta_i}{1-\beta_i}\right) \ln\left(\frac{4(\Delta
      E)^2}{\lambda^2}\right) -2\Li{\frac{2\beta_i}{1+\beta_i}}
    -\frac{1}{2}\ln^2\left(\frac{1-\beta_i}{1+\beta_i}\right)
    \right],
\\
  \mathcal{I}_{ij} &= \ln\left(\frac{\hat s_{ij}^2}{m_i^2m_j^2}\right)
  \ln\left(\frac{4(\Delta E)^2}{\lambda^2}\right) -\frac{1}{2}
  \ln^2\left(\frac{\hat s_{12}}{m_i^2}\right)
  -\frac{1}{2}\ln^2\left(\frac{1-\beta_i}{1+\beta_i}\right) -\frac{\pi^2}{3}
\\
  & -2\Li{1-\frac{2p_i^0p_j^0}{\hat s_{ij}}(1+\beta_j)}
  -2\Li{1-\frac{2p_i^0p_j^0}{\hat s_{ij}}(1-\beta_j)} 
\qquad
  \text{for } \begin{matrix}i=\{1,2\},\\j=\{3,4\},\end{matrix}
\end{split}
\label{eq_Isoft}
\end{align}
with $\hat s_{ij}=2p_i\cdot p_j$, $\beta_i=|p_i|/p_i^0$, 
$v_{ij}=\sqrt{1-4m_im_j/\hat s_{ij}^2}$, $\Delta E =
\sqrt{\hat s_{12}}\delta_s/2$. $\lambda$ is the fictitious photon
mass.

The partonic cross section in the collinear region can be written in
terms of a convolution integral
\bee
  d\hat\sigma_{12\rightarrow 34\gamma}^{2,\,1}(\hat s)\Big|_{\text{coll}} &=&
  \frac{\alpha (e_q^2+e_{q^{\prime}}^2)}{2\pi} \int_{z_0}^{1-\delta_s}
  \text{d}z\; \kappa_{qq}(z,\hat s)\;d\hat\sigma_{12 \rightarrow
    34}^{2,0}(z\hat s),
\eee
where $z_0$ and $\kappa_{qq}$ are given by
\bee
z_0=(m_{\squark}^2+m_{\squark^\prime}^2)/\hat s,~~~~~~  \kappa_{qq}(z,\hat s) &=& \frac{1+z^2}{1-z} \ln \left( \frac{\hat s\delta_{\theta}}{2m_q}
    \right) -\frac{2z}{1-z}. 
\label{eq_coll}
\eee
The upper integration bound is lowered by $\delta_s$ to avoid double
counting of the soft regime.

\subsection{Soft and collinear gluon bremsstrahlung}
 \label{subsec_Asoftgluon}

The bremsstrahlung integrals for soft gluon emission are the same
as those for the photonic case, but in addition one has to take
the color correlation of the amplitudes into account. We use the
following notation to keep track of the color factors.\\
Let $\left| c_1,\ldots,c_m\right>$ denote a complete color
basis. The colored matrix element with $m$ external particles ${\rm P}_i$
carrying momentum $p_i$ and color $c_i$ is then given by
\bee
  \mathcal{M}_{12\rightarrow 3\ldots m}^{i,j\;c_1\ldots c_m} &=& \Big\langle
  c_1,\ldots,c_m \Big| \mathcal{M}^{i,j}_{12\rightarrow 3\ldots m} \Big\rangle.
\eee
For the color structure in the case of real gluon emission it is
convenient to associate a color charge ${\bf T}_i$ with the emission
of a gluon of color $a$ from parton $i$. The action of this color charge
onto the color space is given by 
\bee
  \Big\langle c_1 \ldots c_i \ldots c_m \Big| {\bf T}_i \Big| b_1 \ldots b_i
  \ldots b_m \Big\rangle &=& 
  \delta_{c_1b_1} \ldots T^a_{c_i b_i} \ldots  \delta_{c_mb_m}.
\eee
Explicit expressions for the $T^a_{kl}$ and the color charge algebra
are given in \cite{Catani:1996vz}.
The real gluon emission in the soft limit is then given by:
\bee
d\hat\sigma_{12\rightarrow 34g}\Big|_{\text{soft}} &=&
-\frac{\alphas}{2\pi} \Big\{ \sum_{i,j=1;i<j}^4 \mathcal{I}_{ij}
 \mathcal{F}_{ij} \Big\} \frac{dt}{16\pi \hat s^2},
\eee
where the phase space integrals $\mathcal{I}_{ij}$ are given in 
Eq.~(\ref{eq_Isoft})\footnote{In this case $\lambda$ is the fictitious
  gluon mass.} and 
 $\mathcal{F}_{ij}$ denote color correlated amplitudes.
At $\mathcal{O}(\alphas^2\alpha)$, $\mathcal{F}_{ij}$ are given by
\begin{align}
\begin{split}
\mathcal{F}_{ij} =& \, 2\operatorname{Re} \left\{ \Big\langle
  \mathcal{M}^{0,1}_{12\rightarrow 34} \Big| {\bf T}_i {\bf T}_j \Big|
  \mathcal{M}^{1,0}_{12\rightarrow 34} \Big\rangle \right\} 
\\
  =& \, 2\operatorname{Re} \left\{ \Big\lbrack \mathcal{M}^{0,1\;c_1\ldots
    b_i\ldots b_j\ldots c_4}_{12\rightarrow 34}\Big\rbrack^{*}
  T^a_{b_ic_i}T^a_{b_jc_j} \mathcal{M}^{1,0\;c_1\ldots c_i\ldots
    c_j\ldots c_4}_{12\rightarrow 34}\right\}.
\end{split}
\end{align}
In the case of squark--squark production the tree-level amplitude can
be decomposed according to their color structure as
\begin{align}
\begin{split}
\mathcal{M}^{i,j \; c_1c_2c_3c_4}_{12\rightarrow 34} &=
\delta_{c_1c_3}\delta_{c_2c_4} \mathcal{M}_1^{i,j} + 
\delta_{c_1c_4}\delta_{c_2c_2} \mathcal{M}_2^{i,j}, ~~~~~~ (i,j)=(1,0),(0,1),
\\
\mathcal{M}_1^{0,1} &= \mathcal{M}_T^{0,1},\qquad\quad
\mathcal{M}_1^{1,0} = \frac{1}{2} \left( \mathcal{M}_U^{1,0} -
\frac{1}{3} \mathcal{M}_T^{1,0}\right),
\\
\mathcal{M}_2^{0,1} &= \mathcal{M}_U^{0,1},\qquad\quad
\mathcal{M}_2^{1,0} = \frac{1}{2} \left( \mathcal{M}_T^{1,0} -
\frac{1}{3} \mathcal{M}_U^{1,0}\right),
\end{split}
\end{align}
where $\mathcal{M}_{U,T}^{i,j}$ are the amplitudes corresponding
to the $u$-channel and $t$-channel diagrams, respectively. In this
case the color correlated amplitudes $\mathcal{F}_{ij}$ are
given by
\begin{align}
    \begin{split}
    \mathcal{F}_{12}&=\mathcal{F}_{34} = 4 \left[
    \left(\mathcal{M}_1^{0,1}\right)^* \mathcal{M}_2^{1,0} +
    \mathcal{M}_2^{0,1*}\mathcal{M}_1^{1,0} \right], 
\\
    \mathcal{F}_{13}&=\mathcal{F}_{24} = -12\,
    \left(\mathcal{M}_1^{0,1}\right)^*\mathcal{M}_1^{1,0} -4 \left[
    \left(\mathcal{M}_1^{0,1}\right)^*\mathcal{M}_2^{1,0}
    +\left(\mathcal{M}_2^{0,1}\right)^*\mathcal{M}_1^{1,0} \right],
\\
    \mathcal{F}_{14}&=\mathcal{F}_{23} = -4 \left[
    \left(\mathcal{M}_1^{0,1}\right)^*\mathcal{M}_2^{1,0}
    +\left(\mathcal{M}_2^{0,1}\right)^*\mathcal{M}_1^{1,0} \right]
    -12\,\left(\mathcal{M}_2^{0,1}\right)^*\mathcal{M}_2^{1,0}, 
\\
    \mathcal{F}_{ii} &= 12
    \left[\left(\mathcal{M}_1^{0,1}\right)^*\mathcal{M}_1^{1,0} +
    \left(\mathcal{M}_2^{0,1}\right)^*\mathcal{M}_2^{1,0}\right] + 4 \left[
    \left(\mathcal{M}_1^{0,1}\right)^*\mathcal{M}_2^{1,0} +
    \left(\mathcal{M}_2^{0,1}\right)^*\mathcal{M}_1^{1,0} \right],
\end{split}
\end{align}
where in the last case $i=1,\ldots,4.$

\par The partonic cross section in the collinear region is again
given by a convolution integral, similar to \eqref{eq_coll},
\bee
d\hat\sigma_{12 \rightarrow 34 g}^{2,1}(\hat s)
\Big|_{\text{coll}} &=& \frac{\alphas C_F}{\pi}
\int_{z_0}^{1-\delta_s} dz\;\kappa_{qq}(z,\hat s)\;
d\hat\sigma_{12 \rightarrow 34}^{1,1}(z\hat s),
\eee
with $C_F=4/3$. $z_0$ and $\kappa_{qq}$ are defined in
Eq.~(\ref{eq_coll}).

\subsection{Collinear quark bremsstrahlung}
 \label{subsec_Acollquark}

For real quark emission, only initial-state collinear singularities
arise. The partonic cross section in the collinear region is given by a convolution integral,
\bee
  d\hat\sigma_{12 \rightarrow 34 \bar q}^{2,1}(\hat s)\Big|_{\text{coll}} &=&
  \frac{\alphas T_F}{2\pi} \int_{z_0}^1 \text{d}z\; \kappa_{qg}(z,\hat s)
  d\hat\sigma_{12 \rightarrow 34}^{1,1}(z\hat s),
\eee
with $T_F = 1/2$. $\kappa_{qg}$ is given by
\begin{equation}
  \kappa_{qg} = (z^2+(1-z)^2) \ln\left(\frac{\hat s\delta_\theta
    (1-z)^2}{2m_q^2}\right) +2z(1-z),
\end{equation}
while $z_0$ is defined according to Eq~(\ref{eq_coll}).

\clearpage
\section{Feynman diagrams}
\label{sect_App_feynman}

In this Appendix, we list all parton-level Feynman diagrams for the EW
contributions to the generic process $q q' \to \squarka \squarkb'$
with $q,q'=\{u,d,c,s\}$. The complete list of LO Feynman diagrams is
given in \figref{tab_classes}.\\[2ex]


\FIGURE{
\vspace*{1ex}
\small\unitlength=0.228bp%
\begin{feynartspicture}(300,300)(1,1)
\FADiagram{}
\FAProp(0.,15.)(10.,14.5)(0.,){/Straight}{1}
\FALabel(5.0774,15.8181)[b]{$q$}
\FAProp(0.,5.)(6.5,5.5)(0.,){/Straight}{1}
\FALabel(3.36888,4.18457)[t]{$q^{\prime}$}
\FAProp(20.,15.)(10.,14.5)(0.,){/ScalarDash}{-1}
\FALabel(14.9226,15.8181)[b]{$\tilde q_\alpha$}
\FAProp(20.,5.)(13.5,5.5)(0.,){/ScalarDash}{-1}
\FALabel(16.6311,4.18457)[t]{$\tilde q^{\prime}_\beta$}
\FAProp(10.,14.5)(10.,11.)(0.,){/Straight}{0}
\FALabel(10.82,12.75)[l]{$\tilde g$}
\FAProp(6.5,5.5)(13.5,5.5)(0.,){/Straight}{0}
\FALabel(10.,4.68)[t]{$\tilde \chi_k^0$}
\FAProp(6.5,5.5)(10.,11.)(0.,){/ScalarDash}{1}
\FALabel(7.42232,8.60216)[br]{$\tilde q^{\prime}$}
\FAProp(13.5,5.5)(10.,11.)(0.,){/Straight}{-1}
\FALabel(12.5777,8.60216)[bl]{$q^{\prime}$}
\FAVert(10.,14.5){0}
\FAVert(6.5,5.5){0}
\FAVert(13.5,5.5){0}
\FAVert(10.,11.){0}
\end{feynartspicture}
\begin{feynartspicture}(300,300)(1,1)
\FADiagram{}
\FAProp(0.,15.)(10.,14.5)(0.,){/Straight}{1}
\FALabel(5.0774,15.8181)[b]{$q$}
\FAProp(0.,5.)(6.5,5.5)(0.,){/Straight}{1}
\FALabel(3.36888,4.18457)[t]{$q^{\prime}$}
\FAProp(20.,15.)(10.,14.5)(0.,){/ScalarDash}{-1}
\FALabel(14.9226,15.8181)[b]{$\tilde q_\alpha$}
\FAProp(20.,5.)(13.5,5.5)(0.,){/ScalarDash}{-1}
\FALabel(16.6311,4.18457)[t]{$\tilde q^{\prime}_\beta$}
\FAProp(10.,14.5)(10.,11.)(0.,){/Straight}{0}
\FALabel(10.82,12.75)[l]{$\tilde g$}
\FAProp(6.5,5.5)(13.5,5.5)(0.,){/Straight}{-1}
\FALabel(10.,4.43)[t]{$\tilde \chi_k^\pm$}
\FAProp(6.5,5.5)(10.,11.)(0.,){/ScalarDash}{1}
\FALabel(7.42232,8.60216)[br]{$\tilde Q^{\prime}$}
\FAProp(13.5,5.5)(10.,11.)(0.,){/Straight}{-1}
\FALabel(12.5777,8.60216)[bl]{$Q^{\prime}$}
\FAVert(10.,14.5){0}
\FAVert(6.5,5.5){0}
\FAVert(13.5,5.5){0}
\FAVert(10.,11.){0}
\end{feynartspicture}
\begin{feynartspicture}(300,300)(1,1)
\FADiagram{}
\FAProp(0.,15.)(10.,14.5)(0.,){/Straight}{1}
\FALabel(5.0774,15.8181)[b]{$q$}
\FAProp(0.,5.)(6.5,5.5)(0.,){/Straight}{1}
\FALabel(3.36888,4.18457)[t]{$q^{\prime}$}
\FAProp(20.,15.)(10.,14.5)(0.,){/ScalarDash}{-1}
\FALabel(14.9226,15.8181)[b]{$\tilde q_\alpha$}
\FAProp(20.,5.)(13.5,5.5)(0.,){/ScalarDash}{-1}
\FALabel(16.6311,4.18457)[t]{$\tilde q^{\prime}_\beta$}
\FAProp(10.,14.5)(10.,11.)(0.,){/Straight}{0}
\FALabel(10.82,12.75)[l]{$\tilde g$}
\FAProp(6.5,5.5)(13.5,5.5)(0.,){/Sine}{0}
\FALabel(10.,4.43)[t]{$V^0$}
\FAProp(6.5,5.5)(10.,11.)(0.,){/Straight}{1}
\FALabel(7.42232,8.60216)[br]{$q^{\prime}$}
\FAProp(13.5,5.5)(10.,11.)(0.,){/ScalarDash}{-1}
\FALabel(12.5777,8.60216)[bl]{$\tilde q^{\prime}_\beta$}
\FAVert(10.,14.5){0}
\FAVert(6.5,5.5){0}
\FAVert(13.5,5.5){0}
\FAVert(10.,11.){0}
\end{feynartspicture}
\begin{feynartspicture}(300,300)(1,1)
\FADiagram{}
\FAProp(0.,15.)(10.,14.5)(0.,){/Straight}{1}
\FALabel(5.0774,15.8181)[b]{$q$}
\FAProp(0.,5.)(6.5,5.5)(0.,){/Straight}{1}
\FALabel(3.36888,4.18457)[t]{$q^{\prime}$}
\FAProp(20.,15.)(10.,14.5)(0.,){/ScalarDash}{-1}
\FALabel(14.9226,15.8181)[b]{$\tilde q_\alpha$}
\FALabel(16.6311,4.18457)[t]{$\tilde q^{\prime}_\beta$}
\FAProp(20.,5.)(13.5,5.5)(0.,){/ScalarDash}{-1}
\FAProp(10.,14.5)(10.,11.)(0.,){/Straight}{0}
\FALabel(10.82,12.75)[l]{$\tilde g$}
\FAProp(6.5,5.5)(13.5,5.5)(0.,){/Sine}{-1}
\FALabel(10.,4.43)[t]{$W$}
\FAProp(6.5,5.5)(10.,11.)(0.,){/Straight}{1}
\FALabel(7.42232,8.60216)[br]{$Q^{\prime}$}
\FAProp(13.5,5.5)(10.,11.)(0.,){/ScalarDash}{-1}
\FALabel(12.5777,8.60216)[bl]{$\tilde Q^{\prime}_{\beta}$}
\FAVert(10.,14.5){0}
\FAVert(6.5,5.5){0}
\FAVert(13.5,5.5){0}
\FAVert(10.,11.){0}
\end{feynartspicture}
\begin{feynartspicture}(300,300)(1,1)
\FADiagram{}
\FAProp(0.,15.)(6.5,14.5)(0.,){/Straight}{1}
\FALabel(3.36888,15.8154)[b]{$q$}
\FAProp(0.,5.)(10.,5.5)(0.,){/Straight}{1}
\FALabel(5.0774,4.18193)[t]{$q^{\prime}$}
\FAProp(20.,15.)(13.5,14.5)(0.,){/ScalarDash}{-1}
\FALabel(16.6311,15.8154)[b]{$\tilde q_\alpha$}
\FAProp(20.,5.)(10.,5.5)(0.,){/ScalarDash}{-1}
\FALabel(14.9226,4.18193)[t]{$\tilde q^{\prime}_\beta$}
\FAProp(10.,5.5)(10.,8.5)(0.,){/Straight}{0}
\FALabel(9.18,7.)[r]{$\tilde g$}
\FAProp(6.5,14.5)(13.5,14.5)(0.,){/Straight}{0}
\FALabel(10.,15.32)[b]{$\tilde \chi_k^0$}
\FAProp(6.5,14.5)(10.,8.5)(0.,){/ScalarDash}{1}
\FALabel(7.39114,11.199)[tr]{$\tilde q$}
\FAProp(13.5,14.5)(10.,8.5)(0.,){/Straight}{-1}
\FALabel(12.6089,11.199)[tl]{$q$}
\FAVert(6.5,14.5){0}
\FAVert(10.,5.5){0}
\FAVert(13.5,14.5){0}
\FAVert(10.,8.5){0}
\end{feynartspicture}
\begin{feynartspicture}(300,300)(1,1)
\FADiagram{}
\FAProp(0.,15.)(6.5,14.5)(0.,){/Straight}{1}
\FALabel(3.36888,15.8154)[b]{$q$}
\FAProp(0.,5.)(10.,5.5)(0.,){/Straight}{1}
\FALabel(5.0774,4.18193)[t]{$q^{\prime}$}
\FAProp(20.,15.)(13.5,14.5)(0.,){/ScalarDash}{-1}
\FALabel(16.6311,15.8154)[b]{$\tilde q_\alpha$}
\FAProp(20.,5.)(10.,5.5)(0.,){/ScalarDash}{-1}
\FALabel(14.9226,4.18193)[t]{$\tilde q^{\prime}_\beta$}
\FAProp(10.,5.5)(10.,8.5)(0.,){/Straight}{0}
\FALabel(9.18,7.)[r]{$\tilde g$}
\FAProp(6.5,14.5)(13.5,14.5)(0.,){/Straight}{-1}
\FALabel(10.,15.57)[b]{$\tilde \chi_k$}
\FAProp(6.5,14.5)(10.,8.5)(0.,){/ScalarDash}{1}
\FALabel(7.39114,11.199)[tr]{$\tilde Q$}
\FAProp(13.5,14.5)(10.,8.5)(0.,){/Straight}{-1}
\FALabel(12.6089,11.199)[tl]{$Q$}
\FAVert(6.5,14.5){0}
\FAVert(10.,5.5){0}
\FAVert(13.5,14.5){0}
\FAVert(10.,8.5){0}
\end{feynartspicture}
\begin{feynartspicture}(300,300)(1,1)
\FADiagram{}
\FAProp(0.,15.)(6.5,14.5)(0.,){/Straight}{1}
\FALabel(3.36888,15.8154)[b]{$q$}
\FAProp(0.,5.)(10.,5.5)(0.,){/Straight}{1}
\FALabel(5.0774,4.18193)[t]{$q^{\prime}$}
\FAProp(20.,15.)(13.5,14.5)(0.,){/ScalarDash}{-1}
\FALabel(16.6311,15.8154)[b]{$\tilde q_\alpha$}
\FAProp(20.,5.)(10.,5.5)(0.,){/ScalarDash}{-1}
\FALabel(14.9226,4.18193)[t]{$\tilde q^{\prime}_\beta$}
\FAProp(10.,5.5)(10.,8.5)(0.,){/Straight}{0}
\FALabel(9.18,7.)[r]{$\tilde g$}
\FAProp(6.5,14.5)(13.5,14.5)(0.,){/Sine}{0}
\FALabel(10.,15.57)[b]{$V^0$}
\FAProp(6.5,14.5)(10.,8.5)(0.,){/Straight}{1}
\FALabel(7.39114,11.199)[tr]{$q$}
\FAProp(13.5,14.5)(10.,8.5)(0.,){/ScalarDash}{-1}
\FALabel(12.6089,11.199)[tl]{$\tilde q_\alpha$}
\FAVert(6.5,14.5){0}
\FAVert(10.,5.5){0}
\FAVert(13.5,14.5){0}
\FAVert(10.,8.5){0}
\end{feynartspicture}
\begin{feynartspicture}(300,300)(1,1)
\FADiagram{}
\FAProp(0.,15.)(6.5,14.5)(0.,){/Straight}{1}
\FALabel(3.36888,15.8154)[b]{$q$}
\FAProp(0.,5.)(10.,5.5)(0.,){/Straight}{1}
\FALabel(5.0774,4.18193)[t]{$q^{\prime}$}
\FAProp(20.,15.)(13.5,14.5)(0.,){/ScalarDash}{-1}
\FALabel(16.6311,15.8154)[b]{$\tilde q_\alpha$}
\FAProp(20.,5.)(10.,5.5)(0.,){/ScalarDash}{-1}
\FALabel(14.9226,4.18193)[t]{$\tilde q^{\prime}_\beta$}
\FAProp(10.,5.5)(10.,8.5)(0.,){/Straight}{0}
\FALabel(9.18,7.)[r]{$\tilde g$}
\FAProp(6.5,14.5)(13.5,14.5)(0.,){/Sine}{-1}
\FALabel(10.,15.57)[b]{$W$}
\FAProp(6.5,14.5)(10.,8.5)(0.,){/Straight}{1}
\FALabel(7.39114,11.199)[tr]{$Q$}
\FAProp(13.5,14.5)(10.,8.5)(0.,){/ScalarDash}{-1}
\FALabel(12.6089,11.199)[tl]{$\tilde Q_{\alpha}$}
\FAVert(6.5,14.5){0}
\FAVert(10.,5.5){0}
\FAVert(13.5,14.5){0}
\FAVert(10.,8.5){0}
\end{feynartspicture}
\begin{feynartspicture}(300,300)(1,1)
\FADiagram{}
\FAProp(0.,15.)(6.,13.5)(0.,){/Straight}{1}
\FALabel(3.37593,15.2737)[b]{$q$}
\FAProp(0.,5.)(6.,6.5)(0.,){/Straight}{1}
\FALabel(3.37593,4.72628)[t]{$q^{\prime}$}
\FAProp(20.,15.)(12.,10.)(0.,){/ScalarDash}{-1}
\FALabel(20.,15.8181)[b]{$\tilde q_\alpha$}
\FALabel(20.,4.18457)[t]{$\tilde q^{\prime}_\beta$}
\FAProp(20.,5.)(12.,10.)(0.,){/ScalarDash}{-1}
\FAProp(6.,13.5)(6.,6.5)(0.,){/Straight}{0}
\FALabel(5.18,10.)[r]{$\tilde g$}
\FAProp(6.,13.5)(12.,10.)(0.,){/ScalarDash}{1}
\FALabel(9.301,12.6089)[bl]{$\tilde q$}
\FAProp(6.,6.5)(12.,10.)(0.,){/ScalarDash}{1}
\FALabel(9.301,7.39114)[tl]{$\tilde q^{\prime}$}
\FAVert(6.,13.5){0}
\FAVert(6.,6.5){0}
\FAVert(12.,10.){0}
\end{feynartspicture}
\begin{feynartspicture}(300,300)(1,1)
\FADiagram{}
\FAProp(0.,15.)(6.5,13.5)(0.,){/Straight}{1}
\FALabel(3.59853,15.2803)[b]{$q$}
\FAProp(0.,5.)(6.5,6.5)(0.,){/Straight}{1}
\FALabel(3.59853,4.71969)[t]{$q^{\prime}$}
\FAProp(20.,15.)(13.5,13.5)(0.,){/ScalarDash}{-1}
\FALabel(16.4015,15.2803)[b]{$\tilde q_\alpha$}
\FAProp(20.,5.)(13.5,6.5)(0.,){/ScalarDash}{-1}
\FALabel(16.4015,4.71969)[t]{$\tilde q^{\prime}_\beta$}
\FAProp(6.5,13.5)(6.5,6.5)(0.,){/Straight}{0}
\FALabel(5.68,10.)[r]{$\tilde g$}
\FAProp(6.5,13.5)(13.5,13.5)(0.,){/ScalarDash}{1}
\FALabel(10.,14.57)[b]{$\tilde q$}
\FAProp(6.5,6.5)(13.5,6.5)(0.,){/ScalarDash}{1}
\FALabel(10.,5.43)[t]{$\tilde q^{\prime}$}
\FAProp(13.5,13.5)(13.5,6.5)(0.,){/ScalarDash}{0}
\FALabel(14.32,10.)[l]{$S^0$}
\FAVert(6.5,13.5){0}
\FAVert(6.5,6.5){0}
\FAVert(13.5,13.5){0}
\FAVert(13.5,6.5){0}
\end{feynartspicture}
\begin{feynartspicture}(300,300)(1,1)
\FADiagram{}
\FAProp(0.,15.)(6.5,13.5)(0.,){/Straight}{1}
\FALabel(3.59853,15.2803)[b]{$q$}
\FAProp(0.,5.)(6.5,6.5)(0.,){/Straight}{1}
\FALabel(3.59853,4.71969)[t]{$q^{\prime}$}
\FAProp(20.,15.)(13.5,13.5)(0.,){/ScalarDash}{-1}
\FALabel(16.4015,15.2803)[b]{$\tilde q_\alpha$}
\FAProp(20.,5.)(13.5,6.5)(0.,){/ScalarDash}{-1}
\FALabel(16.4015,4.71969)[t]{$\tilde q^{\prime}_\beta$}
\FAProp(6.5,13.5)(6.5,6.5)(0.,){/Sine}{0}
\FALabel(5.43,10.)[r]{$V^0$}
\FAProp(6.5,13.5)(13.5,13.5)(0.,){/Straight}{1}
\FALabel(10.,14.57)[b]{$q$}
\FAProp(6.5,6.5)(13.5,6.5)(0.,){/Straight}{1}
\FALabel(10.,5.43)[t]{$q^{\prime}$}
\FAProp(13.5,13.5)(13.5,6.5)(0.,){/Straight}{0}
\FALabel(14.32,10.)[l]{$\tilde g$}
\FAVert(6.5,13.5){0}
\FAVert(6.5,6.5){0}
\FAVert(13.5,13.5){0}
\FAVert(13.5,6.5){0}
\end{feynartspicture}
\begin{feynartspicture}(300,300)(1,1)
\FADiagram{}
\FAProp(0.,15.)(6.5,13.5)(0.,){/Straight}{1}
\FALabel(3.59853,15.2803)[b]{$q$}
\FAProp(0.,5.)(6.5,6.5)(0.,){/Straight}{1}
\FALabel(3.59853,4.71969)[t]{$q^{\prime}$}
\FAProp(20.,15.)(13.5,13.5)(0.,){/ScalarDash}{-1}
\FALabel(16.4015,15.2803)[b]{$\tilde q_\alpha$}
\FAProp(20.,5.)(13.5,6.5)(0.,){/ScalarDash}{-1}
\FALabel(16.4015,4.71969)[t]{$\tilde q^{\prime}_\beta$}
\FAProp(6.5,13.5)(6.5,6.5)(0.,){/Straight}{0}
\FALabel(5.68,10.)[r]{$\tilde g$}
\FAProp(6.5,13.5)(13.5,13.5)(0.,){/ScalarDash}{1}
\FALabel(10.,14.57)[b]{$\tilde q_\alpha$}
\FAProp(6.5,6.5)(13.5,6.5)(0.,){/ScalarDash}{1}
\FALabel(10.,5.43)[t]{$\tilde q^{\prime}_\beta$}
\FAProp(13.5,13.5)(13.5,6.5)(0.,){/Sine}{0}
\FALabel(14.57,10.)[l]{$V^0$}
\FAVert(6.5,13.5){0}
\FAVert(6.5,6.5){0}
\FAVert(13.5,13.5){0}
\FAVert(13.5,6.5){0}
\end{feynartspicture}
\\
\begin{feynartspicture}(300,300)(1,1)
\FADiagram{}
\FAProp(0.,15.)(13.5,13.)(0.,){/Straight}{1}
\FALabel(3.41023,15.5279)[b]{$u$}
\FAProp(0.,5.)(6.5,6.)(0.,){/Straight}{1}
\FALabel(3.48569,4.44802)[t]{$d$}
\FAProp(20.,15.)(6.5,13.)(0.,){/ScalarDash}{-1}
\FALabel(16.5898,15.5279)[b]{$\tilde u_\alpha$}
\FAProp(20.,5.)(13.5,6.)(0.,){/ScalarDash}{-1}
\FALabel(16.5143,4.44802)[t]{$\tilde d_\beta$}
\FAProp(13.5,13.)(6.5,13.)(0.,){/Straight}{0}
\FALabel(10.,12.18)[t]{$\tilde g$}
\FAProp(13.5,13.)(13.5,6.)(0.,){/ScalarDash}{1}
\FALabel(14.57,9.5)[l]{$\tilde u_{\beta}$}
\FAProp(6.5,6.)(6.5,13.)(0.,){/Straight}{1}
\FALabel(5.43,9.5)[r]{$u$}
\FAProp(6.5,6.)(13.5,6.)(0.,){/Sine}{1}
\FALabel(10.,4.93)[t]{$W$}
\FAVert(13.5,13.){0}
\FAVert(6.5,6.){0}
\FAVert(6.5,13.){0}
\FAVert(13.5,6.){0}
\end{feynartspicture}
\begin{feynartspicture}(300,300)(1,1)
\FADiagram{}
\FAProp(0.,15.)(13.5,13.)(0.,){/Straight}{1}
\FALabel(3.41023,15.5279)[b]{$u$}
\FAProp(0.,5.)(6.5,6.)(0.,){/Straight}{1}
\FALabel(3.48569,4.44802)[t]{$d$}
\FAProp(20.,15.)(6.5,13.)(0.,){/ScalarDash}{-1}
\FALabel(16.5898,15.5279)[b]{$\tilde u_\alpha$}
\FAProp(20.,5.)(13.5,6.)(0.,){/ScalarDash}{-1}
\FALabel(16.5143,4.44802)[t]{$\tilde d_\beta$}
\FAProp(13.5,13.)(6.5,13.)(0.,){/Sine}{-1}
\FALabel(10.,11.93)[t]{$W$}
\FAProp(13.5,13.)(13.5,6.)(0.,){/Straight}{1}
\FALabel(14.57,9.5)[l]{$d$}
\FAProp(6.5,6.)(6.5,13.)(0.,){/ScalarDash}{1}
\FALabel(5.43,9.5)[r]{$\tilde d_{\alpha}$}
\FAProp(6.5,6.)(13.5,6.)(0.,){/Straight}{0}
\FALabel(10.,5.18)[t]{$\tilde g$}
\FAVert(13.5,13.){0}
\FAVert(6.5,6.){0}
\FAVert(6.5,13.){0}
\FAVert(13.5,6.){0}
\end{feynartspicture}
\begin{feynartspicture}(300,300)(1,1)
\FADiagram{}
\FAProp(0.,15.)(6.5,13.5)(0.,){/Straight}{1}
\FALabel(3.59853,15.2803)[b]{$u$}
\FAProp(0.,5.)(6.5,6.5)(0.,){/Straight}{1}
\FALabel(3.59853,4.71969)[t]{$d$}
\FAProp(20.,15.)(13.5,6.5)(0.,){/ScalarDash}{-1}
\FALabel(20.,15.8181)[b]{$\tilde u_\alpha$}
\FALabel(20.,4.18457)[t]{$\tilde d_\beta$}
\FAProp(20.,5.)(13.5,13.5)(0.,){/ScalarDash}{-1}
\FAProp(6.5,13.5)(6.5,6.5)(0.,){/Straight}{0}
\FALabel(5.68,10.)[r]{$\tilde g$}
\FAProp(6.5,13.5)(13.5,13.5)(0.,){/ScalarDash}{1}
\FALabel(10.,14.57)[b]{$\tilde u$}
\FAProp(6.5,6.5)(13.5,6.5)(0.,){/ScalarDash}{1}
\FALabel(10.,5.43)[t]{$\tilde d$}
\FAProp(13.5,6.5)(13.5,13.5)(0.,){/ScalarDash}{1}
\FALabel(12.43,10.)[r]{$H$}
\FAVert(6.5,13.5){0}
\FAVert(6.5,6.5){0}
\FAVert(13.5,6.5){0}
\FAVert(13.5,13.5){0}
\end{feynartspicture}
\begin{feynartspicture}(300,300)(1,1)
\FADiagram{}
\FAProp(0.,15.)(6.5,13.5)(0.,){/Straight}{1}
\FALabel(3.59853,15.2803)[b]{$u$}
\FAProp(0.,5.)(6.5,6.5)(0.,){/Straight}{1}
\FALabel(3.59853,4.71969)[t]{$d$}
\FAProp(20.,15.)(13.5,6.5)(0.,){/ScalarDash}{-1}
\FALabel(20.,15.8181)[b]{$\tilde u_\alpha$}
\FALabel(20.,4.18457)[t]{$\tilde d_\beta$}
\FAProp(20.,5.)(13.5,13.5)(0.,){/ScalarDash}{-1}
\FAProp(6.5,13.5)(6.5,6.5)(0.,){/Straight}{0}
\FALabel(5.68,10.)[r]{$\tilde g$}
\FAProp(6.5,13.5)(13.5,13.5)(0.,){/ScalarDash}{1}
\FALabel(10.,14.57)[b]{$\tilde u$}
\FAProp(6.5,6.5)(13.5,6.5)(0.,){/ScalarDash}{1}
\FALabel(10.,5.43)[t]{$\tilde d$}
\FAProp(13.5,6.5)(13.5,13.5)(0.,){/ScalarDash}{1}
\FALabel(12.43,10.)[r]{$G$}
\FAVert(6.5,13.5){0}
\FAVert(6.5,6.5){0}
\FAVert(13.5,6.5){0}
\FAVert(13.5,13.5){0}
\end{feynartspicture}
\begin{feynartspicture}(300,300)(1,1)
\FADiagram{}
\FAProp(0.,15.)(6.5,13.5)(0.,){/Straight}{1}
\FALabel(3.59853,15.2803)[b]{$u$}
\FAProp(0.,5.)(6.5,6.5)(0.,){/Straight}{1}
\FALabel(3.59853,4.71969)[t]{$d$}
\FAProp(20.,15.)(13.5,6.5)(0.,){/ScalarDash}{-1}
\FALabel(20.,15.8181)[b]{$\tilde u_\alpha$}
\FALabel(20.,4.18457)[t]{$\tilde d_\beta$}
\FAProp(20.,5.)(13.5,13.5)(0.,){/ScalarDash}{-1}
\FAProp(6.5,13.5)(6.5,6.5)(0.,){/Sine}{-1}
\FALabel(5.43,10.)[r]{$W$}
\FAProp(6.5,13.5)(13.5,13.5)(0.,){/Straight}{1}
\FALabel(10.,14.57)[b]{$d$}
\FAProp(6.5,6.5)(13.5,6.5)(0.,){/Straight}{1}
\FALabel(10.,5.43)[t]{$u$}
\FAProp(13.5,6.5)(13.5,13.5)(0.,){/Straight}{0}
\FALabel(12.68,10.)[r]{$\tilde g$}
\FAVert(6.5,13.5){0}
\FAVert(6.5,6.5){0}
\FAVert(13.5,6.5){0}
\FAVert(13.5,13.5){0}
\end{feynartspicture}
\begin{feynartspicture}(300,300)(1,1)
\FADiagram{}
\FAProp(0.,15.)(6.5,13.5)(0.,){/Straight}{1}
\FALabel(3.59853,15.2803)[b]{$u$}
\FAProp(0.,5.)(6.5,6.5)(0.,){/Straight}{1}
\FALabel(3.59853,4.71969)[t]{$d$}
\FAProp(20.,15.)(13.5,6.5)(0.,){/ScalarDash}{-1}
\FALabel(20.,15.8181)[b]{$\tilde u_\alpha$}
\FALabel(20.,4.18457)[t]{$\tilde d_\beta$}
\FAProp(20.,5.)(13.5,13.5)(0.,){/ScalarDash}{-1}
\FAProp(6.5,13.5)(6.5,6.5)(0.,){/Straight}{0}
\FALabel(5.68,10.)[r]{$\tilde g$}
\FAProp(6.5,13.5)(13.5,13.5)(0.,){/ScalarDash}{1}
\FALabel(10.,14.57)[b]{$\tilde u_{\beta}$}
\FAProp(6.5,6.5)(13.5,6.5)(0.,){/ScalarDash}{1}
\FALabel(10.,5.43)[t]{$\tilde d_{\alpha}$}
\FAProp(13.5,6.5)(13.5,13.5)(0.,){/Sine}{1}
\FALabel(12.43,10.)[r]{$W$}
\FAVert(6.5,13.5){0}
\FAVert(6.5,6.5){0}
\FAVert(13.5,6.5){0}
\FAVert(13.5,13.5){0}
\end{feynartspicture}
\\

\begin{feynartspicture}(300,300)(1,1)
\FADiagram{}
\FAProp(0.,15.)(13.5,13.)(0.,){/Straight}{1}
\FALabel(3.41023,15.5279)[b]{$q$}
\FAProp(0.,5.)(6.5,6.)(0.,){/Straight}{1}
\FALabel(3.48569,4.44802)[t]{$q^{\prime}$}
\FAProp(20.,15.)(6.5,13.)(0.,){/ScalarDash}{-1}
\FALabel(16.5898,15.5279)[b]{$\tilde q_\alpha$}
\FAProp(20.,5.)(13.5,6.)(0.,){/ScalarDash}{-1}
\FALabel(16.5143,4.44802)[t]{$\tilde q^{\prime}_\beta$}
\FAProp(13.5,13.)(6.5,13.)(0.,){/Straight}{1}
\FALabel(10.,11.93)[t]{$q$}
\FAProp(13.5,13.)(13.5,6.)(0.,){/Sine}{0}
\FALabel(14.57,9.5)[l]{$V^0$}
\FAProp(6.5,6.)(6.5,13.)(0.,){/Straight}{0}
\FALabel(5.68,9.5)[r]{$\tilde g$}
\FAProp(6.5,6.)(13.5,6.)(0.,){/ScalarDash}{1}
\FALabel(10.,4.93)[t]{$\tilde q^{\prime}_\beta$}
\FAVert(13.5,13.){0}
\FAVert(6.5,6.){0}
\FAVert(6.5,13.){0}
\FAVert(13.5,6.){0}
\end{feynartspicture}
\begin{feynartspicture}(300,300)(1,1)
\FADiagram{}
\FAProp(0.,15.)(13.5,13.)(0.,){/Straight}{1}
\FALabel(3.41023,15.5279)[b]{$q$}
\FAProp(0.,5.)(6.5,6.)(0.,){/Straight}{1}
\FALabel(3.48569,4.44802)[t]{$q^{\prime}$}
\FAProp(20.,15.)(6.5,13.)(0.,){/ScalarDash}{-1}
\FALabel(16.5898,15.5279)[b]{$\tilde q_\alpha$}
\FAProp(20.,5.)(13.5,6.)(0.,){/ScalarDash}{-1}
\FALabel(16.5143,4.44802)[t]{$\tilde q^{\prime}_\beta$}
\FAProp(13.5,13.)(6.5,13.)(0.,){/ScalarDash}{1}
\FALabel(10.,11.93)[t]{$\tilde q_\alpha$}
\FAProp(13.5,13.)(13.5,6.)(0.,){/Straight}{0}
\FALabel(14.32,9.5)[l]{$\tilde g$}
\FAProp(6.5,6.)(6.5,13.)(0.,){/Sine}{0}
\FALabel(5.43,9.5)[r]{$V^0$}
\FAProp(6.5,6.)(13.5,6.)(0.,){/Straight}{1}
\FALabel(10.,4.93)[t]{$q^{\prime}$}
\FAVert(13.5,13.){0}
\FAVert(6.5,6.){0}
\FAVert(6.5,13.){0}
\FAVert(13.5,6.){0}
\end{feynartspicture}
\begin{feynartspicture}(300,300)(1,1)
\FADiagram{}
\FAProp(0.,15.)(10.,14.)(0.,){/Straight}{1}
\FALabel(5.0774,15.8181)[b]{$q$}
\FAProp(0.,5.)(10.,6.)(0.,){/Straight}{1}
\FALabel(3.36888,4.18457)[t]{$q^{\prime}$}
\FAProp(20.,15.)(10.,14.)(0.,){/ScalarDash}{-1}
\FALabel(16.5898,15.5279)[b]{$\tilde q_\alpha$}
\FAProp(20.,5.)(10.,6.)(0.,){/ScalarDash}{-1}
\FALabel(16.5143,4.44802)[t]{$\tilde q^{\prime}_\beta$}
\FAProp(10.,14.)(10.,6.)(0.,){/Straight}{0}
\FALabel(9.18,10.)[r]{$\tilde g$}
\FAVert(10.,14.){0}
\FAVert(10.,6.){1}
\end{feynartspicture}
\begin{feynartspicture}(300,300)(1,1)
\FADiagram{}
\FAProp(0.,15.)(10.,14.)(0.,){/Straight}{1}
\FALabel(5.0774,15.8181)[b]{$q$}
\FAProp(0.,5.)(10.,6.)(0.,){/Straight}{1}
\FALabel(3.36888,4.18457)[t]{$q^{\prime}$}
\FAProp(20.,15.)(10.,14.)(0.,){/ScalarDash}{-1}
\FALabel(16.5898,15.5279)[b]{$\tilde q_\alpha$}
\FAProp(20.,5.)(10.,6.)(0.,){/ScalarDash}{-1}
\FALabel(16.5143,4.44802)[t]{$\tilde q^{\prime}_\beta$}
\FAProp(10.,14.)(10.,6.)(0.,){/Straight}{0}
\FALabel(9.18,10.)[r]{$\tilde g$}
\FAVert(10.,6.){0}
\FAVert(10.,14.){1}
\end{feynartspicture}

\vspace*{2ex}
\caption{Virtual corrections (I): EW one-loop insertions into QCD Born diagrams.
  $Q$ and $Q'$ denote the SU(2) partner of
  quark $q$ and $q'$, respectively. We use generic labels $V^0 = \gamma, Z$ and 
    $S^0 = h^0, H^0, G^0, A^0$. If the chirality of an internal squark
    is not specified, it can be any.
  The diagram containing the four-squark vertex has to be
  taken at $\mathcal{O}(\alphas\alpha)$. The diagrams in the third
  line contribute only for $u=\{u,c\},\,d=\{d,s\}$. The last two diagrams 
      contain the counter terms, whose renormalization constants have
  to be evaluated at $\mathcal{O}(\alpha)$. For $q=q^\prime$ crossed
  diagrams have to be taken into account.
  \label{fig_EWinsQCD}}
}

\FIGURE{
\vspace*{15ex}
\\
\small\unitlength=0.26bp%
\begin{feynartspicture}(310,310)(1,1)
\FADiagram{}
\FAProp(0.,15.)(10.,14.5)(0.,){/Straight}{1}
\FALabel(5.0774,15.8181)[b]{$q$}
\FAProp(0.,5.)(6.5,5.5)(0.,){/Straight}{1}
\FALabel(3.36888,4.18457)[t]{$q^{\prime}$}
\FAProp(20.,15.)(10.,14.5)(0.,){/ScalarDash}{-1}
\FALabel(14.9226,15.8181)[b]{$\tilde q_\alpha$}
\FAProp(20.,5.)(13.5,5.5)(0.,){/ScalarDash}{-1}
\FALabel(16.6311,4.18457)[t]{$\tilde q^{\prime}_{\beta}$}
\FAProp(10.,14.5)(10.,11.)(0.,){/Straight}{0}
\FALabel(10.82,12.75)[l]{$\tilde \chi_k^0$}
\FAProp(6.5,5.5)(13.5,5.5)(0.,){/Straight}{0}
\FALabel(10.,4.68)[t]{$\tilde g$}
\FAProp(6.5,5.5)(10.,11.)(0.,){/ScalarDash}{1}
\FALabel(7.42232,8.60216)[br]{$\tilde q^{\prime}$}
\FAProp(13.5,5.5)(10.,11.)(0.,){/Straight}{-1}
\FALabel(13.0,6.5)[bl]{$q^{\prime}$}
\FAVert(10.,14.5){0}
\FAVert(6.5,5.5){0}
\FAVert(13.5,5.5){0}
\FAVert(10.,11.){0}
\end{feynartspicture}
\begin{feynartspicture}(310,310)(1,1)
\FADiagram{}
\FAProp(0.,15.)(10.,14.5)(0.,){/Straight}{1}
\FALabel(5.0774,15.8181)[b]{$q$}
\FAProp(0.,5.)(6.5,5.5)(0.,){/Straight}{1}
\FALabel(3.36888,4.18457)[t]{$q^{\prime}$}
\FAProp(20.,15.)(10.,14.5)(0.,){/ScalarDash}{-1}
\FALabel(14.9226,15.8181)[b]{$\tilde q_\alpha$}
\FAProp(20.,5.)(13.5,5.5)(0.,){/ScalarDash}{-1}
\FALabel(16.6311,4.18457)[t]{$\tilde q^{\prime}_{\beta}$}
\FAProp(10.,14.5)(10.,11.)(0.,){/Straight}{0}
\FALabel(10.82,12.75)[l]{$\tilde \chi_k^0$}
\FAProp(6.5,5.5)(13.5,5.5)(0.,){/Cycles}{0}
\FALabel(10.,4.43)[t]{$g$}
\FAProp(6.5,5.5)(10.,11.)(0.,){/Straight}{1}
\FALabel(7.42232,8.60216)[br]{$q^{\prime}$}
\FAProp(13.5,5.5)(10.,11.)(0.,){/ScalarDash}{-1}
\FALabel(13.5,6.0)[bl]{$\tilde q^{\prime}_{\beta}$}
\FAVert(10.,14.5){0}
\FAVert(6.5,5.5){0}
\FAVert(13.5,5.5){0}
\FAVert(10.,11.){0}
\end{feynartspicture}
\begin{feynartspicture}(310,310)(1,1)
\FADiagram{}
\FAProp(0.,15.)(6.5,14.5)(0.,){/Straight}{1}
\FALabel(3.36888,15.8154)[b]{$q$}
\FAProp(0.,5.)(10.,5.5)(0.,){/Straight}{1}
\FALabel(5.0774,4.18193)[t]{$q^{\prime}$}
\FAProp(20.,15.)(13.5,14.5)(0.,){/ScalarDash}{-1}
\FALabel(16.6311,15.8154)[b]{$\tilde q_\alpha$}
\FAProp(20.,5.)(10.,5.5)(0.,){/ScalarDash}{-1}
\FALabel(14.9226,4.18193)[t]{$\tilde q^{\prime}_{\beta}$}
\FAProp(10.,5.5)(10.,8.5)(0.,){/Straight}{0}
\FALabel(9.3,8.3)[r]{$\tilde \chi_k^0$}
\FAProp(6.5,14.5)(13.5,14.5)(0.,){/Straight}{0}
\FALabel(10.,15.32)[b]{$\tilde g$}
\FAProp(6.5,14.5)(10.,8.5)(0.,){/ScalarDash}{1}
\FALabel(7.0,13.0)[tr]{$\tilde q$}
\FAProp(13.5,14.5)(10.,8.5)(0.,){/Straight}{-1}
\FALabel(12.6089,11.199)[tl]{$q$}
\FAVert(6.5,14.5){0}
\FAVert(10.,5.5){0}
\FAVert(13.5,14.5){0}
\FAVert(10.,8.5){0}
\end{feynartspicture}
\begin{feynartspicture}(310,310)(1,1)
\FADiagram{}
\FAProp(0.,15.)(6.5,14.5)(0.,){/Straight}{1}
\FALabel(3.36888,15.8154)[b]{$q$}
\FAProp(0.,5.)(10.,5.5)(0.,){/Straight}{1}
\FALabel(5.0774,4.18193)[t]{$q^{\prime}$}
\FAProp(20.,15.)(13.5,14.5)(0.,){/ScalarDash}{-1}
\FALabel(16.6311,15.8154)[b]{$\tilde q_\alpha$}
\FAProp(20.,5.)(10.,5.5)(0.,){/ScalarDash}{-1}
\FALabel(14.9226,4.18193)[t]{$\tilde q^{\prime}_{\beta}$}
\FAProp(10.,5.5)(10.,8.5)(0.,){/Straight}{0}
\FALabel(9.3,8.3)[r]{$\tilde \chi_k^0$}
\FAProp(6.5,14.5)(13.5,14.5)(0.,){/Cycles}{0}
\FALabel(10.,16.27)[b]{$g$}
\FAProp(6.5,14.5)(10.,8.5)(0.,){/Straight}{1}
\FALabel(7.0,13.0)[tr]{$q$}
\FAProp(13.5,14.5)(10.,8.5)(0.,){/ScalarDash}{-1}
\FALabel(12.6089,11.199)[tl]{$\tilde q_\alpha$}
\FAVert(6.5,14.5){0}
\FAVert(10.,5.5){0}
\FAVert(13.5,14.5){0}
\FAVert(10.,8.5){0}
\end{feynartspicture}
\begin{feynartspicture}(310,310)(1,1)
\FADiagram{}
\FAProp(0.,15.)(6.,13.5)(0.,){/Straight}{1}
\FALabel(3.37593,15.2737)[b]{$q$}
\FAProp(0.,5.)(6.,6.5)(0.,){/Straight}{1}
\FALabel(3.37593,4.72628)[t]{$q^{\prime}$}
\FAProp(20.,15.)(12.,10.)(0.,){/ScalarDash}{-1}
\FAProp(20.,5.)(12.,10.)(0.,){/ScalarDash}{-1}
\FALabel(20.,15.8181)[b]{$\tilde q_\alpha$}
\FALabel(20.,4.18457)[t]{$\tilde q^{\prime}_\beta$}
\FAProp(6.,13.5)(6.,6.5)(0.,){/Straight}{0}
\FALabel(5.18,10.)[r]{$\tilde \chi_k^0$}
\FAProp(6.,13.5)(12.,10.)(0.,){/ScalarDash}{1}
\FALabel(9.301,12.6089)[bl]{$\tilde q$}
\FAProp(6.,6.5)(12.,10.)(0.,){/ScalarDash}{1}
\FALabel(9.301,7.39114)[tl]{$\tilde q^{\prime}$}
\FAVert(6.,13.5){0}
\FAVert(6.,6.5){0}
\FAVert(12.,10.){0}
\end{feynartspicture}
\begin{feynartspicture}(310,310)(1,1)
\FADiagram{}
\FAProp(0.,15.)(6.5,13.5)(0.,){/Straight}{1}
\FALabel(3.59853,15.2803)[b]{$q$}
\FAProp(0.,5.)(6.5,6.5)(0.,){/Straight}{1}
\FALabel(3.59853,4.71969)[t]{$q^{\prime}$}
\FAProp(20.,15.)(13.5,13.5)(0.,){/ScalarDash}{-1}
\FALabel(16.4015,15.2803)[b]{$\tilde q_\alpha$}
\FAProp(20.,5.)(13.5,6.5)(0.,){/ScalarDash}{-1}
\FALabel(16.4015,4.71969)[t]{$\tilde q^{\prime}_{\beta}$}
\FAProp(6.5,13.5)(6.5,6.5)(0.,){/Cycles}{0}
\FALabel(5.43,10.)[r]{$g$}
\FAProp(6.5,13.5)(13.5,13.5)(0.,){/Straight}{1}
\FALabel(10.,14.57)[b]{$q$}
\FAProp(6.5,6.5)(13.5,6.5)(0.,){/Straight}{1}
\FALabel(10.,5.43)[t]{$q^{\prime}$}
\FAProp(13.5,13.5)(13.5,6.5)(0.,){/Straight}{0}
\FALabel(14.32,10.)[l]{$\tilde \chi_k^0$}
\FAVert(6.5,13.5){0}
\FAVert(6.5,6.5){0}
\FAVert(13.5,13.5){0}
\FAVert(13.5,6.5){0}
\end{feynartspicture}
\begin{feynartspicture}(310,310)(1,1)
\FADiagram{}
\FAProp(0.,15.)(6.5,13.5)(0.,){/Straight}{1}
\FALabel(3.59853,15.2803)[b]{$q$}
\FAProp(0.,5.)(6.5,6.5)(0.,){/Straight}{1}
\FALabel(3.59853,4.71969)[t]{$q^{\prime}$}
\FAProp(20.,15.)(13.5,13.5)(0.,){/ScalarDash}{-1}
\FALabel(16.4015,15.2803)[b]{$\tilde q_\alpha$}
\FAProp(20.,5.)(13.5,6.5)(0.,){/ScalarDash}{-1}
\FALabel(16.4015,4.71969)[t]{$\tilde q^{\prime}_{\beta}$}
\FAProp(6.5,13.5)(6.5,6.5)(0.,){/Straight}{0}
\FALabel(5.68,10.)[r]{$\tilde \chi_k^0$}
\FAProp(6.5,13.5)(13.5,13.5)(0.,){/ScalarDash}{1}
\FALabel(10.,14.57)[b]{$\tilde q_\alpha$}
\FAProp(6.5,6.5)(13.5,6.5)(0.,){/ScalarDash}{1}
\FALabel(10.,5.43)[t]{$\tilde q^{\prime}_{\beta}$}
\FAProp(13.5,13.5)(13.5,6.5)(0.,){/Cycles}{0}
\FALabel(15.27,10.)[l]{$g$}
\FAVert(6.5,13.5){0}
\FAVert(6.5,6.5){0}
\FAVert(13.5,13.5){0}
\FAVert(13.5,6.5){0}
\end{feynartspicture}
\begin{feynartspicture}(310,310)(1,1)
\FADiagram{}
\FAProp(0.,15.)(13.5,13.)(0.,){/Straight}{1}
\FALabel(3.41023,15.5279)[b]{$q$}
\FAProp(0.,5.)(6.5,6.)(0.,){/Straight}{1}
\FALabel(3.48569,4.44802)[t]{$q^{\prime}$}
\FAProp(20.,15.)(6.5,13.)(0.,){/ScalarDash}{-1}
\FALabel(16.5898,15.5279)[b]{$\tilde q_\alpha$}
\FAProp(20.,5.)(13.5,6.)(0.,){/ScalarDash}{-1}
\FALabel(16.5143,4.44802)[t]{$\tilde q^{\prime}_{\beta}$}
\FAProp(13.5,13.)(6.5,13.)(0.,){/Straight}{1}
\FALabel(10.,11.93)[t]{$q$}
\FAProp(13.5,13.)(13.5,6.)(0.,){/Cycles}{0}
\FALabel(15.27,9.5)[l]{$g$}
\FAProp(6.5,6.)(6.5,13.)(0.,){/Straight}{0}
\FALabel(5.68,9.5)[r]{$\tilde \chi_k^0$}
\FAProp(6.5,6.)(13.5,6.)(0.,){/ScalarDash}{1}
\FALabel(10.,4.93)[t]{$\tilde q^{\prime}_{\beta}$}
\FAVert(13.5,13.){0}
\FAVert(6.5,6.){0}
\FAVert(6.5,13.){0}
\FAVert(13.5,6.){0}
\end{feynartspicture}
\begin{feynartspicture}(310,310)(1,1)
\FADiagram{}
\FAProp(0.,15.)(13.5,13.)(0.,){/Straight}{1}
\FALabel(3.41023,15.5279)[b]{$q$}
\FAProp(0.,5.)(6.5,6.)(0.,){/Straight}{1}
\FALabel(3.48569,4.44802)[t]{$q^{\prime}$}
\FAProp(20.,15.)(6.5,13.)(0.,){/ScalarDash}{-1}
\FALabel(16.5898,15.5279)[b]{$\tilde q_\alpha$}
\FAProp(20.,5.)(13.5,6.)(0.,){/ScalarDash}{-1}
\FALabel(16.5143,4.44802)[t]{$\tilde q^{\prime}_{\beta}$}
\FAProp(13.5,13.)(6.5,13.)(0.,){/ScalarDash}{1}
\FALabel(10.,11.93)[t]{$\tilde q_\alpha$}
\FAProp(13.5,13.)(13.5,6.)(0.,){/Straight}{0}
\FALabel(14.32,9.5)[l]{$\tilde \chi_k^0$}
\FAProp(6.5,6.)(6.5,13.)(0.,){/Cycles}{0}
\FALabel(4.73,9.5)[r]{$g$}
\FAProp(6.5,6.)(13.5,6.)(0.,){/Straight}{1}
\FALabel(10.,4.93)[t]{$q^{\prime}$}
\FAVert(13.5,13.){0}
\FAVert(6.5,6.){0}
\FAVert(6.5,13.){0}
\FAVert(13.5,6.){0}
\end{feynartspicture} \\
\begin{feynartspicture}(310,310)(1,1)
\FADiagram{}
\FAProp(0.,15.)(6.5,14.5)(0.,){/Straight}{1}
\FALabel(3.36888,15.8154)[b]{$u$}
\FAProp(0.,5.)(10.,5.5)(0.,){/Straight}{1}
\FALabel(5.0774,4.18193)[t]{$d$}
\FAProp(20.,15.)(10.,5.5)(0.,){/ScalarDash}{-1}
\FAProp(20.,5.)(13.5,14.5)(0.,){/ScalarDash}{-1}
\FALabel(20.,15.8181)[b]{$\tilde u_\alpha$}
\FALabel(20.,4.18457)[t]{$\tilde d_\beta$}
\FAProp(10.,5.5)(10.,8.5)(0.,){/Straight}{1}
\FALabel(8.93,8.)[r]{$\tilde \chi_k^\pm$}
\FAProp(6.5,14.5)(13.5,14.5)(0.,){/Straight}{0}
\FALabel(10.,15.32)[b]{$\tilde g$}
\FAProp(6.5,14.5)(10.,8.5)(0.,){/ScalarDash}{1}
\FALabel(7.0,13.0)[tr]{$\tilde u$}
\FAProp(13.5,14.5)(10.,8.5)(0.,){/Straight}{-1}
\FALabel(12.,11.9169)[tl]{$d$}
\FAVert(6.5,14.5){0}
\FAVert(10.,5.5){0}
\FAVert(13.5,14.5){0}
\FAVert(10.,8.5){0}
\end{feynartspicture}
\begin{feynartspicture}(310,310)(1,1)
\FADiagram{}
\FAProp(0.,15.)(6.5,14.5)(0.,){/Straight}{1}
\FALabel(3.36888,15.8154)[b]{$u$}
\FAProp(0.,5.)(10.,5.5)(0.,){/Straight}{1}
\FALabel(5.0774,4.18193)[t]{$d$}
\FAProp(20.,15.)(10.,5.5)(0.,){/ScalarDash}{-1}
\FAProp(20.,5.)(13.5,14.5)(0.,){/ScalarDash}{-1}
\FALabel(20.,15.8181)[b]{$\tilde u_\alpha$}
\FALabel(20.,4.18457)[t]{$\tilde d_\beta$}
\FAProp(10.,5.5)(10.,8.5)(0.,){/Straight}{1}
\FALabel(8.93,8.)[r]{$\tilde \chi_k^\pm$}
\FAProp(6.5,14.5)(13.5,14.5)(0.,){/Cycles}{0}
\FALabel(10.,16.27)[b]{$g$}
\FAProp(6.5,14.5)(10.,8.5)(0.,){/Straight}{1}
\FALabel(7.0,13.0)[tr]{$u$}
\FAProp(13.5,14.5)(10.,8.5)(0.,){/ScalarDash}{-1}
\FALabel(12.,13.)[tl]{$\tilde d_\beta$}
\FAVert(6.5,14.5){0}
\FAVert(10.,5.5){0}
\FAVert(13.5,14.5){0}
\FAVert(10.,8.5){0}
\end{feynartspicture}
\begin{feynartspicture}(310,310)(1,1)
\FADiagram{}
\FAProp(0.,15.)(10.,14.5)(0.,){/Straight}{1}
\FALabel(5.0774,15.8181)[b]{$u$}
\FAProp(0.,5.)(6.5,5.5)(0.,){/Straight}{1}
\FALabel(3.36888,4.18457)[t]{$d$}
\FAProp(20.,15.)(13.5,5.5)(0.,){/ScalarDash}{-1}
\FAProp(20.,5.)(10.,14.5)(0.,){/ScalarDash}{-1}
\FALabel(20.,15.8181)[b]{$\tilde u_\alpha$}
\FALabel(20.,4.18457)[t]{$\tilde d_\beta$}
\FAProp(10.,14.5)(10.,11.5)(0.,){/Straight}{-1}
\FALabel(9.03,13.)[r]{$\tilde \chi_k^\pm$}
\FAProp(6.5,5.5)(13.5,5.5)(0.,){/Straight}{0}
\FALabel(10.,4.68)[t]{$\tilde g$}
\FAProp(6.5,5.5)(10.,11.5)(0.,){/ScalarDash}{1}
\FALabel(7.,6.5)[br]{$\tilde d$}
\FAProp(13.5,5.5)(10.,11.5)(0.,){/Straight}{-1}
\FALabel(12.,7.84012)[bl]{$u$}
\FAVert(10.,14.5){0}
\FAVert(6.5,5.5){0}
\FAVert(13.5,5.5){0}
\FAVert(10.,11.5){0}
\end{feynartspicture}
\begin{feynartspicture}(310,310)(1,1)
\FADiagram{}
\FAProp(0.,15.)(10.,14.5)(0.,){/Straight}{1}
\FALabel(5.0774,15.8181)[b]{$u$}
\FAProp(0.,5.)(6.5,5.5)(0.,){/Straight}{1}
\FALabel(3.36888,4.18457)[t]{$d$}
\FAProp(20.,15.)(13.5,5.5)(0.,){/ScalarDash}{-1}
\FAProp(20.,5.)(10.,14.5)(0.,){/ScalarDash}{-1}
\FALabel(20.,15.8181)[b]{$\tilde u_\alpha$}
\FALabel(20.,4.18457)[t]{$\tilde d_\beta$}
\FAProp(10.,14.5)(10.,11.5)(0.,){/Straight}{-1}
\FALabel(9.03,13.)[r]{$\tilde \chi_k^\pm$}
\FAProp(6.5,5.5)(13.5,5.5)(0.,){/Cycles}{0}
\FALabel(10.,4.43)[t]{$g$}
\FAProp(6.5,5.5)(10.,11.5)(0.,){/Straight}{1}
\FALabel(7.,7.)[br]{$d$}
\FAProp(13.5,5.5)(10.,11.5)(0.,){/ScalarDash}{-1}
\FALabel(12.,7.84012)[bl]{$\tilde u_\alpha$}
\FAVert(10.,14.5){0}
\FAVert(6.5,5.5){0}
\FAVert(13.5,5.5){0}
\FAVert(10.,11.5){0}
\end{feynartspicture}
\begin{feynartspicture}(310,310)(1,1)
\FADiagram{}
\FAProp(0.,15.)(6.,13.5)(0.,){/Straight}{1}
\FALabel(3.37593,15.2737)[b]{$u$}
\FAProp(0.,5.)(6.,6.5)(0.,){/Straight}{1}
\FALabel(3.37593,4.72628)[t]{$d$}
\FAProp(20.,15.)(12.,10.)(0.,){/ScalarDash}{-1}
\FAProp(20.,5.)(12.,10.)(0.,){/ScalarDash}{-1}
\FALabel(20.,15.8181)[b]{$\tilde u_\alpha$}
\FALabel(20.,4.18457)[t]{$\tilde d_\beta$}
\FAProp(6.,13.5)(6.,6.5)(0.,){/Straight}{-1}
\FALabel(4.93,10.)[r]{$\tilde \chi_k^\pm$}
\FAProp(6.,13.5)(12.,10.)(0.,){/ScalarDash}{1}
\FALabel(9.301,12.6089)[bl]{$\tilde d$}
\FAProp(6.,6.5)(12.,10.)(0.,){/ScalarDash}{1}
\FALabel(9.301,7.39114)[tl]{$\tilde u$}
\FAVert(6.,13.5){0}
\FAVert(6.,6.5){0}
\FAVert(12.,10.){0}
\end{feynartspicture}
\begin{feynartspicture}(310,310)(1,1)
\FADiagram{}
\FAProp(0.,15.)(6.5,13.5)(0.,){/Straight}{1}
\FALabel(3.59853,15.2803)[b]{$u$}
\FAProp(0.,5.)(6.5,6.5)(0.,){/Straight}{1}
\FALabel(3.59853,4.71969)[t]{$d$}
\FAProp(20.,15.)(13.5,6.5)(0.,){/ScalarDash}{-1}
\FAProp(20.,5.)(13.5,13.5)(0.,){/ScalarDash}{-1}
\FALabel(20.,15.8181)[b]{$\tilde u_\alpha$}
\FALabel(20.,4.18457)[t]{$\tilde d_\beta$}
\FAProp(6.5,13.5)(6.5,6.5)(0.,){/Cycles}{0}
\FALabel(5.43,10.)[r]{$g$}
\FAProp(6.5,13.5)(13.5,13.5)(0.,){/Straight}{1}
\FALabel(10.,14.57)[b]{$u$}
\FAProp(6.5,6.5)(13.5,6.5)(0.,){/Straight}{1}
\FALabel(10.,5.43)[t]{$d$}
\FAProp(13.5,6.5)(13.5,13.5)(0.,){/Straight}{1}
\FALabel(12.43,10.)[r]{$\tilde \chi_k^\pm$}
\FAVert(6.5,13.5){0}
\FAVert(6.5,6.5){0}
\FAVert(13.5,6.5){0}
\FAVert(13.5,13.5){0}
\end{feynartspicture}
\begin{feynartspicture}(310,310)(1,1)
\FADiagram{}
\FAProp(0.,15.)(6.5,13.5)(0.,){/Straight}{1}
\FALabel(3.59853,15.2803)[b]{$u$}
\FAProp(0.,5.)(6.5,6.5)(0.,){/Straight}{1}
\FALabel(3.59853,4.71969)[t]{$d$}
\FAProp(20.,15.)(13.5,6.5)(0.,){/ScalarDash}{-1}
\FAProp(20.,5.)(13.5,13.5)(0.,){/ScalarDash}{-1}
\FALabel(20.,15.8181)[b]{$\tilde u_\alpha$}
\FALabel(20.,4.18457)[t]{$\tilde d_\beta$}
\FAProp(6.5,13.5)(6.5,6.5)(0.,){/Straight}{-1}
\FALabel(5.43,10.)[r]{$\tilde \chi_k^\pm$}
\FAProp(6.5,13.5)(13.5,13.5)(0.,){/ScalarDash}{1}
\FALabel(10.,14.57)[b]{$\tilde d_\beta$}
\FAProp(6.5,6.5)(13.5,6.5)(0.,){/ScalarDash}{1}
\FALabel(10.,5.43)[t]{$\tilde u_\alpha$}
\FAProp(13.5,6.5)(13.5,13.5)(0.,){/Cycles}{0}
\FALabel(11.73,10.)[r]{$g$}
\FAVert(6.5,13.5){0}
\FAVert(6.5,6.5){0}
\FAVert(13.5,6.5){0}
\FAVert(13.5,13.5){0}
\end{feynartspicture}
\begin{feynartspicture}(310,310)(1,1)
\FADiagram{}
\FAProp(0.,15.)(13.5,13.)(0.,){/Straight}{1}
\FALabel(3.41023,15.5279)[b]{$u$}
\FAProp(0.,5.)(6.5,6.)(0.,){/Straight}{1}
\FALabel(3.48569,4.44802)[t]{$d$}
\FAProp(20.,15.)(6.5,13.)(0.,){/ScalarDash}{-1}
\FAProp(20.,5.)(13.5,6.)(0.,){/ScalarDash}{-1}
\FALabel(16.5898,15.5279)[b]{$\tilde u_\alpha$}
\FALabel(16.5143,4.44802)[t]{$\tilde d_\beta$}
\FAProp(13.5,13.)(6.5,13.)(0.,){/Straight}{-1}
\FALabel(10.,11.93)[t]{$\tilde \chi_k^\pm$}
\FAProp(13.5,13.)(13.5,6.)(0.,){/ScalarDash}{1}
\FALabel(14.57,9.5)[l]{$\tilde d_\beta$}
\FAProp(6.5,6.)(6.5,13.)(0.,){/Straight}{1}
\FALabel(5.43,9.5)[r]{$d$}
\FAProp(6.5,6.)(13.5,6.)(0.,){/Cycles}{0}
\FALabel(10.,4.93)[t]{$g$}
\FAVert(13.5,13.){0}
\FAVert(6.5,6.){0}
\FAVert(6.5,13.){0}
\FAVert(13.5,6.){0}
\end{feynartspicture}
\begin{feynartspicture}(310,310)(1,1)
\FADiagram{}
\FAProp(0.,15.)(13.5,13.)(0.,){/Straight}{1}
\FALabel(3.41023,15.5279)[b]{$u$}
\FAProp(0.,5.)(6.5,6.)(0.,){/Straight}{1}
\FALabel(3.48569,4.44802)[t]{$d$}
\FAProp(20.,15.)(6.5,13.)(0.,){/ScalarDash}{-1}
\FAProp(20.,5.)(13.5,6.)(0.,){/ScalarDash}{-1}
\FALabel(16.5898,15.5279)[b]{$\tilde u_\alpha$}
\FALabel(16.5143,4.44802)[t]{$\tilde d_\beta$}
\FAProp(13.5,13.)(6.5,13.)(0.,){/Cycles}{0}
\FALabel(10.,11.23)[t]{$g$}
\FAProp(13.5,13.)(13.5,6.)(0.,){/Straight}{1}
\FALabel(14.57,9.5)[l]{$u$}
\FAProp(6.5,6.)(6.5,13.)(0.,){/ScalarDash}{1}
\FALabel(5.43,9.5)[r]{$\tilde u_\alpha$}
\FAProp(6.5,6.)(13.5,6.)(0.,){/Straight}{1}
\FALabel(10.,4.93)[t]{$\tilde \chi_k^\pm$}
\FAVert(13.5,13.){0}
\FAVert(6.5,6.){0}
\FAVert(6.5,13.){0}
\FAVert(13.5,6.){0}
\end{feynartspicture}
\\
\begin{feynartspicture}(310,310)(1,1)
\FADiagram{}
\FAProp(0.,15.)(10.,14.)(0.,){/Straight}{1}
\FALabel(4.84577,13.4377)[t]{$q$}
\FAProp(0.,5.)(10.,6.)(0.,){/Straight}{1}
\FALabel(5.15423,4.43769)[t]{$q^{\prime}$}
\FAProp(20.,15.)(10.,14.)(0.,){/ScalarDash}{-1}
\FALabel(16.5,15.5623)[b]{$\tilde q_\alpha$}
\FAProp(20.,5.)(10.,6.)(0.,){/ScalarDash}{-1}
\FALabel(16.5,4.44802)[t]{$\tilde q^{\prime}_{\beta}$}
\FAProp(10.,14.)(10.,6.)(0.,){/Straight}{0}
\FALabel(9.18,10.)[r]{$\tilde \chi_k^0$}
\FAVert(10.,14.){0}
\FAVert(10.,6.){1}
\end{feynartspicture}
\begin{feynartspicture}(310,310)(1,1)
\FADiagram{}
\FAProp(0.,15.)(10.,14.)(0.,){/Straight}{1}
\FALabel(4.84577,13.4377)[t]{$q$}
\FAProp(0.,5.)(10.,6.)(0.,){/Straight}{1}
\FALabel(5.15423,4.43769)[t]{$q^{\prime}$}
\FAProp(20.,15.)(10.,14.)(0.,){/ScalarDash}{-1}
\FALabel(16.5,15.5623)[b]{$\tilde q_\alpha$}
\FAProp(20.,5.)(10.,6.)(0.,){/ScalarDash}{-1}
\FALabel(16.5,4.44802)[t]{$\tilde q^{\prime}_{\beta}$}
\FAProp(10.,14.)(10.,6.)(0.,){/Straight}{0}
\FALabel(9.18,10.)[r]{$\tilde \chi_k^0$}
\FAVert(10.,6.){0}
\FAVert(10.,14.){1}
\end{feynartspicture}
\begin{feynartspicture}(310,310)(1,1)
\FADiagram{}
\FAProp(0.,15.)(10.,14.)(0.,){/Straight}{1}
\FALabel(4.84577,13.4377)[t]{$u$}
\FAProp(0.,5.)(10.,6.)(0.,){/Straight}{1}
\FALabel(5.15423,4.43769)[t]{$d$}
\FAProp(20.,15.)(10.,14.)(0.,){/ScalarDash}{-1}
\FALabel(16.5,15.5623)[b]{$\tilde u_{\alpha}$}
\FAProp(20.,5.)(10.,6.)(0.,){/ScalarDash}{-1}
\FALabel(16.5,4.44802)[t]{$\tilde d_{\beta}$}
\FAProp(10.,14.)(10.,6.)(0.,){/Straight}{0}
\FALabel(9.18,10.)[r]{$\tilde \chi_k^{\pm}$}
\FAVert(10.,14.){0}
\FAVert(10.,6.){1}
\end{feynartspicture}
\begin{feynartspicture}(310,310)(1,1)
\FADiagram{}
\FAProp(0.,15.)(10.,14.)(0.,){/Straight}{1}
\FALabel(4.84577,13.4377)[t]{$u$}
\FAProp(0.,5.)(10.,6.)(0.,){/Straight}{1}
\FALabel(5.15423,4.43769)[t]{$d$}
\FAProp(20.,15.)(10.,14.)(0.,){/ScalarDash}{-1}
\FALabel(16.5,15.5623)[b]{$\tilde u_\alpha$}
\FAProp(20.,5.)(10.,6.)(0.,){/ScalarDash}{-1}
\FALabel(16.5,4.44802)[t]{$\tilde d_{\beta}$}
\FAProp(10.,14.)(10.,6.)(0.,){/Straight}{0}
\FALabel(9.18,10.)[r]{$\tilde \chi_k^{\pm}$}
\FAVert(10.,6.){0}
\FAVert(10.,14.){1}
\end{feynartspicture}

\vspace*{5ex}
\caption{Virtual corrections (II): QCD one-loop insertions into EW Born diagrams. For $q=q^\prime$ crossed
  diagrams have to be taken into account. The diagrams containing the
  four squark interaction vertex have to be evaluated at
  $\mathcal{O}(\alphas\alpha)$. The chargino-mediated diagrams 
   only contribute for $u=\{u,c\},\,d=\{d,s\}$. 
  The renormalization constants appearing in the
  counterterm diagrams (last line) have to be evaluated at
  $\mathcal{O}(\alphas)$.
\vspace*{5ex}
\label{fig_QCDinsEW}}
}
\clearpage

\FIGURE{
\small\unitlength=0.24bp%
\begin{feynartspicture}(310,310)(1,1)
\FADiagram{}
\FAProp(0.,15.)(10.,14.5)(0.,){/Straight}{1}
\FALabel(5.0774,15.8181)[b]{$q$}
\FAProp(0.,5.)(6.5,5.5)(0.,){/Straight}{1}
\FALabel(3.36888,4.18457)[t]{$q^{\prime}$}
\FAProp(20.,15.)(10.,14.5)(0.,){/ScalarDash}{-1}
\FALabel(14.9226,15.8181)[b]{$\tilde q_{\alpha}$}
\FAProp(20.,5.)(13.5,5.5)(0.,){/ScalarDash}{-1}
\FALabel(16.6311,4.18457)[t]{$\tilde q^{\prime}_{\beta}$}
\FAProp(10.,14.5)(10.,11.)(0.,){/Straight}{0}
\FALabel(10.82,12.75)[l]{$\tilde g$}
\FAProp(6.5,5.5)(13.5,5.5)(0.,){/Straight}{0}
\FALabel(10.,4.68)[t]{$\tilde g$}
\FAProp(6.5,5.5)(10.,11.)(0.,){/ScalarDash}{1}
\FALabel(7.42232,8.60216)[br]{$\tilde q^{\prime}$}
\FAProp(13.5,5.5)(10.,11.)(0.,){/Straight}{-1}
\FALabel(12.5777,8.60216)[bl]{$q^{\prime}$}
\FAVert(10.,14.5){0}
\FAVert(6.5,5.5){0}
\FAVert(13.5,5.5){0}
\FAVert(10.,11.){0}
\end{feynartspicture}
\begin{feynartspicture}(310,310)(1,1)
\FADiagram{}
\FAProp(0.,15.)(10.,14.5)(0.,){/Straight}{1}
\FALabel(5.0774,15.8181)[b]{$q$}
\FAProp(0.,5.)(6.5,5.5)(0.,){/Straight}{1}
\FALabel(3.36888,4.18457)[t]{$q^{\prime}$}
\FAProp(20.,15.)(10.,14.5)(0.,){/ScalarDash}{-1}
\FALabel(14.9226,15.8181)[b]{$\tilde q_{\alpha}$}
\FAProp(20.,5.)(13.5,5.5)(0.,){/ScalarDash}{-1}
\FALabel(16.6311,4.18457)[t]{$\tilde q^{\prime}_{\beta}$}
\FAProp(10.,14.5)(10.,11.)(0.,){/Straight}{0}
\FALabel(10.82,12.75)[l]{$\tilde g$}
\FAProp(6.5,5.5)(13.5,5.5)(0.,){/Straight}{1}
\FALabel(10.,4.43)[t]{$q^{\prime}$}
\FAProp(6.5,5.5)(10.,11.)(0.,){/Cycles}{0}
\FALabel(6.83176,8.97797)[br]{$g$}
\FAProp(13.5,5.5)(10.,11.)(0.,){/Straight}{0}
\FALabel(12.3668,8.46794)[bl]{$\tilde g$}
\FAVert(10.,14.5){0}
\FAVert(6.5,5.5){0}
\FAVert(13.5,5.5){0}
\FAVert(10.,11.){0}
\end{feynartspicture}
\begin{feynartspicture}(310,310)(1,1)
\FADiagram{}
\FAProp(0.,15.)(10.,14.5)(0.,){/Straight}{1}
\FALabel(5.0774,15.8181)[b]{$q$}
\FAProp(0.,5.)(6.5,5.5)(0.,){/Straight}{1}
\FALabel(3.36888,4.18457)[t]{$q^{\prime}$}
\FAProp(20.,15.)(10.,14.5)(0.,){/ScalarDash}{-1}
\FALabel(14.9226,15.8181)[b]{$\tilde q_{\alpha}$}
\FAProp(20.,5.)(13.5,5.5)(0.,){/ScalarDash}{-1}
\FALabel(16.6311,4.18457)[t]{$\tilde q^{\prime}_{\beta}$}
\FAProp(10.,14.5)(10.,11.)(0.,){/Straight}{0}
\FALabel(10.82,12.75)[l]{$\tilde g$}
\FAProp(6.5,5.5)(13.5,5.5)(0.,){/ScalarDash}{1}
\FALabel(10.,4.43)[t]{$\tilde q^{\prime}_{\beta}$}
\FAProp(6.5,5.5)(10.,11.)(0.,){/Straight}{0}
\FALabel(7.63324,8.46794)[br]{$\tilde g$}
\FAProp(13.5,5.5)(10.,11.)(0.,){/Cycles}{0}
\FALabel(12.5777,8.60216)[bl]{$g$}
\FAVert(10.,14.5){0}
\FAVert(6.5,5.5){0}
\FAVert(13.5,5.5){0}
\FAVert(10.,11.){0}
\end{feynartspicture}
\begin{feynartspicture}(310,310)(1,1)
\FADiagram{}
\FAProp(0.,15.)(10.,14.5)(0.,){/Straight}{1}
\FALabel(5.0774,15.8181)[b]{$q$}
\FAProp(0.,5.)(6.5,5.5)(0.,){/Straight}{1}
\FALabel(3.36888,4.18457)[t]{$q^{\prime}$}
\FAProp(20.,15.)(10.,14.5)(0.,){/ScalarDash}{-1}
\FALabel(14.9226,15.8181)[b]{$\tilde q_{\alpha}$}
\FAProp(20.,5.)(13.5,5.5)(0.,){/ScalarDash}{-1}
\FALabel(16.6311,4.18457)[t]{$\tilde q^{\prime}_{\beta}$}
\FAProp(10.,14.5)(10.,11.)(0.,){/Straight}{0}
\FALabel(10.82,12.75)[l]{$\tilde g$}
\FAProp(6.5,5.5)(13.5,5.5)(0.,){/Cycles}{0}
\FALabel(10.,4.43)[t]{$g$}
\FAProp(6.5,5.5)(10.,11.)(0.,){/Straight}{1}
\FALabel(7.42232,8.60216)[br]{$q^{\prime}$}
\FAProp(13.5,5.5)(10.,11.)(0.,){/ScalarDash}{-1}
\FALabel(12.5777,8.60216)[bl]{$\tilde q^{\prime}_{\beta}$}
\FAVert(10.,14.5){0}
\FAVert(6.5,5.5){0}
\FAVert(13.5,5.5){0}
\FAVert(10.,11.){0}
\end{feynartspicture}
\begin{feynartspicture}(310,310)(1,1)
\FADiagram{}
\FAProp(0.,15.)(6.5,14.5)(0.,){/Straight}{1}
\FALabel(3.36888,15.8154)[b]{$q$}
\FAProp(0.,5.)(10.,5.5)(0.,){/Straight}{1}
\FALabel(5.0774,4.18193)[t]{$q^{\prime}$}
\FAProp(20.,15.)(13.5,14.5)(0.,){/ScalarDash}{-1}
\FALabel(16.6311,15.8154)[b]{$\tilde q_{\alpha}$}
\FAProp(20.,5.)(10.,5.5)(0.,){/ScalarDash}{-1}
\FALabel(14.9226,4.18193)[t]{$\tilde q^{\prime}_{\beta}$}
\FAProp(10.,5.5)(10.,8.5)(0.,){/Straight}{0}
\FALabel(9.18,7.)[r]{$\tilde g$}
\FAProp(6.5,14.5)(13.5,14.5)(0.,){/Straight}{0}
\FALabel(10.,15.32)[b]{$\tilde g$}
\FAProp(6.5,14.5)(10.,8.5)(0.,){/ScalarDash}{1}
\FALabel(7.39114,11.199)[tr]{$\tilde q$}
\FAProp(13.5,14.5)(10.,8.5)(0.,){/Straight}{-1}
\FALabel(12.6089,11.199)[tl]{$q$}
\FAVert(6.5,14.5){0}
\FAVert(10.,5.5){0}
\FAVert(13.5,14.5){0}
\FAVert(10.,8.5){0}
\end{feynartspicture}\\[-1ex]
\begin{feynartspicture}(310,310)(1,1)
\FADiagram{}
\FAProp(0.,15.)(6.5,14.5)(0.,){/Straight}{1}
\FALabel(3.36888,15.8154)[b]{$q$}
\FAProp(0.,5.)(10.,5.5)(0.,){/Straight}{1}
\FALabel(5.0774,4.18193)[t]{$q^{\prime}$}
\FAProp(20.,15.)(13.5,14.5)(0.,){/ScalarDash}{-1}
\FALabel(16.6311,15.8154)[b]{$\tilde q_{\alpha}$}
\FAProp(20.,5.)(10.,5.5)(0.,){/ScalarDash}{-1}
\FALabel(14.9226,4.18193)[t]{$\tilde q^{\prime}_{\beta}$}
\FAProp(10.,5.5)(10.,8.5)(0.,){/Straight}{0}
\FALabel(9.18,7.)[r]{$\tilde g$}
\FAProp(6.5,14.5)(13.5,14.5)(0.,){/Straight}{1}
\FALabel(10.,15.57)[b]{$q$}
\FAProp(6.5,14.5)(10.,8.5)(0.,){/Cycles}{0}
\FALabel(7.39114,11.199)[tr]{$g$}
\FAProp(13.5,14.5)(10.,8.5)(0.,){/Straight}{0}
\FALabel(12.3929,11.325)[tl]{$\tilde g$}
\FAVert(6.5,14.5){0}
\FAVert(10.,5.5){0}
\FAVert(13.5,14.5){0}
\FAVert(10.,8.5){0}
\end{feynartspicture}
\begin{feynartspicture}(310,310)(1,1)
\FADiagram{}
\FAProp(0.,15.)(6.5,14.5)(0.,){/Straight}{1}
\FALabel(3.36888,15.8154)[b]{$q$}
\FAProp(0.,5.)(10.,5.5)(0.,){/Straight}{1}
\FALabel(5.0774,4.18193)[t]{$q^{\prime}$}
\FAProp(20.,15.)(13.5,14.5)(0.,){/ScalarDash}{-1}
\FALabel(16.6311,15.8154)[b]{$\tilde q_{\alpha}$}
\FAProp(20.,5.)(10.,5.5)(0.,){/ScalarDash}{-1}
\FALabel(14.9226,4.18193)[t]{$\tilde q^{\prime}_{\beta}$}
\FAProp(10.,5.5)(10.,8.5)(0.,){/Straight}{0}
\FALabel(9.18,7.)[r]{$\tilde g$}
\FAProp(6.5,14.5)(13.5,14.5)(0.,){/ScalarDash}{1}
\FALabel(10.,15.57)[b]{$\tilde q_{\alpha}$}
\FAProp(6.5,14.5)(10.,8.5)(0.,){/Straight}{0}
\FALabel(7.60709,11.325)[tr]{$\tilde g$}
\FAProp(13.5,14.5)(10.,8.5)(0.,){/Cycles}{0}
\FALabel(13.2135,10.8463)[tl]{$g$}
\FAVert(6.5,14.5){0}
\FAVert(10.,5.5){0}
\FAVert(13.5,14.5){0}
\FAVert(10.,8.5){0}
\end{feynartspicture}
\begin{feynartspicture}(310,310)(1,1)
\FADiagram{}
\FAProp(0.,15.)(6.5,14.5)(0.,){/Straight}{1}
\FALabel(3.36888,15.8154)[b]{$q$}
\FAProp(0.,5.)(10.,5.5)(0.,){/Straight}{1}
\FALabel(5.0774,4.18193)[t]{$q^{\prime}$}
\FAProp(20.,15.)(13.5,14.5)(0.,){/ScalarDash}{-1}
\FALabel(16.6311,15.8154)[b]{$\tilde q_{\alpha}$}
\FAProp(20.,5.)(10.,5.5)(0.,){/ScalarDash}{-1}
\FALabel(14.9226,4.18193)[t]{$\tilde q^{\prime}_{\beta}$}
\FAProp(10.,5.5)(10.,8.5)(0.,){/Straight}{0}
\FALabel(9.18,7.)[r]{$\tilde g$}
\FAProp(6.5,14.5)(13.5,14.5)(0.,){/Cycles}{0}
\FALabel(10.,16.27)[b]{$g$}
\FAProp(6.5,14.5)(10.,8.5)(0.,){/Straight}{1}
\FALabel(7.39114,11.199)[tr]{$q$}
\FAProp(13.5,14.5)(10.,8.5)(0.,){/ScalarDash}{-1}
\FALabel(12.6089,11.199)[tl]{$\tilde q_{\alpha}$}
\FAVert(6.5,14.5){0}
\FAVert(10.,5.5){0}
\FAVert(13.5,14.5){0}
\FAVert(10.,8.5){0}
\end{feynartspicture}
\begin{feynartspicture}(310,310)(1,1)
\FADiagram{}
\FAProp(0.,15.)(6.,13.5)(0.,){/Straight}{1}
\FALabel(3.37593,15.2737)[b]{$q$}
\FAProp(0.,5.)(6.,6.5)(0.,){/Straight}{1}
\FALabel(3.37593,4.72628)[t]{$q^{\prime}$}
\FAProp(20.,15.)(12.,10.)(0.,){/ScalarDash}{-1}
\FAProp(20.,5.)(12.,10.)(0.,){/ScalarDash}{-1}
\FALabel(20.,15.8181)[b]{$\tilde q_\alpha$}
\FALabel(20.,4.18457)[t]{$\tilde q^{\prime}_\beta$}
\FAProp(6.,13.5)(6.,6.5)(0.,){/Straight}{0}
\FALabel(5.18,10.)[r]{$\tilde g$}
\FAProp(6.,13.5)(12.,10.)(0.,){/ScalarDash}{1}
\FALabel(9.301,12.6089)[bl]{$\tilde q$}
\FAProp(6.,6.5)(12.,10.)(0.,){/ScalarDash}{1}
\FALabel(9.301,7.39114)[tl]{$\tilde q^{\prime}$}
\FAVert(6.,13.5){0}
\FAVert(6.,6.5){0}
\FAVert(12.,10.){0}
\end{feynartspicture}
\begin{feynartspicture}(310,310)(1,1)
\FADiagram{}
\FAProp(0.,15.)(6.5,13.5)(0.,){/Straight}{1}
\FALabel(3.59853,15.2803)[b]{$q$}
\FAProp(0.,5.)(6.5,6.5)(0.,){/Straight}{1}
\FALabel(3.59853,4.71969)[t]{$q^{\prime}$}
\FAProp(20.,15.)(13.5,13.5)(0.,){/ScalarDash}{-1}
\FALabel(16.4015,15.2803)[b]{$\tilde q_{\alpha}$}
\FAProp(20.,5.)(13.5,6.5)(0.,){/ScalarDash}{-1}
\FALabel(16.4015,4.71969)[t]{$\tilde q^{\prime}_{\beta}$}
\FAProp(6.5,13.5)(6.5,6.5)(0.,){/Cycles}{0}
\FALabel(5.43,10.)[r]{$g$}
\FAProp(6.5,13.5)(13.5,13.5)(0.,){/Straight}{1}
\FALabel(10.,14.57)[b]{$q$}
\FAProp(6.5,6.5)(13.5,6.5)(0.,){/Straight}{1}
\FALabel(10.,5.43)[t]{$q^{\prime}$}
\FAProp(13.5,13.5)(13.5,6.5)(0.,){/Straight}{0}
\FALabel(14.32,10.)[l]{$\tilde g$}
\FAVert(6.5,13.5){0}
\FAVert(6.5,6.5){0}
\FAVert(13.5,13.5){0}
\FAVert(13.5,6.5){0}
\end{feynartspicture}\\[-1ex]
\begin{feynartspicture}(310,310)(1,1)
\FADiagram{}
\FAProp(0.,15.)(6.5,13.5)(0.,){/Straight}{1}
\FALabel(3.59853,15.2803)[b]{$q$}
\FAProp(0.,5.)(6.5,6.5)(0.,){/Straight}{1}
\FALabel(3.59853,4.71969)[t]{$q^{\prime}$}
\FAProp(20.,15.)(13.5,13.5)(0.,){/ScalarDash}{-1}
\FALabel(16.4015,15.2803)[b]{$\tilde q_{\alpha}$}
\FAProp(20.,5.)(13.5,6.5)(0.,){/ScalarDash}{-1}
\FALabel(16.4015,4.71969)[t]{$\tilde q^{\prime}_{\beta}$}
\FAProp(6.5,13.5)(6.5,6.5)(0.,){/Straight}{0}
\FALabel(5.68,10.)[r]{$\tilde g$}
\FAProp(6.5,13.5)(13.5,13.5)(0.,){/ScalarDash}{1}
\FALabel(10.,14.57)[b]{$\tilde q_{\alpha}$}
\FAProp(6.5,6.5)(13.5,6.5)(0.,){/ScalarDash}{1}
\FALabel(10.,5.43)[t]{$\tilde q^{\prime}_{\beta}$}
\FAProp(13.5,13.5)(13.5,6.5)(0.,){/Cycles}{0}
\FALabel(15.27,10.)[l]{$g$}
\FAVert(6.5,13.5){0}
\FAVert(6.5,6.5){0}
\FAVert(13.5,13.5){0}
\FAVert(13.5,6.5){0}
\end{feynartspicture}
\begin{feynartspicture}(310,310)(1,1)
\FADiagram{}
\FAProp(0.,15.)(13.5,13.)(0.,){/Straight}{1}
\FALabel(3.41023,15.5279)[b]{$q$}
\FAProp(0.,5.)(6.5,6.)(0.,){/Straight}{1}
\FALabel(3.48569,4.44802)[t]{$q^{\prime}$}
\FAProp(20.,15.)(6.5,13.)(0.,){/ScalarDash}{-1}
\FALabel(16.5898,15.5279)[b]{$\tilde q_{\alpha}$}
\FAProp(20.,5.)(13.5,6.)(0.,){/ScalarDash}{-1}
\FALabel(16.5143,4.44802)[t]{$\tilde q^{\prime}_{\beta}$}
\FAProp(13.5,13.)(6.5,13.)(0.,){/Straight}{1}
\FALabel(10.,11.93)[t]{$q$}
\FAProp(13.5,13.)(13.5,6.)(0.,){/Cycles}{0}
\FALabel(15.27,9.5)[l]{$g$}
\FAProp(6.5,6.)(6.5,13.)(0.,){/Straight}{0}
\FALabel(5.68,9.5)[r]{$\tilde g$}
\FAProp(6.5,6.)(13.5,6.)(0.,){/ScalarDash}{1}
\FALabel(10.,4.93)[t]{$\tilde q^{\prime}_{\beta}$}
\FAVert(13.5,13.){0}
\FAVert(6.5,6.){0}
\FAVert(6.5,13.){0}
\FAVert(13.5,6.){0}
\end{feynartspicture}
\begin{feynartspicture}(310,310)(1,1)
\FADiagram{}
\FAProp(0.,15.)(13.5,13.)(0.,){/Straight}{1}
\FALabel(3.41023,15.5279)[b]{$q$}
\FAProp(0.,5.)(6.5,6.)(0.,){/Straight}{1}
\FALabel(3.48569,4.44802)[t]{$q^{\prime}$}
\FAProp(20.,15.)(6.5,13.)(0.,){/ScalarDash}{-1}
\FALabel(16.5898,15.5279)[b]{$\tilde q_{\alpha}$}
\FAProp(20.,5.)(13.5,6.)(0.,){/ScalarDash}{-1}
\FALabel(16.5143,4.44802)[t]{$\tilde q^{\prime}_{\beta}$}
\FAProp(13.5,13.)(6.5,13.)(0.,){/ScalarDash}{1}
\FALabel(10.,11.93)[t]{$\tilde q_{\alpha}$}
\FAProp(13.5,13.)(13.5,6.)(0.,){/Straight}{0}
\FALabel(14.32,9.5)[l]{$\tilde g$}
\FAProp(6.5,6.)(6.5,13.)(0.,){/Cycles}{0}
\FALabel(4.73,9.5)[r]{$g$}
\FAProp(6.5,6.)(13.5,6.)(0.,){/Straight}{1}
\FALabel(10.,4.93)[t]{$q^{\prime}$}
\FAVert(13.5,13.){0}
\FAVert(6.5,6.){0}
\FAVert(6.5,13.){0}
\FAVert(13.5,6.){0}
\end{feynartspicture}
\begin{feynartspicture}(310,310)(1,1)
\FADiagram{}
\FAProp(0.,15.)(10.,14.5)(0.,){/Straight}{1}
\FALabel(5.0774,15.8181)[b]{$q$}
\FAProp(0.,5.)(10.,5.5)(0.,){/Straight}{1}
\FALabel(5.0774,4.18193)[t]{$q^{\prime}$}
\FAProp(20.,15.)(10.,14.5)(0.,){/ScalarDash}{-1}
\FALabel(14.9226,15.8181)[b]{$\tilde q_{\alpha}$}
\FAProp(20.,5.)(10.,5.5)(0.,){/ScalarDash}{-1}
\FALabel(14.9226,4.18193)[t]{$\tilde q^{\prime}_{\beta}$}
\FAProp(10.,14.5)(10.,12.)(0.,){/Straight}{0}
\FALabel(10.82,13.25)[l]{$\tilde g$}
\FAProp(10.,5.5)(10.,8.)(0.,){/Straight}{0}
\FALabel(9.18,6.75)[r]{$\tilde g$}
\FAProp(10.,12.)(10.,8.)(1.,){/Straight}{-1}
\FALabel(6.93,10.)[r]{$Q_i$}
\FAProp(10.,12.)(10.,8.)(-1.,){/ScalarDash}{1}
\FALabel(13.07,10.)[l]{$\tilde Q_{i}$}
\FAVert(10.,14.5){0}
\FAVert(10.,5.5){0}
\FAVert(10.,12.){0}
\FAVert(10.,8.){0}
\end{feynartspicture}
\begin{feynartspicture}(310,310)(1,1)
\FADiagram{}
\FAProp(0.,15.)(10.,14.5)(0.,){/Straight}{1}
\FALabel(5.0774,15.8181)[b]{$q$}
\FAProp(0.,5.)(10.,5.5)(0.,){/Straight}{1}
\FALabel(5.0774,4.18193)[t]{$q^{\prime}$}
\FAProp(20.,15.)(10.,14.5)(0.,){/ScalarDash}{-1}
\FALabel(14.9226,15.8181)[b]{$\tilde q_{\alpha}$}
\FAProp(20.,5.)(10.,5.5)(0.,){/ScalarDash}{-1}
\FALabel(14.9226,4.18193)[t]{$\tilde q^{\prime}_{\beta}$}
\FAProp(10.,14.5)(10.,12.)(0.,){/Straight}{0}
\FALabel(10.82,13.25)[l]{$\tilde g$}
\FAProp(10.,5.5)(10.,8.)(0.,){/Straight}{0}
\FALabel(9.18,6.75)[r]{$\tilde g$}
\FAProp(10.,12.)(10.,8.)(1.,){/Straight}{1}
\FALabel(6.93,10.)[r]{$Q_i$}
\FAProp(10.,12.)(10.,8.)(-1.,){/ScalarDash}{-1}
\FALabel(13.07,10.)[l]{$\tilde Q_{i}$}
\FAVert(10.,14.5){0}
\FAVert(10.,5.5){0}
\FAVert(10.,12.){0}
\FAVert(10.,8.){0}
\end{feynartspicture}\\[-2ex]
\begin{feynartspicture}(310,310)(1,1)
\FADiagram{}
\FAProp(0.,15.)(10.,14.5)(0.,){/Straight}{1}
\FALabel(5.0774,15.8181)[b]{$q$}
\FAProp(0.,5.)(10.,5.5)(0.,){/Straight}{1}
\FALabel(5.0774,4.18193)[t]{$q^{\prime}$}
\FAProp(20.,15.)(10.,14.5)(0.,){/ScalarDash}{-1}
\FALabel(14.9226,15.8181)[b]{$\tilde q_{\alpha}$}
\FAProp(20.,5.)(10.,5.5)(0.,){/ScalarDash}{-1}
\FALabel(14.9226,4.18193)[t]{$\tilde q^{\prime}_{\beta}$}
\FAProp(10.,14.5)(10.,12.)(0.,){/Straight}{0}
\FALabel(10.82,13.25)[l]{$\tilde g$}
\FAProp(10.,5.5)(10.,8.)(0.,){/Straight}{0}
\FALabel(9.18,6.75)[r]{$\tilde g$}
\FAProp(10.,12.)(10.,8.)(1.,){/Straight}{0}
\FALabel(7.18,10.)[r]{$\tilde g$}
\FAProp(10.,12.)(10.,8.)(-1.,){/Cycles}{0}
\FALabel(13.07,10.)[l]{$g$}
\FAVert(10.,14.5){0}
\FAVert(10.,5.5){0}
\FAVert(10.,12.){0}
\FAVert(10.,8.){0}
\end{feynartspicture}
\begin{feynartspicture}(310,310)(1,1)
\FADiagram{}
\FAProp(0.,15.)(10.,14.)(0.,){/Straight}{1}
\FALabel(4.84577,13.4377)[t]{$q$}
\FAProp(0.,5.)(10.,6.)(0.,){/Straight}{1}
\FALabel(5.15423,4.43769)[t]{$q^{\prime}$}
\FAProp(20.,15.)(10.,14.)(0.,){/ScalarDash}{-1}
\FALabel(14.8458,15.5623)[b]{$\tilde q_{\alpha}$}
\FAProp(20.,5.)(10.,6.)(0.,){/ScalarDash}{-1}
\FALabel(15.1542,6.56231)[b]{$\tilde q^{\prime}_{\beta}$}
\FAProp(10.,14.)(10.,6.)(0.,){/Straight}{0}
\FALabel(9.18,10.)[r]{$\tilde g$}
\FAVert(10.,14.){0}
\FAVert(10.,6.){1}
\end{feynartspicture}
\begin{feynartspicture}(310,310)(1,1)
\FADiagram{}
\FAProp(0.,15.)(10.,14.)(0.,){/Straight}{1}
\FALabel(4.84577,13.4377)[t]{$q$}
\FAProp(0.,5.)(10.,6.)(0.,){/Straight}{1}
\FALabel(5.15423,4.43769)[t]{$q^{\prime}$}
\FAProp(20.,15.)(10.,14.)(0.,){/ScalarDash}{-1}
\FALabel(14.8458,15.5623)[b]{$\tilde q_{\alpha}$}
\FAProp(20.,5.)(10.,6.)(0.,){/ScalarDash}{-1}
\FALabel(15.1542,6.56231)[b]{$\tilde q^{\prime}_{\beta}$}
\FAProp(10.,14.)(10.,6.)(0.,){/Straight}{0}
\FALabel(9.18,10.)[r]{$\tilde g$}
\FAVert(10.,6.){0}
\FAVert(10.,14.){1}
\end{feynartspicture}
\begin{feynartspicture}(310,310)(1,1)
\FADiagram{}
\FAProp(0.,15.)(10.,14.)(0.,){/Straight}{1}
\FALabel(4.84577,13.4377)[t]{$q$}
\FAProp(0.,5.)(10.,6.)(0.,){/Straight}{1}
\FALabel(5.15423,4.43769)[t]{$q^{\prime}$}
\FAProp(20.,15.)(10.,14.)(0.,){/ScalarDash}{-1}
\FALabel(14.8458,15.5623)[b]{$\tilde q_{\alpha}$}
\FAProp(20.,5.)(10.,6.)(0.,){/ScalarDash}{-1}
\FALabel(15.1542,6.56231)[b]{$\tilde q^{\prime}_{\beta}$}
\FAProp(10.,10.)(10.,14.)(0.,){/Straight}{0}
\FALabel(10.82,12.)[l]{$\tilde g$}
\FAProp(10.,10.)(10.,6.)(0.,){/Straight}{0}
\FALabel(9.18,8.)[r]{$\tilde g$}
\FAVert(10.,14.){0}
\FAVert(10.,6.){0}
\FAVert(10.,10.){1}
\end{feynartspicture}
\vspace*{-1ex}
\caption{Virtual corrections (III): QCD one-loop insertions into QCD Born diagrams. For $q=q^\prime$ crossed
  diagrams have to be taken into account. Here, $Q_i$ can be any of the six quark flavors. The diagram containing the
  four squark vertex has to be evaluated at
  $\mathcal{O}(\alpha^2_s)$. The renormalization constants appearing in
  the counter term diagrams (last three diagrams) have to be evaluated at
  $\mathcal{O}(\alphas)$, i.e. the strong sector has to be
  renormalized.
\vspace*{-3ex}
\label{fig_QCDinsQCD}}
}

\FIGURE{
\small
\unitlength=0.24bp%
\begin{feynartspicture}(310,310)(1,1)
\FADiagram{}
\FAProp(0.,15.)(10.,14.5)(0.,){/Straight}{1}
\FALabel(5.0774,15.8181)[b]{$q$}
\FAProp(0.,5.)(10.,5.5)(0.,){/Straight}{1}
\FALabel(5.0774,4.18193)[t]{$q^\prime$}
\FAProp(20.,17.)(10.,14.5)(0.,){/ScalarDash}{-1}
\FALabel(14.6241,16.7737)[b]{$\tilde q_\alpha$}
\FAProp(20.,10.)(10.,10.)(0.,){/ScalarDash}{-1}
\FALabel(15.,11.07)[b]{$\tilde q^\prime_\beta$}
\FAProp(20.,3.)(10.,5.5)(0.,){/Sine}{0}
\FALabel(15.3759,5.27372)[b]{$\gamma$}
\FAProp(10.,14.5)(10.,10.)(0.,){/Straight}{0}
\FALabel(9.18,12.25)[r]{$\tilde g$}
\FAProp(10.,5.5)(10.,10.)(0.,){/Straight}{1}
\FALabel(8.93,7.75)[r]{$q^\prime$}
\FAVert(10.,14.5){0}
\FAVert(10.,5.5){0}
\FAVert(10.,10.){0}
\end{feynartspicture}
\begin{feynartspicture}(310,310)(1,1)
\FADiagram{}
\FAProp(0.,15.)(10.,14.5)(0.,){/Straight}{1}
\FALabel(5.0774,15.8181)[b]{$q$}
\FAProp(0.,5.)(10.,7.)(0.,){/Straight}{1}
\FALabel(5.30398,4.9601)[t]{$q^\prime$}
\FAProp(20.,17.)(10.,14.5)(0.,){/ScalarDash}{-1}
\FALabel(14.6241,16.7737)[b]{$\tilde q_\alpha$}
\FAProp(20.,10.)(15.5,6.5)(0.,){/ScalarDash}{-1}
\FALabel(17.2784,8.9935)[br]{$\tilde q^\prime_\beta$}
\FAProp(20.,3.)(15.5,6.5)(0.,){/Sine}{0}
\FALabel(17.2784,4.0065)[tr]{$\gamma$}
\FAProp(10.,14.5)(10.,7.)(0.,){/Straight}{0}
\FALabel(9.18,10.75)[r]{$\tilde g$}
\FAProp(10.,7.)(15.5,6.5)(0.,){/ScalarDash}{1}
\FALabel(12.6097,5.68637)[t]{$\tilde q^\prime_\beta$}
\FAVert(10.,14.5){0}
\FAVert(10.,7.){0}
\FAVert(15.5,6.5){0}
\end{feynartspicture}
\begin{feynartspicture}(310,310)(1,1)
\FADiagram{}
\FAProp(0.,15.)(10.,14.5)(0.,){/Straight}{1}
\FALabel(5.0774,15.8181)[b]{$q$}
\FAProp(0.,5.)(10.,5.5)(0.,){/Straight}{1}
\FALabel(5.0774,4.18193)[t]{$q^\prime$}
\FAProp(20.,17.)(10.,10.)(0.,){/ScalarDash}{-1}
\FALabel(15.8366,15.2248)[br]{$\tilde q_\alpha$}
\FAProp(20.,10.)(10.,5.5)(0.,){/ScalarDash}{-1}
\FALabel(17.7693,10.1016)[br]{$\tilde q^\prime_\beta$}
\FAProp(20.,3.)(10.,14.5)(0.,){/Sine}{0}
\FALabel(16.7913,4.81596)[tr]{$\gamma$}
\FAProp(10.,14.5)(10.,10.)(0.,){/Straight}{1}
\FALabel(8.93,12.25)[r]{$q$}
\FAProp(10.,5.5)(10.,10.)(0.,){/Straight}{0}
\FALabel(9.18,7.75)[r]{$\tilde g$}
\FAVert(10.,14.5){0}
\FAVert(10.,5.5){0}
\FAVert(10.,10.){0}
\end{feynartspicture}
\begin{feynartspicture}(310,310)(1,1)
\FADiagram{}
\FAProp(0.,15.)(10.,14.5)(0.,){/Straight}{1}
\FALabel(5.0774,15.8181)[b]{$q$}
\FAProp(0.,5.)(10.,5.5)(0.,){/Straight}{1}
\FALabel(5.0774,4.18193)[t]{$q^\prime$}
\FAProp(20.,17.)(15.5,14.)(0.,){/ScalarDash}{-1}
\FALabel(17.3702,16.3097)[br]{$\tilde q_\alpha$}
\FAProp(20.,10.)(10.,5.5)(0.,){/ScalarDash}{-1}
\FALabel(13.8364,8.43358)[br]{$\tilde q^\prime_\beta$}
\FAProp(20.,3.)(15.5,14.)(0.,){/Sine}{0}
\FALabel(18.2486,4.40534)[r]{$\gamma$}
\FAProp(10.,14.5)(10.,5.5)(0.,){/Straight}{0}
\FALabel(9.18,10.)[r]{$\tilde g$}
\FAProp(10.,14.5)(15.5,14.)(0.,){/ScalarDash}{1}
\FALabel(12.8903,15.3136)[b]{$\tilde q_\alpha$}
\FAVert(10.,14.5){0}
\FAVert(10.,5.5){0}
\FAVert(15.5,14.){0}
\end{feynartspicture}\\[-1ex]
\begin{feynartspicture}(310,310)(1,1)
\FADiagram{}
\FAProp(0.,15.)(10.,14.5)(0.,){/Straight}{1}
\FALabel(5.0774,15.8181)[b]{$q$}
\FAProp(0.,5.)(10.,5.5)(0.,){/Straight}{1}
\FALabel(5.0774,4.18193)[t]{$q^\prime$}
\FAProp(20.,17.)(10.,14.5)(0.,){/ScalarDash}{-1}
\FALabel(14.6241,16.7737)[b]{$\tilde q_\alpha$}
\FAProp(20.,10.)(10.,10.)(0.,){/ScalarDash}{-1}
\FALabel(15.,11.07)[b]{$\tilde q^\prime_\beta$}
\FAProp(20.,3.)(10.,5.5)(0.,){/Sine}{0}
\FALabel(15.3759,5.27372)[b]{$\gamma$}
\FAProp(10.,14.5)(10.,10.)(0.,){/Straight}{0}
\FALabel(9.18,12.25)[r]{$\tilde \chi_k^0$}
\FAProp(10.,5.5)(10.,10.)(0.,){/Straight}{1}
\FALabel(8.93,7.75)[r]{$q^\prime$}
\FAVert(10.,14.5){0}
\FAVert(10.,5.5){0}
\FAVert(10.,10.){0}
\end{feynartspicture}
\begin{feynartspicture}(310,310)(1,1)
\FADiagram{}
\FAProp(0.,15.)(10.,14.5)(0.,){/Straight}{1}
\FALabel(5.0774,15.8181)[b]{$q$}
\FAProp(0.,5.)(10.,7.)(0.,){/Straight}{1}
\FALabel(5.30398,4.9601)[t]{$q^\prime$}
\FAProp(20.,17.)(10.,14.5)(0.,){/ScalarDash}{-1}
\FALabel(14.6241,16.7737)[b]{$\tilde q_\alpha$}
\FAProp(20.,10.)(15.5,6.5)(0.,){/ScalarDash}{-1}
\FALabel(17.2784,8.9935)[br]{$\tilde q^\prime_\beta$}
\FAProp(20.,3.)(15.5,6.5)(0.,){/Sine}{0}
\FALabel(17.2784,4.0065)[tr]{$\gamma$}
\FAProp(10.,14.5)(10.,7.)(0.,){/Straight}{0}
\FALabel(9.18,10.75)[r]{$\tilde \chi_k^0$}
\FAProp(10.,7.)(15.5,6.5)(0.,){/ScalarDash}{1}
\FALabel(12.6097,5.68637)[t]{$\tilde q^\prime_\beta$}
\FAVert(10.,14.5){0}
\FAVert(10.,7.){0}
\FAVert(15.5,6.5){0}
\end{feynartspicture}
\begin{feynartspicture}(310,310)(1,1)
\FADiagram{}
\FAProp(0.,15.)(10.,14.5)(0.,){/Straight}{1}
\FALabel(5.0774,15.8181)[b]{$q$}
\FAProp(0.,5.)(10.,5.5)(0.,){/Straight}{1}
\FALabel(5.0774,4.18193)[t]{$q^\prime$}
\FAProp(20.,17.)(10.,10.)(0.,){/ScalarDash}{-1}
\FALabel(15.8366,15.2248)[br]{$\tilde q_\alpha$}
\FAProp(20.,10.)(10.,5.5)(0.,){/ScalarDash}{-1}
\FALabel(17.7693,10.1016)[br]{$\tilde q^\prime_\beta$}
\FAProp(20.,3.)(10.,14.5)(0.,){/Sine}{0}
\FALabel(16.7913,4.81596)[tr]{$\gamma$}
\FAProp(10.,14.5)(10.,10.)(0.,){/Straight}{1}
\FALabel(8.93,12.25)[r]{$q$}
\FAProp(10.,5.5)(10.,10.)(0.,){/Straight}{0}
\FALabel(9.18,7.75)[r]{$\tilde \chi_k^0$}
\FAVert(10.,14.5){0}
\FAVert(10.,5.5){0}
\FAVert(10.,10.){0}
\end{feynartspicture}
\begin{feynartspicture}(310,310)(1,1)
\FADiagram{}
\FAProp(0.,15.)(10.,14.5)(0.,){/Straight}{1}
\FALabel(5.0774,15.8181)[b]{$q$}
\FAProp(0.,5.)(10.,5.5)(0.,){/Straight}{1}
\FALabel(5.0774,4.18193)[t]{$q^\prime$}
\FAProp(20.,17.)(15.5,14.)(0.,){/ScalarDash}{-1}
\FALabel(17.3702,16.3097)[br]{$\tilde q_\alpha$}
\FAProp(20.,10.)(10.,5.5)(0.,){/ScalarDash}{-1}
\FALabel(13.8364,8.43358)[br]{$\tilde q^\prime_\beta$}
\FAProp(20.,3.)(15.5,14.)(0.,){/Sine}{0}
\FALabel(18.2486,4.40534)[r]{$\gamma$}
\FAProp(10.,14.5)(10.,5.5)(0.,){/Straight}{0}
\FALabel(9.18,10.)[r]{$\tilde \chi_k^0$}
\FAProp(10.,14.5)(15.5,14.)(0.,){/ScalarDash}{1}
\FALabel(12.8903,15.3136)[b]{$\tilde q_\alpha$}
\FAVert(10.,14.5){0}
\FAVert(10.,5.5){0}
\FAVert(15.5,14.){0}
\end{feynartspicture}\\[-1ex]
\begin{feynartspicture}(310,310)(1,1)
\FADiagram{}
\FAProp(0.,15.)(10.,14.5)(0.,){/Straight}{1}
\FALabel(5.0774,15.8181)[b]{$u$}
\FAProp(0.,5.)(10.,5.5)(0.,){/Straight}{1}
\FALabel(5.0774,4.18193)[t]{$d$}
\FAProp(20.,17.)(10.,5.5)(0.,){/ScalarDash}{-1}
\FALabel(17.2913,15.184)[br]{$\tilde u_\alpha$}
\FAProp(20.,10.)(10.,14.5)(0.,){/ScalarDash}{-1}
\FALabel(19.8274,10.002)[tr]{$\tilde d_\beta$}
\FAProp(20.,3.)(10.,10.)(0.,){/Sine}{0}
\FALabel(14.5911,5.71019)[tr]{$\gamma$}
\FAProp(10.,14.5)(10.,10.)(0.,){/Straight}{-1}
\FALabel(8.93,12.25)[r]{$\tilde \chi_k^\pm$}
\FAProp(10.,5.5)(10.,10.)(0.,){/Straight}{1}
\FALabel(8.93,7.75)[r]{$\tilde \chi_k^\pm$}
\FAVert(10.,14.5){0}
\FAVert(10.,5.5){0}
\FAVert(10.,10.){0}
\end{feynartspicture}
\begin{feynartspicture}(310,310)(1,1)
\FADiagram{}
\FAProp(0.,15.)(10.,5.5)(0.,){/Straight}{1}
\FALabel(3.19219,13.2012)[bl]{$u$}
\FAProp(0.,5.)(10.,14.5)(0.,){/Straight}{1}
\FALabel(3.10297,6.58835)[tl]{$d$}
\FAProp(20.,17.)(10.,14.5)(0.,){/ScalarDash}{-1}
\FALabel(14.6241,16.7737)[b]{$\tilde u_\alpha$}
\FAProp(20.,10.)(10.,10.)(0.,){/ScalarDash}{-1}
\FALabel(17.95,11.07)[b]{$\tilde d_\beta$}
\FAProp(20.,3.)(10.,5.5)(0.,){/Sine}{0}
\FALabel(14.6241,3.22628)[t]{$\gamma$}
\FAProp(10.,5.5)(10.,10.)(0.,){/Straight}{1}
\FALabel(11.07,7.75)[l]{$u$}
\FAProp(10.,14.5)(10.,10.)(0.,){/Straight}{1}
\FALabel(11.07,12.25)[l]{$\tilde \chi_k^\pm$}
\FAVert(10.,5.5){0}
\FAVert(10.,14.5){0}
\FAVert(10.,10.){0}
\end{feynartspicture}
\begin{feynartspicture}(310,310)(1,1)
\FADiagram{}
\FAProp(0.,15.)(10.,7.)(0.,){/Straight}{1}
\FALabel(2.82617,13.9152)[bl]{$u$}
\FAProp(0.,5.)(10.,14.5)(0.,){/Straight}{1}
\FALabel(3.04219,6.54875)[tl]{$d$}
\FAProp(20.,17.)(10.,14.5)(0.,){/ScalarDash}{-1}
\FALabel(14.6241,16.7737)[b]{$\tilde u_\alpha$}
\FAProp(20.,10.)(15.5,6.5)(0.,){/ScalarDash}{-1}
\FALabel(19.2784,8.9935)[br]{$\tilde d_\beta$}
\FAProp(20.,3.)(15.5,6.5)(0.,){/Sine}{0}
\FALabel(17.2784,4.0065)[tr]{$\gamma$}
\FAProp(10.,7.)(10.,14.5)(0.,){/Straight}{-1}
\FALabel(11.07,10.75)[l]{$\tilde \chi_k^\pm$}
\FAProp(10.,7.)(15.5,6.5)(0.,){/ScalarDash}{1}
\FALabel(12.6097,5.68637)[t]{$\tilde d_\beta$}
\FAVert(10.,7.){0}
\FAVert(10.,14.5){0}
\FAVert(15.5,6.5){0}
\end{feynartspicture}
\begin{feynartspicture}(310,310)(1,1)
\FADiagram{}
\FAProp(0.,15.)(10.,14.5)(0.,){/Straight}{1}
\FALabel(5.0774,15.8181)[b]{$u$}
\FAProp(0.,5.)(10.,5.5)(0.,){/Straight}{1}
\FALabel(5.0774,4.18193)[t]{$d$}
\FAProp(20.,17.)(10.,10.)(0.,){/ScalarDash}{-1}
\FALabel(17.4388,16.3076)[br]{$\tilde u_\alpha$}
\FAProp(20.,10.)(10.,14.5)(0.,){/ScalarDash}{-1}
\FALabel(17.2774,10.302)[tr]{$\tilde d_\beta$}
\FAProp(20.,3.)(10.,5.5)(0.,){/Sine}{0}
\FALabel(14.6241,3.22628)[t]{$\gamma$}
\FAProp(10.,14.5)(10.,10.)(0.,){/Straight}{-1}
\FALabel(8.93,12.25)[r]{$\tilde \chi_k^\pm$}
\FAProp(10.,5.5)(10.,10.)(0.,){/Straight}{1}
\FALabel(8.93,7.75)[r]{$d$}
\FAVert(10.,14.5){0}
\FAVert(10.,5.5){0}
\FAVert(10.,10.){0}
\end{feynartspicture}
\begin{feynartspicture}(310,310)(1,1)
\FADiagram{}
\FAProp(0.,15.)(10.,14.5)(0.,){/Straight}{1}
\FALabel(5.0774,15.8181)[b]{$u$}
\FAProp(0.,5.)(10.,5.5)(0.,){/Straight}{1}
\FALabel(5.0774,4.18193)[t]{$d$}
\FAProp(20.,17.)(15.5,8.)(0.,){/ScalarDash}{-1}
\FALabel(24.0564,14.0818)[br]{$\tilde u_\alpha$}
\FAProp(20.,10.)(10.,14.5)(0.,){/ScalarDash}{-1}
\FALabel(12.6226,14.248)[bl]{$\tilde d_\beta$}
\FAProp(20.,3.)(15.5,8.)(0.,){/Sine}{0}
\FALabel(18.4221,6.0569)[bl]{$\gamma$}
\FAProp(10.,14.5)(10.,5.5)(0.,){/Straight}{-1}
\FALabel(8.93,10.)[r]{$\tilde \chi_k^\pm$}
\FAProp(10.,5.5)(15.5,8.)(0.,){/ScalarDash}{1}
\FALabel(12.9114,5.81893)[tl]{$\tilde u_\alpha$}
\FAVert(10.,14.5){0}
\FAVert(10.,5.5){0}
\FAVert(15.5,8.){0}
\end{feynartspicture}
\vspace*{-2ex}
\caption{Feynman diagrams for real photon emission. For
  $q=q^\prime$ crossed diagrams have to be taken into
  account. Diagrams in the last row only contribute
for $u=\{u,c\},\,d=\{d,s\}$.
\vspace*{-5ex}
\label{fig_realphoton}}
}

\clearpage

\FIGURE{
\vspace*{2ex}
\small
\unitlength=0.24bp%
\begin{feynartspicture}(310,310)(1,1)
\FADiagram{}
\FAProp(0.,15.)(10.,14.5)(0.,){/Straight}{1}
\FALabel(5.0774,15.8181)[b]{$q$}
\FAProp(0.,5.)(10.,5.5)(0.,){/Straight}{1}
\FALabel(5.0774,4.18193)[t]{$q^\prime$}
\FAProp(20.,17.)(10.,14.5)(0.,){/ScalarDash}{-1}
\FALabel(14.6241,16.7737)[b]{$\tilde q_\alpha$}
\FAProp(20.,10.)(10.,5.5)(0.,){/ScalarDash}{-1}
\FALabel(17.2909,9.58086)[br]{$\tilde q^\prime_\beta$}
\FAProp(20.,3.)(10.,10.)(0.,){/Cycles}{0}
\FALabel(16.9352,3.20173)[tr]{$g$}
\FAProp(10.,14.5)(10.,10.)(0.,){/Straight}{0}
\FALabel(9.18,12.25)[r]{$\tilde g$}
\FAProp(10.,5.5)(10.,10.)(0.,){/Straight}{0}
\FALabel(9.18,7.75)[r]{$\tilde g$}
\FAVert(10.,14.5){0}
\FAVert(10.,5.5){0}
\FAVert(10.,10.){0}
\end{feynartspicture}
\begin{feynartspicture}(310,310)(1,1)
\FADiagram{}
\FAProp(0.,15.)(10.,14.5)(0.,){/Straight}{1}
\FALabel(5.0774,15.8181)[b]{$q$}
\FAProp(0.,5.)(10.,5.5)(0.,){/Straight}{1}
\FALabel(5.0774,4.18193)[t]{$q^\prime$}
\FAProp(20.,17.)(10.,14.5)(0.,){/ScalarDash}{-1}
\FALabel(14.6241,16.7737)[b]{$\tilde q_\alpha$}
\FAProp(20.,10.)(10.,10.)(0.,){/ScalarDash}{-1}
\FALabel(15.,11.07)[b]{$\tilde q^\prime_\beta$}
\FAProp(20.,3.)(10.,5.5)(0.,){/Cycles}{0}
\FALabel(15.3759,5.27372)[b]{$g$}
\FAProp(10.,14.5)(10.,10.)(0.,){/Straight}{0}
\FALabel(9.18,12.25)[r]{$\tilde g$}
\FAProp(10.,5.5)(10.,10.)(0.,){/Straight}{1}
\FALabel(8.93,7.75)[r]{$q^\prime$}
\FAVert(10.,14.5){0}
\FAVert(10.,5.5){0}
\FAVert(10.,10.){0}
\end{feynartspicture}
\begin{feynartspicture}(310,310)(1,1)
\FADiagram{}
\FAProp(0.,15.)(10.,14.5)(0.,){/Straight}{1}
\FALabel(5.0774,15.8181)[b]{$q$}
\FAProp(0.,5.)(10.,7.)(0.,){/Straight}{1}
\FALabel(5.30398,4.9601)[t]{$q^\prime$}
\FAProp(20.,17.)(10.,14.5)(0.,){/ScalarDash}{-1}
\FALabel(14.6241,16.7737)[b]{$\tilde q_\alpha$}
\FAProp(20.,10.)(15.5,6.5)(0.,){/ScalarDash}{-1}
\FALabel(17.2784,8.9935)[br]{$\tilde q^\prime_\beta$}
\FAProp(20.,3.)(15.5,6.5)(0.,){/Cycles}{0}
\FALabel(16.8486,3.45396)[tr]{$g$}
\FAProp(10.,14.5)(10.,7.)(0.,){/Straight}{0}
\FALabel(9.18,10.75)[r]{$\tilde g$}
\FAProp(10.,7.)(15.5,6.5)(0.,){/ScalarDash}{1}
\FALabel(12.6097,5.68637)[t]{$\tilde q^\prime_\beta$}
\FAVert(10.,14.5){0}
\FAVert(10.,7.){0}
\FAVert(15.5,6.5){0}
\end{feynartspicture}
\begin{feynartspicture}(310,310)(1,1)
\FADiagram{}
\FAProp(0.,15.)(10.,14.5)(0.,){/Straight}{1}
\FALabel(5.0774,15.8181)[b]{$q$}
\FAProp(0.,5.)(10.,5.5)(0.,){/Straight}{1}
\FALabel(5.0774,4.18193)[t]{$q^\prime$}
\FAProp(20.,17.)(10.,10.)(0.,){/ScalarDash}{-1}
\FALabel(15.8366,15.2248)[br]{$\tilde q_\alpha$}
\FAProp(20.,10.)(10.,5.5)(0.,){/ScalarDash}{-1}
\FALabel(17.7693,10.1016)[br]{$\tilde q^\prime_\beta$}
\FAProp(20.,3.)(10.,14.5)(0.,){/Cycles}{0}
\FALabel(16.2631,4.35663)[tr]{$g$}
\FAProp(10.,14.5)(10.,10.)(0.,){/Straight}{1}
\FALabel(8.93,12.25)[r]{$q$}
\FAProp(10.,5.5)(10.,10.)(0.,){/Straight}{0}
\FALabel(9.18,7.75)[r]{$\tilde g$}
\FAVert(10.,14.5){0}
\FAVert(10.,5.5){0}
\FAVert(10.,10.){0}
\end{feynartspicture}
\begin{feynartspicture}(310,310)(1,1)
\FADiagram{}
\FAProp(0.,15.)(10.,14.5)(0.,){/Straight}{1}
\FALabel(5.0774,15.8181)[b]{$q$}
\FAProp(0.,5.)(10.,5.5)(0.,){/Straight}{1}
\FALabel(5.0774,4.18193)[t]{$q^\prime$}
\FAProp(20.,17.)(15.5,14.)(0.,){/ScalarDash}{-1}
\FALabel(17.3702,16.3097)[br]{$\tilde q_\alpha$}
\FAProp(20.,10.)(10.,5.5)(0.,){/ScalarDash}{-1}
\FALabel(13.8364,8.43358)[br]{$\tilde q^\prime_\beta$}
\FAProp(20.,3.)(15.5,14.)(0.,){/Cycles}{0}
\FALabel(17.6007,4.1403)[r]{$g$}
\FAProp(10.,14.5)(10.,5.5)(0.,){/Straight}{0}
\FALabel(9.18,10.)[r]{$\tilde g$}
\FAProp(10.,14.5)(15.5,14.)(0.,){/ScalarDash}{1}
\FALabel(12.8903,15.3136)[b]{$\tilde q_\alpha$}
\FAVert(10.,14.5){0}
\FAVert(10.,5.5){0}
\FAVert(15.5,14.){0}
\end{feynartspicture}\\
\begin{feynartspicture}(310,310)(1,1)
\FADiagram{}
\FAProp(0.,15.)(10.,14.5)(0.,){/Straight}{1}
\FALabel(5.0774,15.8181)[b]{$q$}
\FAProp(0.,5.)(10.,5.5)(0.,){/Straight}{1}
\FALabel(5.0774,4.18193)[t]{$q^\prime$}
\FAProp(20.,17.)(10.,14.5)(0.,){/ScalarDash}{-1}
\FALabel(14.6241,16.7737)[b]{$\tilde q_\alpha$}
\FAProp(20.,10.)(10.,10.)(0.,){/ScalarDash}{-1}
\FALabel(15.,11.07)[b]{$\tilde q^\prime_\beta$}
\FAProp(20.,3.)(10.,5.5)(0.,){/Cycles}{0}
\FALabel(15.3759,5.27372)[b]{$g$}
\FAProp(10.,14.5)(10.,10.)(0.,){/Straight}{0}
\FALabel(9.18,12.25)[r]{$\tilde \chi_k^0$}
\FAProp(10.,5.5)(10.,10.)(0.,){/Straight}{1}
\FALabel(8.93,7.75)[r]{$q^\prime$}
\FAVert(10.,14.5){0}
\FAVert(10.,5.5){0}
\FAVert(10.,10.){0}
\end{feynartspicture}
\begin{feynartspicture}(310,310)(1,1)
\FADiagram{}
\FAProp(0.,15.)(10.,14.5)(0.,){/Straight}{1}
\FALabel(5.0774,15.8181)[b]{$q$}
\FAProp(0.,5.)(10.,7.)(0.,){/Straight}{1}
\FALabel(5.30398,4.9601)[t]{$q^\prime$}
\FAProp(20.,17.)(10.,14.5)(0.,){/ScalarDash}{-1}
\FALabel(14.6241,16.7737)[b]{$\tilde q_\alpha$}
\FAProp(20.,10.)(15.5,6.5)(0.,){/ScalarDash}{-1}
\FALabel(17.2784,8.9935)[br]{$\tilde q^\prime_\beta$}
\FAProp(20.,3.)(15.5,6.5)(0.,){/Cycles}{0}
\FALabel(16.8486,3.45396)[tr]{$g$}
\FAProp(10.,14.5)(10.,7.)(0.,){/Straight}{0}
\FALabel(9.18,10.75)[r]{$\tilde \chi_k^0$}
\FAProp(10.,7.)(15.5,6.5)(0.,){/ScalarDash}{1}
\FALabel(12.6097,5.68637)[t]{$\tilde q^\prime_\beta$}
\FAVert(10.,14.5){0}
\FAVert(10.,7.){0}
\FAVert(15.5,6.5){0}
\end{feynartspicture}
\begin{feynartspicture}(310,310)(1,1)
\FADiagram{}
\FAProp(0.,15.)(10.,14.5)(0.,){/Straight}{1}
\FALabel(5.0774,15.8181)[b]{$q$}
\FAProp(0.,5.)(10.,5.5)(0.,){/Straight}{1}
\FALabel(5.0774,4.18193)[t]{$q^\prime$}
\FAProp(20.,17.)(10.,10.)(0.,){/ScalarDash}{-1}
\FALabel(15.8366,15.2248)[br]{$\tilde q_\alpha$}
\FAProp(20.,10.)(10.,5.5)(0.,){/ScalarDash}{-1}
\FALabel(17.7693,10.1016)[br]{$\tilde q^\prime_\beta$}
\FAProp(20.,3.)(10.,14.5)(0.,){/Cycles}{0}
\FALabel(16.2631,4.35663)[tr]{$g$}
\FAProp(10.,14.5)(10.,10.)(0.,){/Straight}{1}
\FALabel(8.93,12.25)[r]{$q$}
\FAProp(10.,5.5)(10.,10.)(0.,){/Straight}{0}
\FALabel(9.18,7.75)[r]{$\tilde \chi_k^0$}
\FAVert(10.,14.5){0}
\FAVert(10.,5.5){0}
\FAVert(10.,10.){0}
\end{feynartspicture}
\begin{feynartspicture}(310,310)(1,1)
\FADiagram{}
\FAProp(0.,15.)(10.,14.5)(0.,){/Straight}{1}
\FALabel(5.0774,15.8181)[b]{$q$}
\FAProp(0.,5.)(10.,5.5)(0.,){/Straight}{1}
\FALabel(5.0774,4.18193)[t]{$q^\prime$}
\FAProp(20.,17.)(15.5,14.)(0.,){/ScalarDash}{-1}
\FALabel(17.3702,16.3097)[br]{$\tilde q_\alpha$}
\FAProp(20.,10.)(10.,5.5)(0.,){/ScalarDash}{-1}
\FALabel(13.8364,8.43358)[br]{$\tilde q^\prime_\beta$}
\FAProp(20.,3.)(15.5,14.)(0.,){/Cycles}{0}
\FALabel(17.6007,4.1403)[r]{$g$}
\FAProp(10.,14.5)(10.,5.5)(0.,){/Straight}{0}
\FALabel(9.18,10.)[r]{$\tilde \chi_k^0$}
\FAProp(10.,14.5)(15.5,14.)(0.,){/ScalarDash}{1}
\FALabel(12.8903,15.3136)[b]{$\tilde q_\alpha$}
\FAVert(10.,14.5){0}
\FAVert(10.,5.5){0}
\FAVert(15.5,14.){0}
\end{feynartspicture}\\
\begin{feynartspicture}(310,310)(1,1)
\FADiagram{}
\FAProp(0.,15.)(10.,5.5)(0.,){/Straight}{1}
\FALabel(3.19219,13.2012)[bl]{$u$}
\FAProp(0.,5.)(10.,14.5)(0.,){/Straight}{1}
\FALabel(3.10297,6.58835)[tl]{$d$}
\FAProp(20.,17.)(10.,14.5)(0.,){/ScalarDash}{-1}
\FALabel(14.6241,16.7737)[b]{$\tilde u_\alpha$}
\FAProp(20.,10.)(10.,10.)(0.,){/ScalarDash}{-1}
\FALabel(17.95,11.07)[b]{$\tilde d_\beta$}
\FAProp(20.,3.)(10.,5.5)(0.,){/Cycles}{0}
\FALabel(14.4543,2.54718)[t]{$g$}
\FAProp(10.,5.5)(10.,10.)(0.,){/Straight}{1}
\FALabel(11.07,7.75)[l]{$u$}
\FAProp(10.,14.5)(10.,10.)(0.,){/Straight}{1}
\FALabel(11.07,12.25)[l]{$\tilde \chi_k^\pm$}
\FAVert(10.,5.5){0}
\FAVert(10.,14.5){0}
\FAVert(10.,10.){0}
\end{feynartspicture}
\begin{feynartspicture}(310,310)(1,1)
\FADiagram{}
\FAProp(0.,15.)(10.,7.)(0.,){/Straight}{1}
\FALabel(2.82617,13.9152)[bl]{$u$}
\FAProp(0.,5.)(10.,14.5)(0.,){/Straight}{1}
\FALabel(3.04219,6.54875)[tl]{$d$}
\FAProp(20.,17.)(10.,14.5)(0.,){/ScalarDash}{-1}
\FALabel(14.6241,16.7737)[b]{$\tilde u_\alpha$}
\FAProp(20.,10.)(15.5,6.5)(0.,){/ScalarDash}{-1}
\FALabel(19.2784,9.9935)[br]{$\tilde d_\beta$}
\FAProp(20.,3.)(15.5,6.5)(0.,){/Cycles}{0}
\FALabel(16.8486,3.45396)[tr]{$g$}
\FAProp(10.,7.)(10.,14.5)(0.,){/Straight}{-1}
\FALabel(11.07,10.75)[l]{$\tilde \chi_k^\pm$}
\FAProp(10.,7.)(15.5,6.5)(0.,){/ScalarDash}{1}
\FALabel(12.6097,5.68637)[t]{$\tilde d_\beta$}
\FAVert(10.,7.){0}
\FAVert(10.,14.5){0}
\FAVert(15.5,6.5){0}
\end{feynartspicture}
\begin{feynartspicture}(310,310)(1,1)
\FADiagram{}
\FAProp(0.,15.)(10.,14.5)(0.,){/Straight}{1}
\FALabel(5.0774,15.8181)[b]{$u$}
\FAProp(0.,5.)(10.,5.5)(0.,){/Straight}{1}
\FALabel(5.0774,4.18193)[t]{$d$}
\FAProp(20.,17.)(10.,10.)(0.,){/ScalarDash}{-1}
\FALabel(17.4388,16.3076)[br]{$\tilde u_\alpha$}
\FAProp(20.,10.)(10.,14.5)(0.,){/ScalarDash}{-1}
\FALabel(17.2774,10.302)[tr]{$\tilde d_\beta$}
\FAProp(20.,3.)(10.,5.5)(0.,){/Cycles}{0}
\FALabel(14.4543,2.54718)[t]{$g$}
\FAProp(10.,14.5)(10.,10.)(0.,){/Straight}{-1}
\FALabel(8.93,12.25)[r]{$\tilde \chi_k^\pm$}
\FAProp(10.,5.5)(10.,10.)(0.,){/Straight}{1}
\FALabel(8.93,7.75)[r]{$d$}
\FAVert(10.,14.5){0}
\FAVert(10.,5.5){0}
\FAVert(10.,10.){0}
\end{feynartspicture}
\begin{feynartspicture}(310,310)(1,1)
\FADiagram{}
\FAProp(0.,15.)(10.,14.5)(0.,){/Straight}{1}
\FALabel(5.0774,15.8181)[b]{$u$}
\FAProp(0.,5.)(10.,5.5)(0.,){/Straight}{1}
\FALabel(5.0774,4.18193)[t]{$d$}
\FAProp(20.,17.)(15.5,8.)(0.,){/ScalarDash}{-1}
\FALabel(24.5564,14.0818)[br]{$\tilde u_\alpha$}
\FAProp(20.,10.)(10.,14.5)(0.,){/ScalarDash}{-1}
\FALabel(12.6226,14.248)[bl]{$\tilde d_\beta$}
\FAProp(20.,3.)(15.5,8.)(0.,){/Cycles}{0}
\FALabel(18.4221,6.0569)[bl]{$g$}
\FAProp(10.,14.5)(10.,5.5)(0.,){/Straight}{-1}
\FALabel(8.93,10.)[r]{$\tilde \chi_k^\pm$}
\FAProp(10.,5.5)(15.5,8.)(0.,){/ScalarDash}{1}
\FALabel(12.9114,5.81893)[tl]{$\tilde u_\alpha$}
\FAVert(10.,14.5){0}
\FAVert(10.,5.5){0}
\FAVert(15.5,8.){0}
\end{feynartspicture}

\caption{Feynman diagrams for real gluon emission. For
  $q=q^\prime$ crossed diagrams have to be taken into
  account. Diagrams in the last row only contribute 
for $u=\{u,c\},\,d=\{d,s\}$.
\vspace*{3ex}
\label{fig_realgluon}}
}

\FIGURE{
\small
\unitlength=0.24bp%
\begin{feynartspicture}(310,310)(1,1)
\FADiagram{}
\FAProp(0.,15.)(10.,14.5)(0.,){/Straight}{1}
\FALabel(5.0774,15.8181)[b]{$q$}
\FAProp(0.,5.)(10.,5.5)(0.,){/Cycles}{0}
\FALabel(5.0774,4.18193)[t]{$g$}
\FAProp(20.,17.)(10.,14.5)(0.,){/ScalarDash}{-1}
\FALabel(14.6241,16.7737)[b]{$\tilde q_\alpha$}
\FAProp(20.,10.)(10.,5.5)(0.,){/ScalarDash}{-1}
\FALabel(17.2909,9.58086)[br]{$\tilde q_\beta^\prime$}
\FAProp(20.,3.)(10.,10.)(0.,){/Straight}{1}
\FALabel(17.3366,3.77519)[tr]{$q^\prime$}
\FAProp(10.,14.5)(10.,10.)(0.,){/Straight}{0}
\FALabel(9.18,12.25)[r]{$\tilde g$}
\FAProp(10.,5.5)(10.,10.)(0.,){/ScalarDash}{-1}
\FALabel(8.93,7.75)[r]{$\tilde q_\beta^\prime$}
\FAVert(10.,14.5){0}
\FAVert(10.,5.5){0}
\FAVert(10.,10.){0}
\end{feynartspicture}
\begin{feynartspicture}(310,310)(1,1)
\FADiagram{}
\FAProp(0.,15.)(10.,14.5)(0.,){/Straight}{1}
\FALabel(5.0774,15.8181)[b]{$q$}
\FAProp(0.,5.)(10.,5.5)(0.,){/Cycles}{0}
\FALabel(5.0774,4.18193)[t]{$g$}
\FAProp(20.,17.)(10.,14.5)(0.,){/ScalarDash}{-1}
\FALabel(14.6241,16.7737)[b]{$\tilde q_\alpha$}
\FAProp(20.,10.)(10.,10.)(0.,){/ScalarDash}{-1}
\FALabel(15.,11.07)[b]{$\tilde q_\beta^\prime$}
\FAProp(20.,3.)(10.,5.5)(0.,){/Straight}{1}
\FALabel(15.3759,5.27372)[b]{$q^\prime$}
\FAProp(10.,14.5)(10.,10.)(0.,){/Straight}{0}
\FALabel(9.18,12.25)[r]{$\tilde g$}
\FAProp(10.,5.5)(10.,10.)(0.,){/Straight}{1}
\FALabel(8.93,7.75)[r]{$q^\prime$}
\FAVert(10.,14.5){0}
\FAVert(10.,5.5){0}
\FAVert(10.,10.){0}
\end{feynartspicture}
\begin{feynartspicture}(310,310)(1,1)
\FADiagram{}
\FAProp(0.,15.)(5.5,10.)(0.,){/Straight}{1}
\FALabel(2.18736,11.8331)[tr]{$q$}
\FAProp(0.,5.)(5.5,10.)(0.,){/Cycles}{0}
\FALabel(3.31264,6.83309)[tl]{$g$}
\FAProp(20.,17.)(11.5,10.)(0.,){/ScalarDash}{-1}
\FALabel(15.2447,14.2165)[br]{$\tilde q_\alpha$}
\FAProp(20.,10.)(15.5,6.5)(0.,){/ScalarDash}{-1}
\FALabel(17.2784,8.9935)[br]{$\tilde q_\beta^\prime$}
\FAProp(20.,3.)(15.5,6.5)(0.,){/Straight}{1}
\FALabel(18.2216,5.4935)[bl]{$q^\prime$}
\FAProp(5.5,10.)(11.5,10.)(0.,){/Straight}{1}
\FALabel(8.5,11.07)[b]{$q$}
\FAProp(11.5,10.)(15.5,6.5)(0.,){/Straight}{0}
\FALabel(13.1239,7.75165)[tr]{$\tilde g$}
\FAVert(5.5,10.){0}
\FAVert(11.5,10.){0}
\FAVert(15.5,6.5){0}
\end{feynartspicture}
\begin{feynartspicture}(310,310)(1,1)
\FADiagram{}
\FAProp(0.,15.)(10.,14.5)(0.,){/Straight}{1}
\FALabel(5.0774,15.8181)[b]{$q$}
\FAProp(0.,5.)(10.,7.)(0.,){/Cycles}{0}
\FALabel(5.30398,4.9601)[t]{$g$}
\FAProp(20.,17.)(10.,14.5)(0.,){/ScalarDash}{-1}
\FALabel(14.6241,16.7737)[b]{$\tilde q_\alpha$}
\FAProp(20.,10.)(15.5,6.5)(0.,){/ScalarDash}{-1}
\FALabel(17.2784,8.9935)[br]{$\tilde q_\beta^\prime$}
\FAProp(20.,3.)(15.5,6.5)(0.,){/Straight}{1}
\FALabel(17.2784,4.0065)[tr]{$q^\prime$}
\FAProp(10.,14.5)(10.,7.)(0.,){/Straight}{0}
\FALabel(9.18,10.75)[r]{$\tilde g$}
\FAProp(10.,7.)(15.5,6.5)(0.,){/Straight}{0}
\FALabel(12.6323,5.93534)[t]{$\tilde g$}
\FAVert(10.,14.5){0}
\FAVert(10.,7.){0}
\FAVert(15.5,6.5){0}
\end{feynartspicture}\\
\begin{feynartspicture}(310,310)(1,1)
\FADiagram{}
\FAProp(0.,15.)(10.,7.)(0.,){/Straight}{1}
\FALabel(2.82617,13.9152)[bl]{$q$}
\FAProp(0.,5.)(10.,14.5)(0.,){/Cycles}{0}
\FALabel(3.04219,6.54875)[tl]{$g$}
\FAProp(20.,17.)(10.,14.5)(0.,){/ScalarDash}{-1}
\FALabel(14.6241,16.7737)[b]{$\tilde q_\alpha$}
\FAProp(20.,10.)(15.5,6.5)(0.,){/ScalarDash}{-1}
\FALabel(19.2784,9.9935)[br]{$\tilde q_\beta^\prime$}
\FAProp(20.,3.)(15.5,6.5)(0.,){/Straight}{1}
\FALabel(17.2784,4.0065)[tr]{$q^\prime$}
\FAProp(10.,7.)(10.,14.5)(0.,){/ScalarDash}{1}
\FALabel(11.07,10.75)[l]{$\tilde q_\alpha$}
\FAProp(10.,7.)(15.5,6.5)(0.,){/Straight}{0}
\FALabel(12.6323,5.93534)[t]{$\tilde g$}
\FAVert(10.,7.){0}
\FAVert(10.,14.5){0}
\FAVert(15.5,6.5){0}
\end{feynartspicture}
\begin{feynartspicture}(310,310)(1,1)
\FADiagram{}
\FAProp(0.,15.)(5.5,10.)(0.,){/Straight}{1}
\FALabel(2.18736,11.8331)[tr]{$q$}
\FAProp(0.,5.)(5.5,10.)(0.,){/Cycles}{0}
\FALabel(3.31264,6.83309)[tl]{$g$}
\FAProp(20.,17.)(11.5,10.)(0.,){/ScalarDash}{-1}
\FALabel(15.2447,14.2165)[br]{$\tilde q_\alpha$}
\FAProp(20.,10.)(15.5,6.5)(0.,){/ScalarDash}{-1}
\FALabel(21.2784,9.9935)[br]{$\tilde q_\beta^\prime$}
\FAProp(20.,3.)(15.5,6.5)(0.,){/Straight}{1}
\FALabel(18.2216,5.4935)[bl]{$q^\prime$}
\FAProp(5.5,10.)(11.5,10.)(0.,){/Straight}{1}
\FALabel(8.5,11.07)[b]{$q$}
\FAProp(11.5,10.)(15.5,6.5)(0.,){/Straight}{0}
\FALabel(13.1239,7.75165)[tr]{$\tilde \chi_k^0$}
\FAVert(5.5,10.){0}
\FAVert(11.5,10.){0}
\FAVert(15.5,6.5){0}
\end{feynartspicture}
\begin{feynartspicture}(310,310)(1,1)
\FADiagram{}
\FAProp(0.,15.)(10.,7.)(0.,){/Straight}{1}
\FALabel(2.82617,13.9152)[bl]{$q$}
\FAProp(0.,5.)(10.,14.5)(0.,){/Cycles}{0}
\FALabel(3.04219,6.54875)[tl]{$g$}
\FAProp(20.,17.)(10.,14.5)(0.,){/ScalarDash}{-1}
\FALabel(14.6241,16.7737)[b]{$\tilde q_\alpha$}
\FAProp(20.,10.)(15.5,6.5)(0.,){/ScalarDash}{-1}
\FALabel(21.2784,9.9935)[br]{$\tilde q_\beta^\prime$}
\FAProp(20.,3.)(15.5,6.5)(0.,){/Straight}{1}
\FALabel(17.2784,4.0065)[tr]{$q^\prime$}
\FAProp(10.,7.)(10.,14.5)(0.,){/ScalarDash}{1}
\FALabel(11.07,10.75)[l]{$\tilde q_\alpha$}
\FAProp(10.,7.)(15.5,6.5)(0.,){/Straight}{0}
\FALabel(12.6323,5.93534)[t]{$\tilde \chi_k^0$}
\FAVert(10.,7.){0}
\FAVert(10.,14.5){0}
\FAVert(15.5,6.5){0}
\end{feynartspicture}\\
\begin{feynartspicture}(310,310)(1,1)
\FADiagram{}
\FAProp(0.,15.)(10.,14.5)(0.,){/Straight}{1}
\FALabel(5.0774,15.8181)[b]{$u$}
\FAProp(0.,5.)(10.,5.5)(0.,){/Cycles}{0}
\FALabel(5.0774,4.18193)[t]{$g$}
\FAProp(20.,17.)(10.,5.5)(0.,){/ScalarDash}{-1}
\FALabel(17.2913,15.184)[br]{$\tilde u_\alpha$}
\FAProp(20.,10.)(10.,14.5)(0.,){/ScalarDash}{-1}
\FALabel(20.8274,10.002)[tr]{$\tilde d_\beta$}
\FAProp(20.,3.)(10.,10.)(0.,){/Straight}{1}
\FALabel(14.5911,5.71019)[tr]{$d$}
\FAProp(10.,14.5)(10.,10.)(0.,){/Straight}{-1}
\FALabel(8.93,12.25)[r]{$\tilde \chi_k^\pm$}
\FAProp(10.,5.5)(10.,10.)(0.,){/ScalarDash}{-1}
\FALabel(8.93,7.75)[r]{$\tilde u_\alpha$}
\FAVert(10.,14.5){0}
\FAVert(10.,5.5){0}
\FAVert(10.,10.){0}
\end{feynartspicture}
\begin{feynartspicture}(310,310)(1,1)
\FADiagram{}
\FAProp(0.,15.)(10.,14.5)(0.,){/Straight}{1}
\FALabel(5.0774,15.8181)[b]{$u$}
\FAProp(0.,5.)(10.,5.5)(0.,){/Cycles}{0}
\FALabel(5.0774,4.18193)[t]{$g$}
\FAProp(20.,17.)(10.,10.)(0.,){/ScalarDash}{-1}
\FALabel(17.4388,16.3076)[br]{$\tilde u_\alpha$}
\FAProp(20.,10.)(10.,14.5)(0.,){/ScalarDash}{-1}
\FALabel(19.2774,10.302)[tr]{$\tilde d_\beta$}
\FAProp(20.,3.)(10.,5.5)(0.,){/Straight}{1}
\FALabel(14.6241,3.22628)[t]{$d$}
\FAProp(10.,14.5)(10.,10.)(0.,){/Straight}{-1}
\FALabel(8.93,12.25)[r]{$\tilde \chi_k^\pm$}
\FAProp(10.,5.5)(10.,10.)(0.,){/Straight}{1}
\FALabel(8.93,7.75)[r]{$d$}
\FAVert(10.,14.5){0}
\FAVert(10.,5.5){0}
\FAVert(10.,10.){0}
\end{feynartspicture}
\begin{feynartspicture}(310,310)(1,1)
\FADiagram{}
\FAProp(0.,15.)(4.5,10.)(0.,){/Straight}{1}
\FALabel(1.57789,11.9431)[tr]{$u$}
\FAProp(0.,5.)(4.5,10.)(0.,){/Cycles}{0}
\FALabel(2.92211,6.9431)[tl]{$g$}
\FAProp(20.,17.)(13.,14.5)(0.,){/ScalarDash}{-1}
\FALabel(15.9787,16.7297)[b]{$\tilde u_\alpha$}
\FAProp(20.,10.)(10.85,8.4)(0.,){/ScalarDash}{-1}
\FALabel(18.4569,10.7663)[b]{$\tilde d_\beta$}
\FAProp(20.,3.)(13.,14.5)(0.,){/Straight}{1}
\FALabel(17.7665,4.80001)[tr]{$d$}
\FAProp(4.5,10.)(10.85,8.4)(0.,){/Straight}{1}
\FALabel(7.29629,8.17698)[t]{$u$}
\FAProp(13.,14.5)(10.85,8.4)(0.,){/Straight}{1}
\FALabel(10.9431,11.9652)[r]{$\tilde \chi_k^\pm$}
\FAVert(4.5,10.){0}
\FAVert(13.,14.5){0}
\FAVert(10.85,8.4){0}
\end{feynartspicture}
\begin{feynartspicture}(310,310)(1,1)
\FADiagram{}
\FAProp(0.,15.)(10.,14.5)(0.,){/Straight}{1}
\FALabel(5.0774,15.8181)[b]{$u$}
\FAProp(0.,5.)(10.,5.5)(0.,){/Cycles}{0}
\FALabel(5.0774,4.18193)[t]{$g$}
\FAProp(20.,17.)(15.5,14.)(0.,){/ScalarDash}{-1}
\FALabel(22.3702,17.3097)[br]{$\tilde u_\alpha$}
\FAProp(20.,10.)(10.,5.5)(0.,){/ScalarDash}{-1}
\FALabel(22.8364,9.43358)[br]{$\tilde d_\beta$}
\FAProp(20.,3.)(15.5,14.)(0.,){/Straight}{1}
\FALabel(18.2486,4.40534)[r]{$d$}
\FAProp(10.,14.5)(10.,5.5)(0.,){/ScalarDash}{1}
\FALabel(8.93,10.)[r]{$\tilde d_\beta$}
\FAProp(10.,14.5)(15.5,14.)(0.,){/Straight}{-1}
\FALabel(12.8903,15.3136)[b]{$\tilde \chi_k^\pm$}
\FAVert(10.,14.5){0}
\FAVert(10.,5.5){0}
\FAVert(15.5,14.){0}
\end{feynartspicture}

\caption{Feynman diagrams for real quark emission. For $q\ne
  q^\prime$ diagrams with $q$ and $q^\prime$ exchanged have to be
  considered, too. For $q=q^\prime$ crossed diagrams have to be taken
  into account. Diagrams in the last row only contribute 
for $u=\{u,c\},\,d=\{d,s\}$.
\label{fig_realquark}}
}

\bibliographystyle{JHEP}
\bibliography{references,%
	      references_SQGL}

\end{document}